\documentclass[twocolumn,showpacs,prb,a4paper,floatfix]{revtex4-1}
\usepackage{amssymb}
\usepackage{graphicx}
\usepackage{amsmath}
\usepackage{graphicx}
\usepackage{dcolumn}
\usepackage{bm}
\usepackage{color}

\newcommand{\bi}{\begin{itemize}}
\newcommand{\ei}{\end{itemize}}
\newcommand{\be}{\begin{equation}}
\newcommand{\ee}{\end{equation}}
\newcommand{\bea}{\begin{eqnarray}}
\newcommand{\eea}{\end{eqnarray}}
\begin{document}

\title{Excitation spectrum and high energy plasmons in single- and
multi-layer graphene}
\author{Shengjun Yuan, Rafael Rold\'an and Mikhail I. Katsnelson}
\affiliation{Institute for Molecules and Materials, Radboud
University Nijmegen, Heyendaalseweg 135, 6525 AJ Nijmegen, The Netherlands}
\date{\today }

\begin{abstract}
In this paper we study the excitation spectrum of single- and multi-layer
graphene beyond the Dirac cone approximation. The dynamical polarizability
of graphene is computed using a full $\pi$-band tight-binding model,
considering the possibility of inter-layer hopping in the calculation. The
effect of electron-electron interaction is considered within the random
phase approximation. We further discuss the effect of disorder in the
spectrum, which leads to a smearing of the absorption peaks. Our results
show a redshift of the $\pi$-plasmon dispersion of single-layer graphene
with respect to graphite, in agreement with experimental results. The
inclusion of inter-layer hopping in the kinetic Hamiltonian of multi-layer
graphene is found to be very important to properly capture the low energy
region of the excitation spectrum.
\end{abstract}

\maketitle


\section{Introduction}

One of the main issues in the understanding of the physics of graphene is
the role played by electron-electron interaction.\cite{KG10} Several
collective modes as low and high energy plasmons, as well as plasmarons, are
a consequence of electronic correlations and have been measured in this
material. The high energy $\pi$-plasmons have been observed in electron
energy-loss spectroscopy (EELS),\cite{KP08,EB08,GG08} inelastic X-ray
scattering (IXS)\cite{RA10} or optical conductivity.\cite{MSH11} Recently, a
plasmaron mode (which is a result of coupling between electrons and
plasmons) has been measured in angle-resoved photoemission spectroscopy
(ARPES).\cite{BR10}

At low energies, long range Coulomb interaction leads, in doped graphene, to
a gapless plasmon mode which disperses as $\omega_{pl}\sim \sqrt{q}$,\cite%
{LS08} and which can be described theoretically within the random phase
approximation (RPA).\cite{S86,WSSG06,HS07,PM08,P09,G11b} The low energy linear
dispersion relation of graphene is at the origin of a new series of
collective modes predicted for this material and which do not exist for
other two-dimensional electron gases (2DEG), as inter-valley plasmons\cite%
{TM10} or linear magneto-plasmons,\cite{RFG09} which can be described within
the RPA as well. For undoped graphene, the inclusion of ladder diagrams in
the polarization can lead to a new class of collective modes\cite{GFM08} as
well as to an excitonic instability.\cite{K01,AKT07,GGG09,WFM10,G10}

However, much less is known about the high energy $\pi$-plasmon which, in
the long wavelength limit, has an energy of the order of 5-6 eV, and which
is due to the presence of Van Hove singularities in the band dispersion. For
single-layer graphene (SLG), this mode has been studied by Stauber et al.%
\cite{SSP10} and by Hill et al.\cite{HMZ09} in the RPA. Yang et al. have
included excitonic effects and found a redshift of the absorption peak,\cite%
{YL09} leading to a better agreement with the experimental results.\cite%
{EB08} Here we extend those previous works and study the excitation spectrum
of SLG and multi-layer graphene (MLG) from a tight-binding model on a
honeycomb lattice. By means of the Kubo formula, the non-interacting
polarization function $\Pi(\mathbf{q},\omega)$ is obtained from the
numerical solution of the time-dependent Schr\"odinger equation. Coulomb
interactions are considered in the RPA, the validity of which is discussed.
We also consider the effect of disorder in the system, which lead to a
considerable smearing of the Van Hove singularities in the spectrum. Our
results show a redshift of the $\pi$-plasmon mode in graphene with respect
to graphite, as it has been observed in the experiments.\cite{KP08,EB08}
Furthermore, the inclusion of inter-layer hopping is found to be very
important to capture the low energy region of the spectrum in MLG.

The paper is organized as follows. In Sec. \ref{Sec:Method} we describe the
method that we use to compute the dynamical polarization function of SLG and
MLG. In Sec. \ref{Sec:SLG} we give results for the excitation spectrum of
SLG, considering the effect of disorder and electron-electron interaction.
The spectrum of MLG is described in Sec. \ref{Sec:MLG}. In Sec. \ref%
{Sec:ComparExp} we compare our results to recent experimental data. Our main
conclusions are summarized in Sec. \ref{Sec:Conclusions}.

\section{Description of the method}

\label{Sec:Method}

\begin{figure}[t]
\begin{center}
\mbox{
\includegraphics[width=4cm]{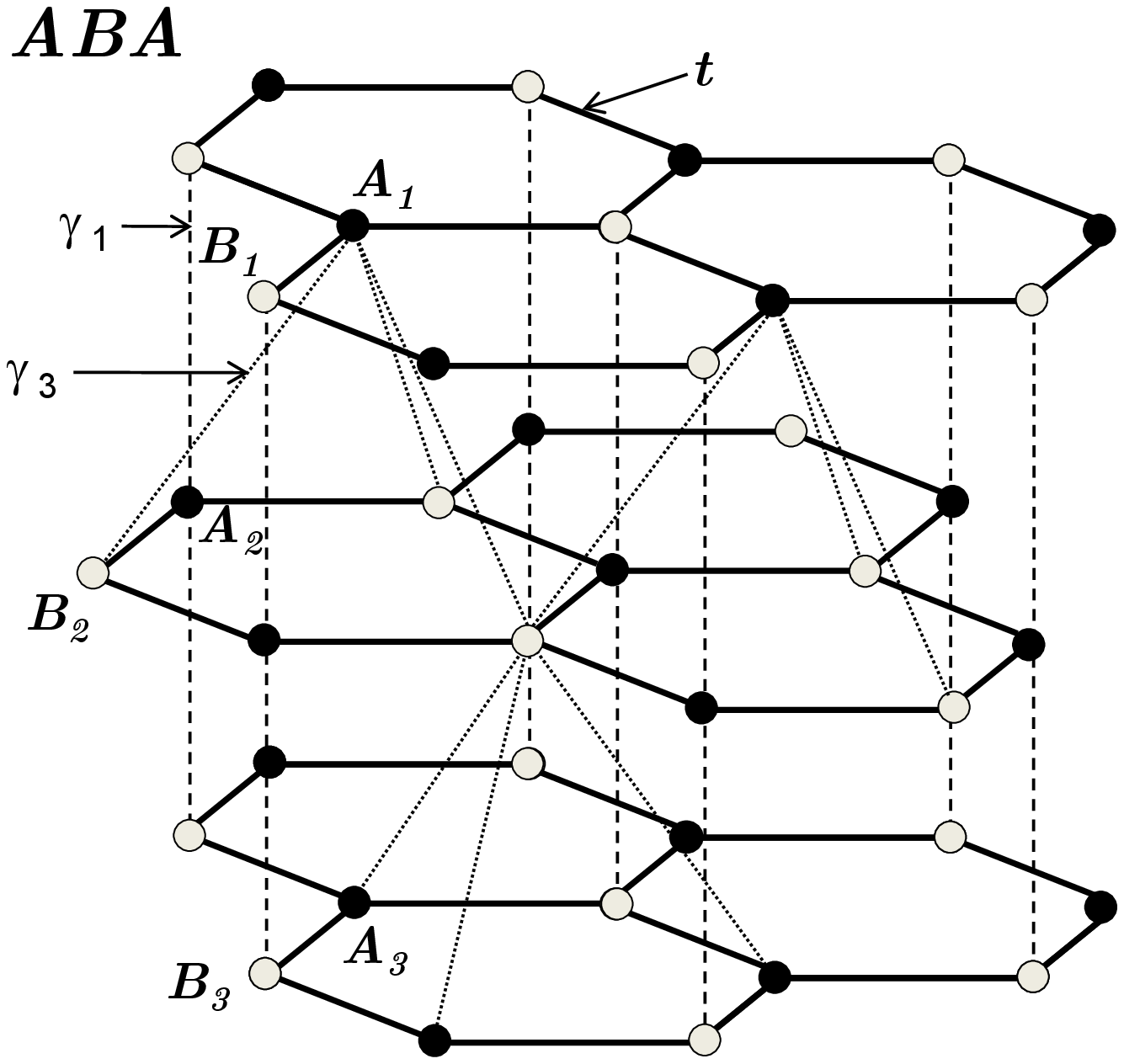}
\includegraphics[width=4cm]{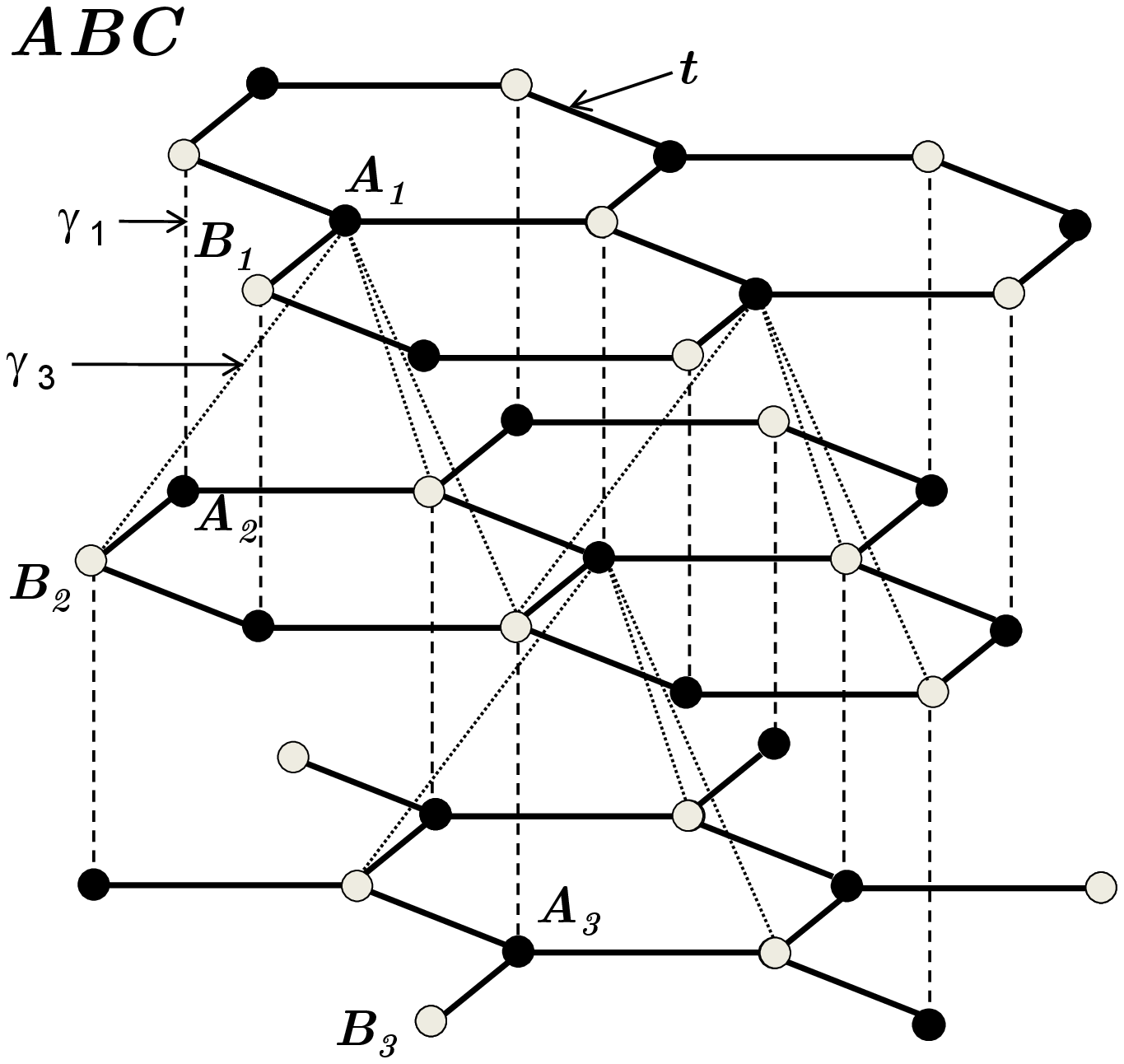}
}
\end{center}
\caption{Atomic structure of ABA- and ABC-stacked multilayer graphene. The
intra-layer ($t$) and inter-layer ($\protect\gamma_1$ and $\protect\gamma_3$%
) hopping amplitudes are considered, as explained in the text.}
\label{Fig:Stacking}
\end{figure}

The tight-binding Hamiltonian of a MLG is given by 
\begin{equation}
H=\sum_{l=1}^{N_{layer}}H_{l}+\sum_{l=1}^{N_{layer}-1}H_{l}^{\prime },
\label{Hamiltonian}
\end{equation}%
where $H_{l}$ is the Hamiltonian of the $l$'th layer of graphene,%
\begin{equation}
H_{l}=-\sum_{<i,j>}(t_{l,ij}a_{l,i}^{\dagger }b_{l,j}+\mathrm{h.c}%
)+\sum_{i}v_{l,i}c_{l,i}^{\dagger }c_{l,i},  \label{Hamiltonian_SLG}
\end{equation}%
where $a_{l,i}^{\dagger }$ ($b_{l,i}$) creates (annihilates) an electron on
sublattice A (B) of the $l$'th layer, and $t_{l,ij}$ is the nearest neighbor
hopping parameter. The second term of $H_l$ accounts for the effect of an
on-site potential $v_{l,i}$, where $n_{l,i}=c^{\dagger}_{l, i}c_{l, i}$ is
the occupation number operator. In the second term of the Hamiltonian Eq. (%
\ref{Hamiltonian}), $H_{l}^{\prime }$ describes the hopping of electrons
between layers $l$ and $l+1$. In nature there are two known forms of stable
stacking sequence in bulk graphite, namely ABA (Bernal) and ABC
(rhombohedral) stacking, and they are schematically shown in Fig. \ref%
{Fig:Stacking}. For a MLG with an ABA stacking, $H_{l}^{\prime }$ is given by%
\begin{equation}
H_{l}^{\prime }=-\gamma _{1}\sum_{j}\left[ a_{l,j}^{\dagger }b_{l+1,j}+%
\mathrm{h.c.}\right] -\gamma _{3}\sum_{j,j^{\prime }}\left[ b_{l,j}^{\dagger
}a_{l+1,j^{\prime }}+\mathrm{h.c.}\right] ,  \label{Eq:H-interlayer}
\end{equation}%
where the inter-layer hopping terms $\gamma_1$ and $\gamma_3$ are shown in
Fig. \ref{Fig:Stacking}. Thus, all the even layers ($l+1$) are rotated with
respect to the odd layers ($l$) by $+120^{\circ }$. The difference between
ABA and ABC stacking is that, the third layer(s) is rotated with respect to
the second layer by $-120^{\circ }$ (then it will be exactly under the first
layer) in ABA stacking, but by $+120^{\circ }$ in ABC stacking.\cite{K10} In
this paper we use the hopping amplitudes $t=3$~eV, $\gamma _{1}=0.4$~eV and $%
\gamma _{3}=0.3$~eV.\cite{CG07} The spin degree of freedom contributes only
through a degeneracy factor and is omitted for simplicity in Eq.~(\ref%
{Hamiltonian}). In our numerical calculations, we use periodic boundary
conditions in the plane ($XY$) of graphene layers, and open boundary
conditions in the stacking direction ($Z$).

\begin{figure}[t]
\begin{center}
\includegraphics[width=6.5cm]{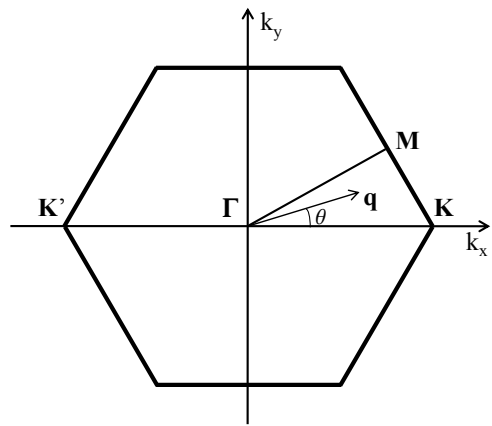}
\end{center}
\caption{2D Brillouin zone of SLG. For undoped graphene, the valence and
conduction bands touch each other at the vertices of the hexagon, the so
called Dirac points (K and K'). The Van Hove singularity lies at the M
point, and we have defined $\protect\theta$ as the angle between the
wave-vector $q$ and the $k_x$-axis.}
\label{Fig:BZ}
\end{figure}

The dynamical polarization can be obtained from the Kubo formula \cite{K57}
as%
\begin{equation}
\Pi \left( \mathbf{q},\omega \right) =\frac{i}{V}\int_{0}^{\infty }d\tau
e^{i\omega \tau}\left\langle \left[ \rho \left( \mathbf{q},\tau\right) ,\rho
\left( -\mathbf{q},0\right) \right] \right\rangle ,  \label{Eq:Kubo}
\end{equation}%
where $V$ denotes the volume (or area in 2D) of the unit cell, $\rho \left( 
\mathbf{q}\right) $ is the density operator given by%
\begin{equation}
\rho \left( \mathbf{q}\right)
=\sum_{l=1}^{N_{layer}}\sum_{i}c_{l,i}^{\dagger}c_{l,i}\exp \left( i\mathbf{%
q\cdot r}_{l,i}\right) ,
\end{equation}%
and the average is taken over the canonical ensemble. For the case of the
single-particle Hamiltonian, Eq.~(\ref{Eq:Kubo}) can be written as\cite%
{YRK10}%
\begin{eqnarray}
&&\Pi \left( \mathbf{q},\omega \right) =-\frac{2}{V}\int_{0}^{\infty }d\tau
e^{i\omega \tau}  \notag  \label{Eq:Kubo2} \\
&&\times \text{Im}\left\langle \varphi \right\vert n_{F}\left( H\right)
e^{iH\tau}\rho \left( \mathbf{q}\right) e^{-iH\tau}\left[ 1-n_{F}\left(
H\right) \right] \rho \left( -\mathbf{q}\right) \left\vert \varphi
\right\rangle ,  \notag \\
&&
\end{eqnarray}%
where $n_{F}\left( H\right) =\frac{1}{e^{\beta \left( H-\mu \right) }+1}$ is
the Fermi-Dirac distribution operator, $\beta =1/k_{B}T$ where $T$ is the
temperature and $k_{B}$ is the Boltzmann constant, and $\mu $ is the
chemical potential. In the numerical simulations, we use units such that $%
\hbar =1 $, and the average in Eq.~(\ref{Eq:Kubo2}) is performed over a
random phase superposition of all the basis states in the real space, i.e.,%
\cite{HR00,YRK10} 
\begin{equation}
\left\vert \varphi \right\rangle =\sum_{l,i}a_{l,i}c_{l,i}^{\dagger
}\left\vert 0\right\rangle ,  \label{Eq:phi0}
\end{equation}%
where $a_{l,i}$ are random complex numbers normalized as $%
\sum_{l,i}\left\vert a_{l,i}\right\vert ^{2}=1$. By introducing the time
evolution of two wave functions 
\begin{eqnarray}
\left\vert \varphi _{1}\left( \mathbf{q,}\tau\right) \right\rangle
&=&e^{-iH\tau}\left[ 1-n_{F}\left( H\right) \right] \rho \left( -\mathbf{q}%
\right) \left\vert \varphi \right\rangle , \\
\left\vert \varphi _{2}\left( \tau\right) \right\rangle
&=&e^{-iH\tau}n_{F}\left( H\right) \left\vert \varphi \right\rangle ,
\end{eqnarray}%
we get the real and imaginary part of the dynamical polarization as 
\begin{eqnarray}  \label{Eq:RePi-ImPi}
\text{Re}\Pi \left( \mathbf{q},\omega \right) &=&-\frac{2}{V}%
\int_{0}^{\infty }d\tau\cos (\omega \tau)~\text{Im}\left\langle \varphi
_{2}\left( \tau\right) \left\vert \rho \left( \mathbf{q}\right) \right\vert
\varphi _{1}\left( \tau\right) \right\rangle ,  \notag \\
\text{Im}\Pi \left( \mathbf{q},\omega \right) &=&-\frac{2}{V}%
\int_{0}^{\infty }d\tau\sin (\omega \tau)~\text{Im}\left\langle \varphi
_{2}\left( \tau\right) \left\vert \rho \left( \mathbf{q}\right) \right\vert
\varphi _{1}\left( \tau\right) \right\rangle ,  \notag \\
&&
\end{eqnarray}%
The Fermi-Dirac distribution operator $n_{F}\left( H\right) $ and the time
evolution operator $e^{-iH\tau}$ can be obtained by the standard Chebyshev
polynomial decomposition.\cite{YRK10}

For the case of SLG, we will further compare the polarization function
obtained from the Kubo formula Eq. (\ref{Eq:Kubo}), to the one obtained from
the usual Lindhard function.\cite{GV05} Notice that this method can be used to
calculate the optical conductivity of graphene beyond the Dirac cone approximation.
\cite{SPG08,YRK10} For pristine graphene, the dynamical
polarization obtained from the Lindhard function using the full $\pi $-band
tight-binding model is\cite{S86,HMZ09,SSP10} 
\begin{eqnarray}  \label{Eq:Lindhard}
&&\Pi \left( \mathbf{q},\omega \right) =-\frac{g_{s}}{\left( 2\pi \right)
^{2}}\int_{BZ}d^{2}\mathbf{k}  \notag  \label{Eq:PolarizationFull} \\
&&\times \sum_{s,s^{\prime }=\pm }f_{s\cdot s^{\prime }}\left( \mathbf{k},%
\mathbf{q}\right) \frac{n_{F}\left[ E^{s}\left( \mathbf{k}\right) \right]
-n_{F}\left[ E^{s^{\prime }}\left( \mathbf{k}+\mathbf{q}\right) \right] }{%
E^{s}\left( \mathbf{k}\right) -E^{s^{\prime }}\left( \mathbf{k}+\mathbf{q}%
\right) +\omega+i\delta },  \notag \\
&&
\end{eqnarray}%
where the integral is over the Brillouin zone, $g_{s}=2$ is the spin
degeneracy, $E^{\pm }\left( \mathbf{k}\right) =\pm t\left\vert \phi _{%
\mathbf{k}}\right\vert -\mu $ is the energy dispersion with respect to the
chemical potential, where 
\begin{equation}
\phi _{\mathbf{k}}=1+2e^{i3k_{x}a/2}\cos \left( \frac{\sqrt{3}}{2}%
k_{y}a\right) ,
\end{equation}%
$a=1.42$\AA ~ being the in-plane carbon-carbon distance, and the overlap
between the wave-functions of the electron and the hole is given by 
\begin{equation}
f_{\pm }\left( \mathbf{k},\mathbf{q}\right) =\frac{1}{2}\left( 1\pm \text{Re}%
\left[ e^{iq_{x}a}\frac{\phi _{\mathbf{k}}}{\left\vert \phi _{\mathbf{k}%
}\right\vert }\frac{\phi _{\mathbf{k+q}}^{\ast }}{\left\vert \phi _{\mathbf{%
k+q}}\right\vert }\right] \right) .
\end{equation}

In the RPA, the response function of SLG due to electron-electron
interactions can be calculated as 
\begin{equation}
\chi \left( \mathbf{q},\omega \right) =\frac{\Pi \left( \mathbf{q},\omega
\right) }{\mathbf{1}-V\left( q\right) \Pi \left( \mathbf{q},\omega \right) },
\label{Eq:chi}
\end{equation}%
where $V\left( q\right) =\frac{2\pi e^{2}}{\kappa q}$ is the Fourier
component of the Coulomb interaction in two dimensions, in terms of the
background dielectric constant $\kappa $, and 
\begin{equation}
\mathbf{\varepsilon }\left( \mathbf{q},\omega \right) =\mathbf{1}-V\left(
q\right) \Pi \left( \mathbf{q},\omega \right)
\end{equation}%
is the dielectric function of the system. We will be interested on the
collective modes of the system, which are defined from the zeroes of the
dielectric function [$\varepsilon (\mathbf{q},\omega )=0$]. The dispersion
relation of the collective modes is defined from 
\begin{equation}
\mathrm{Re}~\varepsilon (\mathbf{q},\omega _{pl})=1-V(q)\Pi (\mathbf{q}%
,\omega _{pl})=0,  \label{Eq:Plasmons}
\end{equation}%
which leads to poles in the response function (\ref{Eq:chi}). The damping $%
\gamma $ of the mode is proportional to $\mathrm{Im}~\Pi (\mathbf{q},\omega
_{pl})$, and it is given by 
\begin{equation}
\gamma =\frac{\mathrm{Im}~\Pi (\mathbf{q},\omega _{pl})}{\left. \frac{%
\partial }{\partial \omega }\mathrm{Re}~\Pi (\mathbf{q},\omega )\right\vert
_{\omega =\omega _{pl}}}.  \label{Eq:Damping}
\end{equation}

For MLG, the response function is calculated as (we use $q_{z}=0$)\cite{S86} 
\begin{equation}  \label{Eq:chi3D}
\chi _{3D}\left( \mathbf{q},\omega \right) =\frac{\Pi _{3D}\left( \mathbf{q}%
,\omega \right) }{\mathbf{1}-V\left( q\right) F\left( q\right) \Pi
_{3D}\left( \mathbf{q},\omega \right) d},
\end{equation}%
where $d=3.35$\AA ~ is the inter-layer separation. Because we use open
boundary conditions in the stacking direction, we define the form factor $%
F\left( q\right) $ as 
\begin{equation}
F\left( q\right) =\frac{1}{N_{layer}}\sum_{l,l^{\prime
}=1}^{N_{layer}}e^{-q\left\vert l-l^{\prime }\right\vert d}.
\label{Eq:FormFactor}
\end{equation}%
The expression (\ref{Eq:chi3D}) assumes that the polarization of each layer
is the same, and it is exact in two different limits: bilayer graphene and
graphite. Notice that a similar effective form factor has been used to study
the loss function of multiwall carbon nanotubes.\cite{SL00} Eq. (\ref%
{Eq:FormFactor}) coincides with the commonly used form factor for a
multi-layer system with an infinite number of layers:\cite{SQ82} 
\begin{equation}
F\left( q\right)|_{N_{layer}\rightarrow \infty} =\sum_{l^{\prime
}}e^{-q\left\vert l-l^{\prime }\right\vert d},
\end{equation}%
where in this last case the periodicity ensures that $F\left( q\right) $ is
independent of layer index $l$, with the asymptotic behavior $%
F(q)=\sinh(qd)/[\cosh(qd)-1]$.\cite{SQ82}

A crucial issue is the value of the dielectric constant $\kappa $ for each
of the cases considered, because it encodes the screening due to high energy
($\sigma $) bands which are not explicitly considered in our calculation. A
good estimation for it can be obtained from the expression\cite{WB11} 
\begin{equation}
\kappa \left( \mathbf{q}\right) =\frac{\kappa _{1}+1-\left( \kappa
_{1}-1\right) e^{-qL}}{\kappa _{1}+1+\left( \kappa _{1}-1\right) e^{-qL}}%
\kappa _{1},  \label{Eq:kappa}
\end{equation}%
where $\kappa _{1}\approx 2.4$ is the dielectric constant of graphite, $%
L=d_{m}+\left( N_{layer}-1\right) d$ is the total height of the multi-layer
system in terms of the number of layers $N_{layer}$ and the height of a
monolayer graphene $d_{m}\approx 2.8$~\AA . As expected, Eq. (\ref{Eq:kappa}%
) gives $\kappa =1$ for SLG at $q\rightarrow 0$ and $\kappa =\kappa _{1}$
for graphite.

We notice that the accuracy of the numerical results for the polarization
function Eq. (\ref{Eq:RePi-ImPi}) is mainly determined by three factors: the
time interval of the propagation, the total number of time steps, and the
size of the sample. The maximum time interval of the propagation in the time
evolution operator is determined by the Nyquist sampling theorem. This
implies that employing a sampling interval $\Delta \tau=\pi
/\max_{i}\left\vert E_{i}\right\vert $, where $E_{i}$ are the eigenenergies,
is sufficient to cover the full \textit{range} of energy eigenvalues. On the
other hand, the \textit{accuracy} of the energy eigenvalues is determined by
the total number of the propagation time steps ($N_{\tau}$) that is the
number of the data items used in the fast Fourier transform (FFT).
Eigenvalues that differ less than $\Delta E=\pi /N_{\tau}\Delta \tau$ cannot
be identified properly. However, since $\Delta E$ is proportional to $%
N_{\tau}^{-1}$ we only have to double the length of the calculation to
increase the accuracy by the same factor. The statistic error of our
numerical method is inversely proportional to the dimension of the Hilbert
space,\cite{HR00} and in our case (the single particle representation), it
is the number of sites in the sample. A sample with more sites in the real
space will have more random coefficients ($a_{l,i}$) in the initial state $%
\left\vert \varphi \right\rangle $, providing a better statistical
representation of the superposition of all energy eigenstates.\cite{YRK10}

Similar algorithm has been successfully used in the numerical calculation of
the electronic structure and transport properties of single- and multi-layer
graphene, such as the density of states (DOS), or \textit{dc} and \textit{ac}
conductivities.\cite{WK10,YRK10,YRK10b} The main advantage of our algorithm
is that different kinds of disorders and boundary conditions can be easily
introduced in the Hamiltonian, and the computer memory and CPU time is
linearly proportional to the size of the sample, which allows us to do the
calculations on a sample containing tens of million sites.

\section{Excitation spectrum of single-layer graphene}

\label{Sec:SLG}

\begin{figure}[t]
\begin{center}
\mbox{
\includegraphics[width=4.2cm]{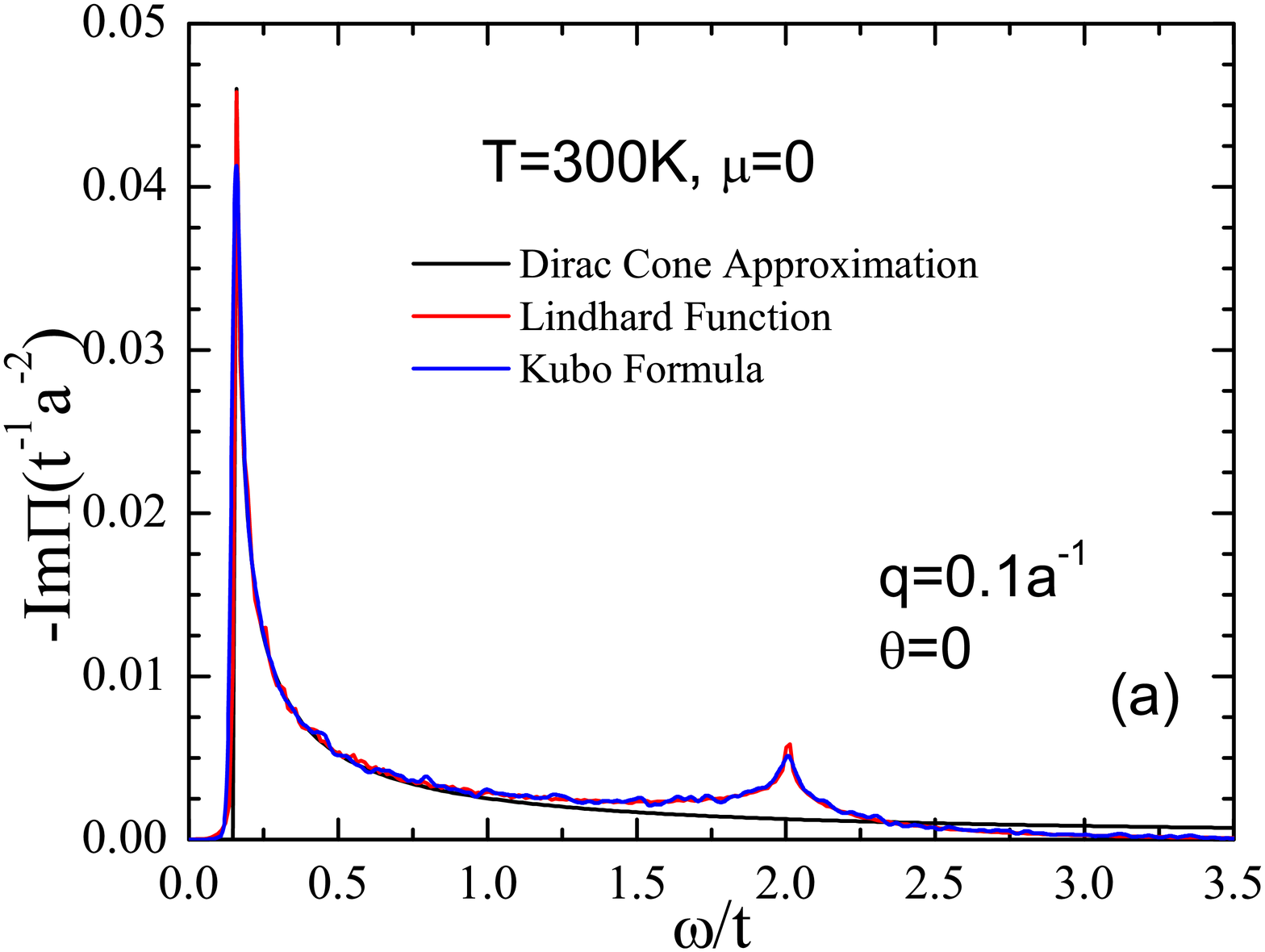}
\includegraphics[width=4.2cm]{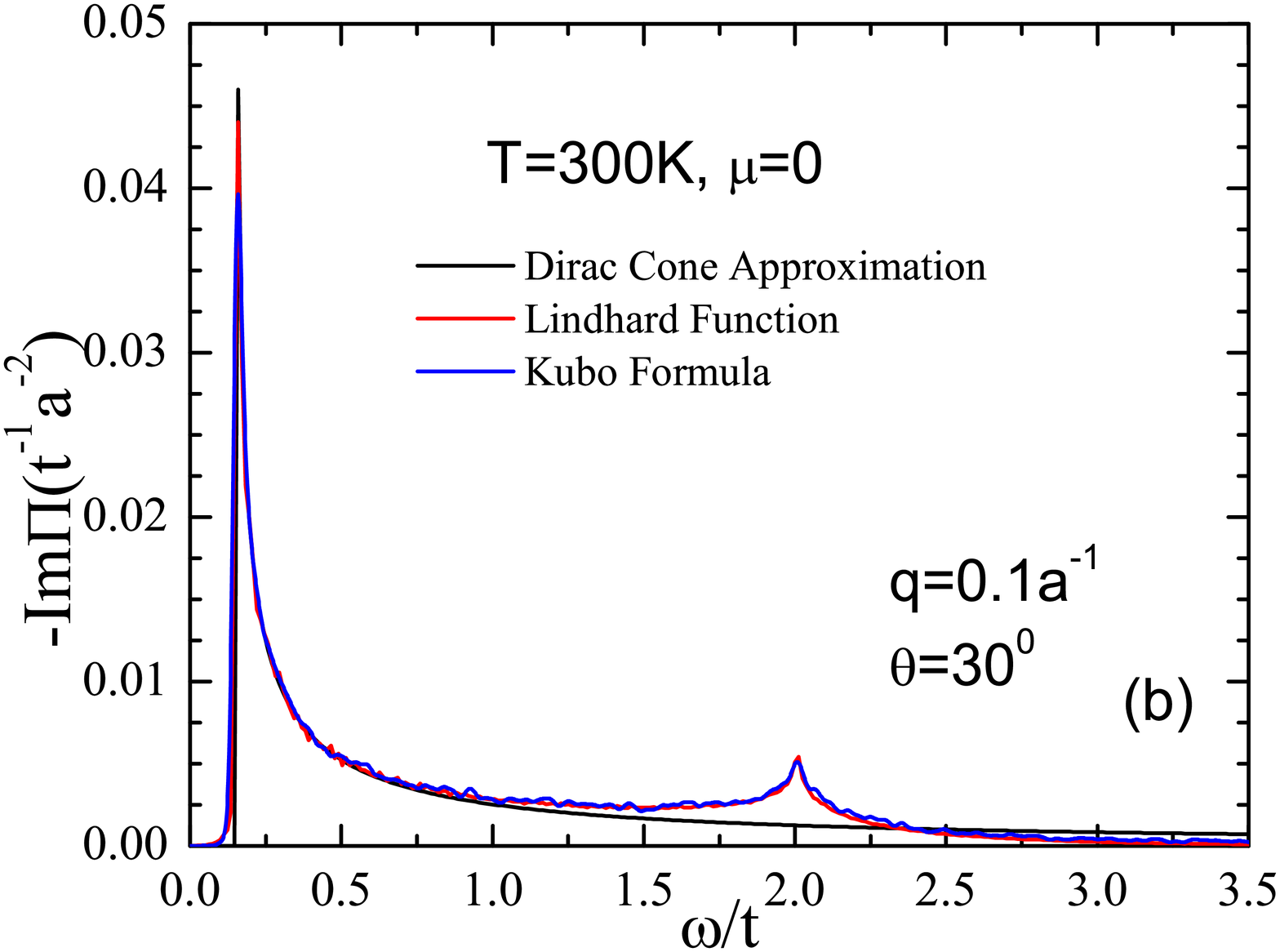}
} 
\mbox{
\includegraphics[width=4.2cm]{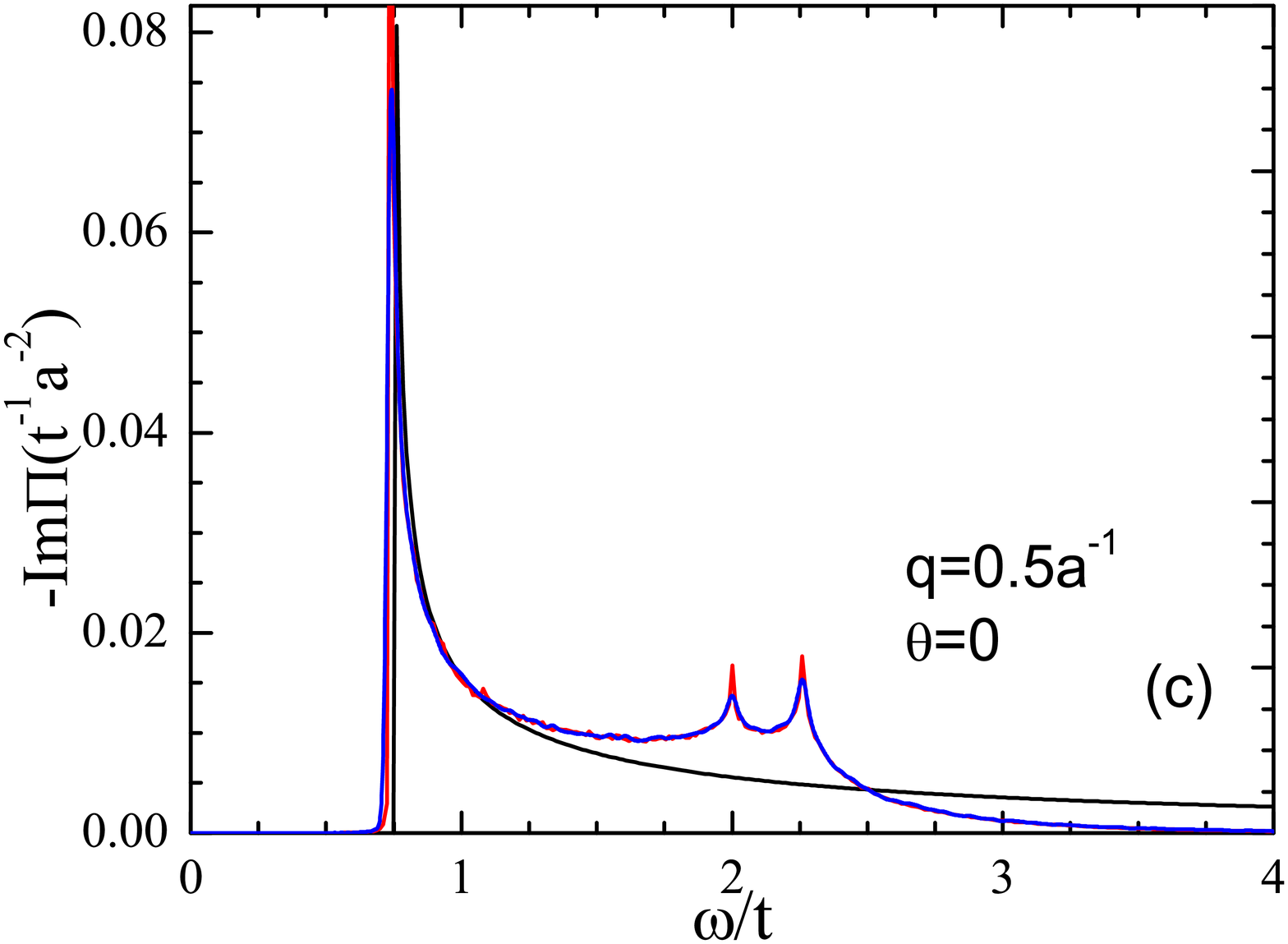}
\includegraphics[width=4.2cm]{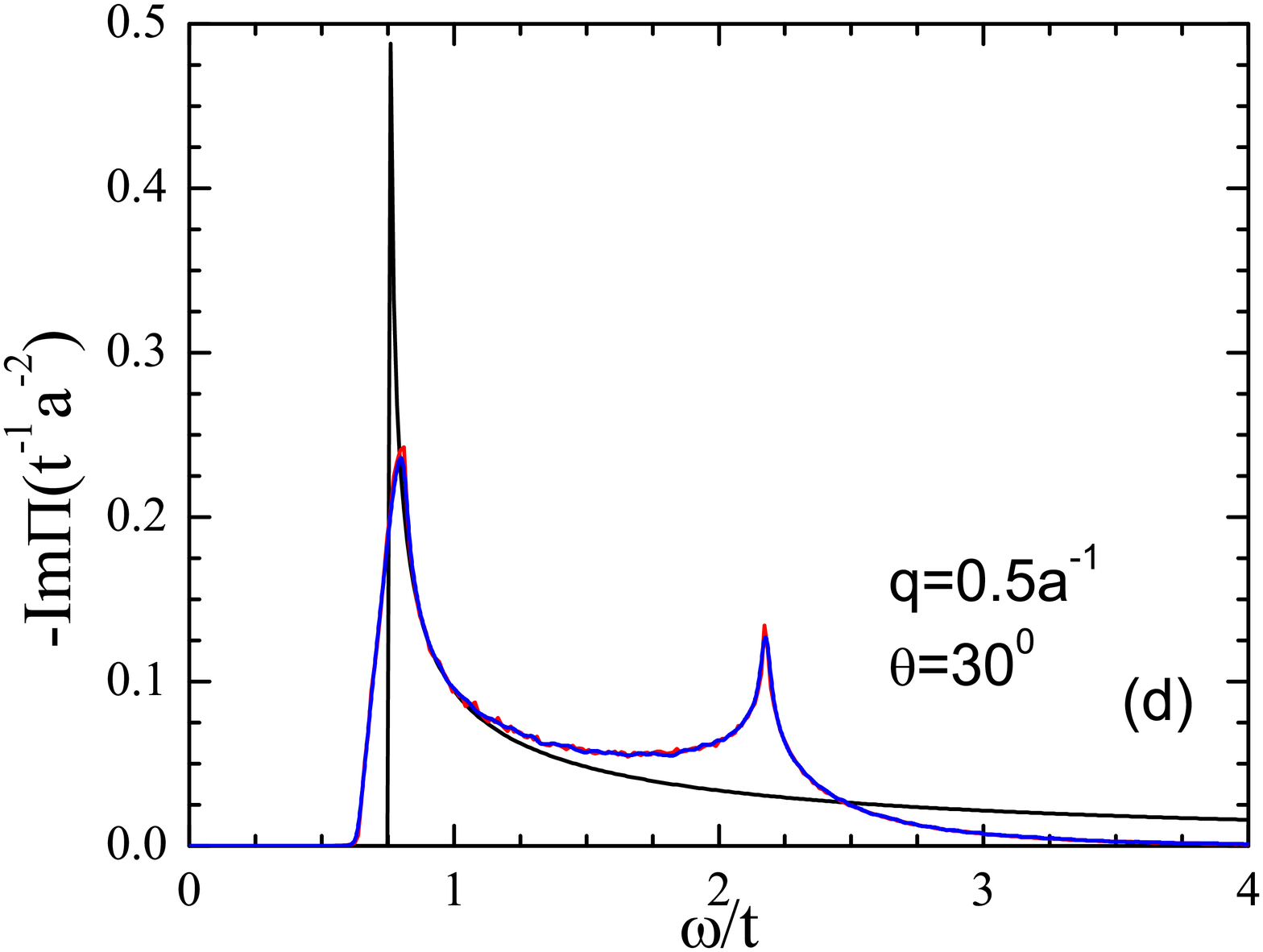}
} 
\mbox{
\includegraphics[width=4.2cm]{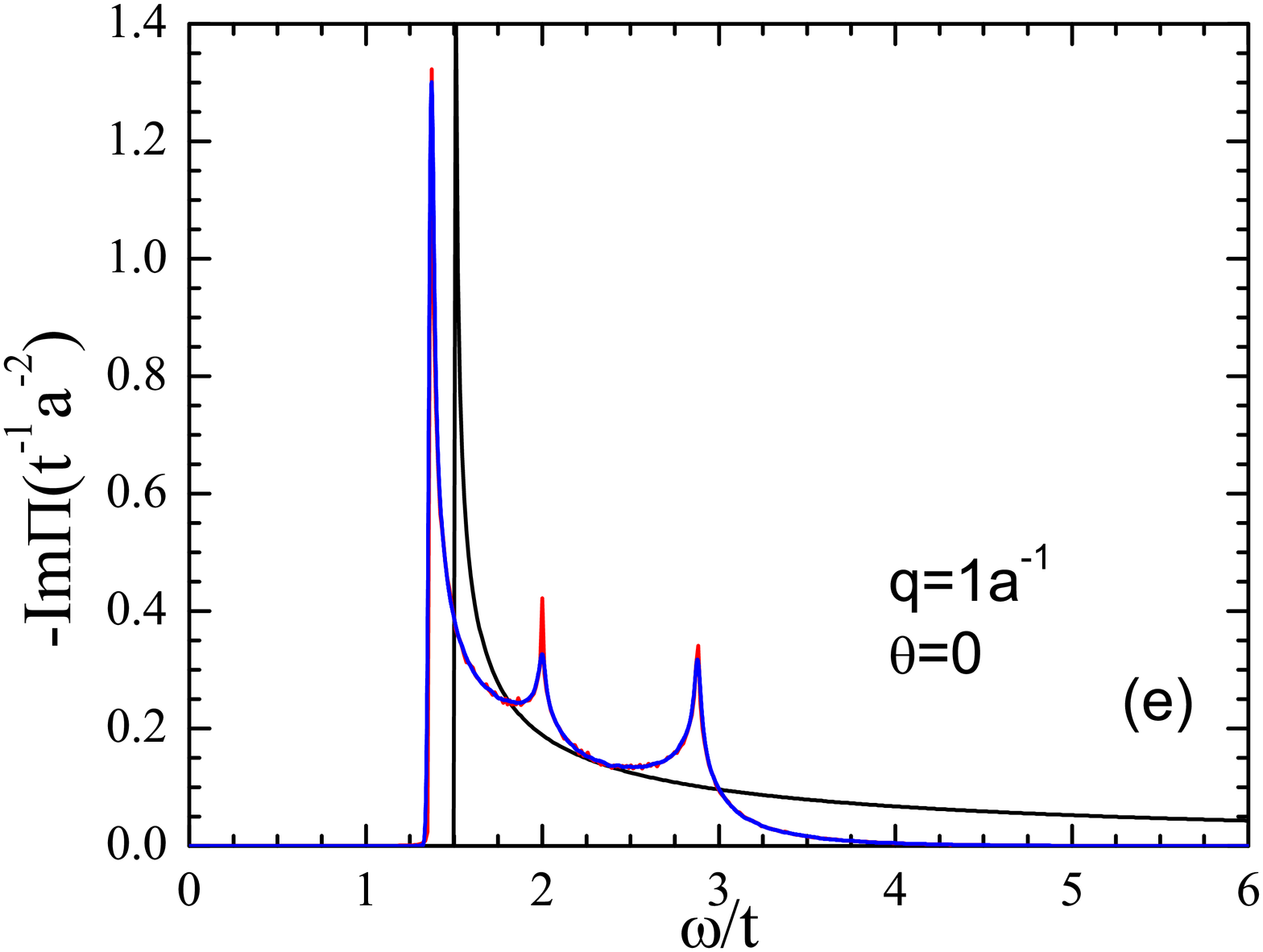}
\includegraphics[width=4.2cm]{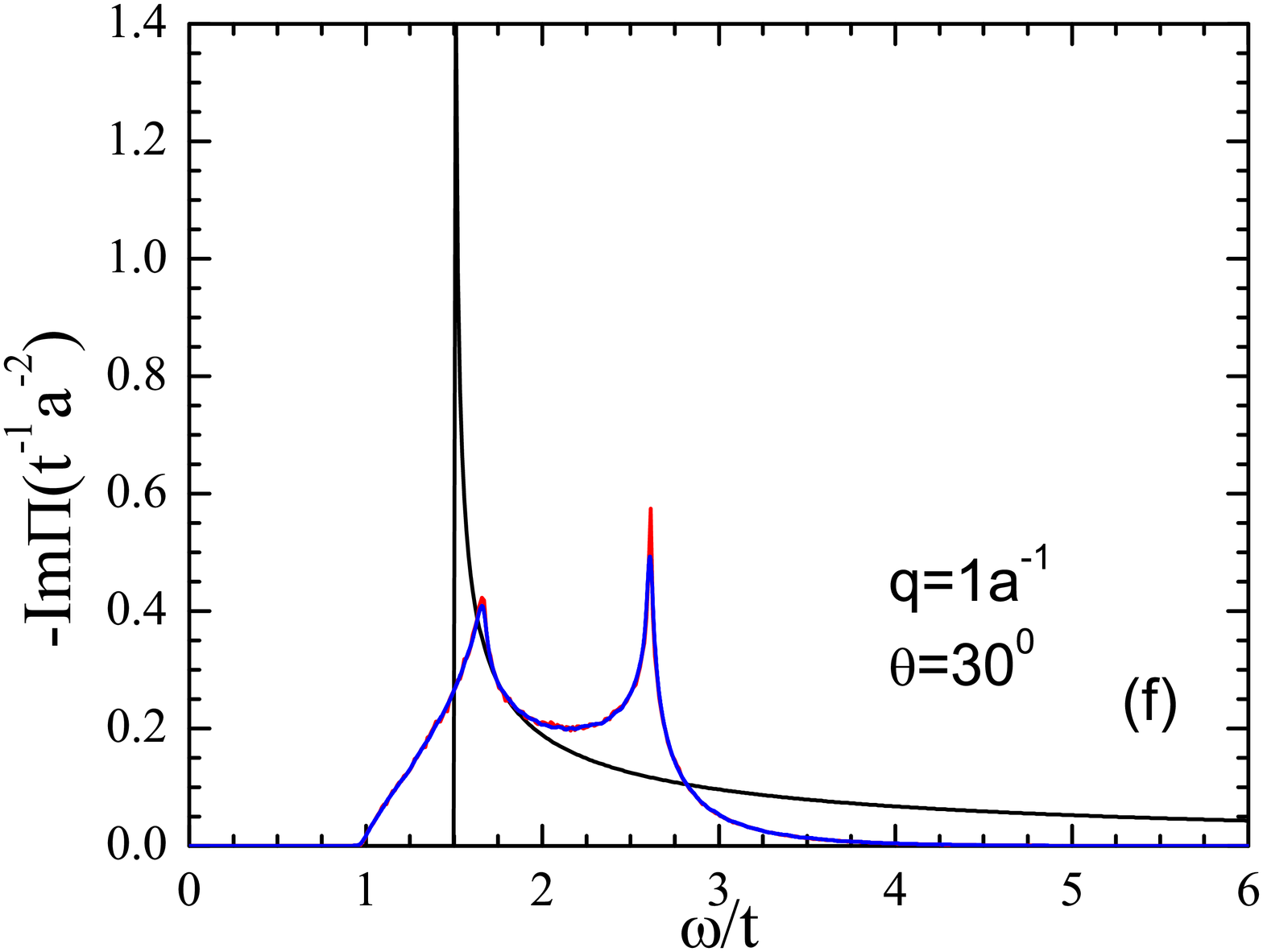}
}
\end{center}
\caption{(Color online) $-\mathrm{Im}~\Pi (\mathbf{q},\protect\omega )$ for
SLG in the clean limit for different values of wave-vector $\mathbf{q}$.
Plots (a), (c) and (e) correspond to a wave-vector $\mathbf{q}$ along $%
\Gamma $-K, whereas (b), (d) and (f) are of a $\mathbf{q}$ parallel to the $%
\Gamma $-M direction. The angle $\protect\theta $ is defined in Fig. \protect
\ref{Fig:BZ}. In the numerical integration of Lindhard function in Eq. (%
\protect\ref{Eq:PolarizationFull}), we use $2\times 10^{8}$ Monte Carlo
points (k) in the first Brilion. The sample size of SLG used in the
numerical calculation of Kubo formula in Eq. (\protect\ref{Eq:Kubo2}) is $%
4096\times 4096.$}
\label{Fig:ImPiSLG}
\end{figure}

\begin{figure}[t]
\begin{center}
\includegraphics[width=6.5cm]{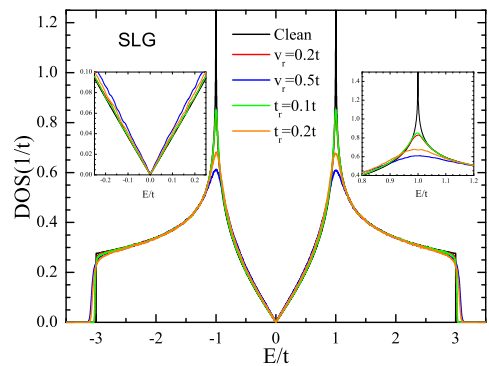}
\end{center}
\caption{(Color online) Density of states for SLG considering different kind
of disorder. The left inset shows a zoom of the DOS near the Dirac point ($%
E=0$), whereas the right hand side inset shows the disorder broadening of
the Van Hove singularity at $E=t$. The numerical method used in the
calculation of DOS is discribed in Ref. \onlinecite{YRK10}, and the sample
size of SLG is $4096\times 4096.$}
\label{Fig:DOS-disorder}
\end{figure}

The particle-hole excitation spectrum is the region of the energy-momentum
space which is available for particle-hole excitations. For non-interacting
electrons, it is defined as the region where $\mathrm{Im}~\Pi (\mathbf{q}%
,\omega )$, as given by Eq. (\ref{Eq:Kubo}) or (\ref{Eq:Lindhard}), is
non-zero.\cite{GV05} The linear low energy dispersion relation of graphene
as well as the possibility for inter-band transitions lead to a rather
peculiar excitation spectrum for SLG as compared to the one of a
two-dimensional electron gas (2DEG) with a parabolic band dispersion.\cite%
{RGF10} Here we focus on undoped graphene ($\mu =0$), for which only
inter-band transitions are allowed. In Fig. \ref{Fig:ImPiSLG} we plot $%
\mathrm{Im}~\Pi (\mathbf{q},\omega )$ for different wave-vectors at $T=300K$
(which is the temperature that we will use from here on in our results). The
first thing one observes is the good agreement between the results obtained
from the Kubo formula Eq. (\ref{Eq:Kubo}), as compared to Lindhard function
Eq. (\ref{Eq:Lindhard}), what proofs the validity of our numerical method.
Furthermore, for the small wave-vector used in Fig. \ref{Fig:ImPiSLG}%
(a)-(b), the results are well described by the Dirac cone approximation,\cite%
{WSSG06,HS07} but only at low energies, around $\omega \sim v_{\mathrm{F}}q$%
, where $v_{\mathrm{F}}=3at/2$ is the Fermi velocity near the Dirac points.
In particular, the continuum approximation cannot capture the peaks of $%
\mathrm{Im}\Pi (\mathbf{q},\omega )$ around $\omega \approx 2t$.
These peaks are related to particle-hole excitations between states of 
the valence band with energy $E\approx -t$ and states of the conduction band 
with energy $E\approx t$, which contribute to the polarization with a strong 
spectral weight due to the enhanced density of states at the Van Hove singularities
of the $\pi $-bands (see Fig. \ref{Fig:DOS-disorder}).

\begin{figure*}[t]
\begin{center}
\mbox{
\includegraphics[width=7cm]{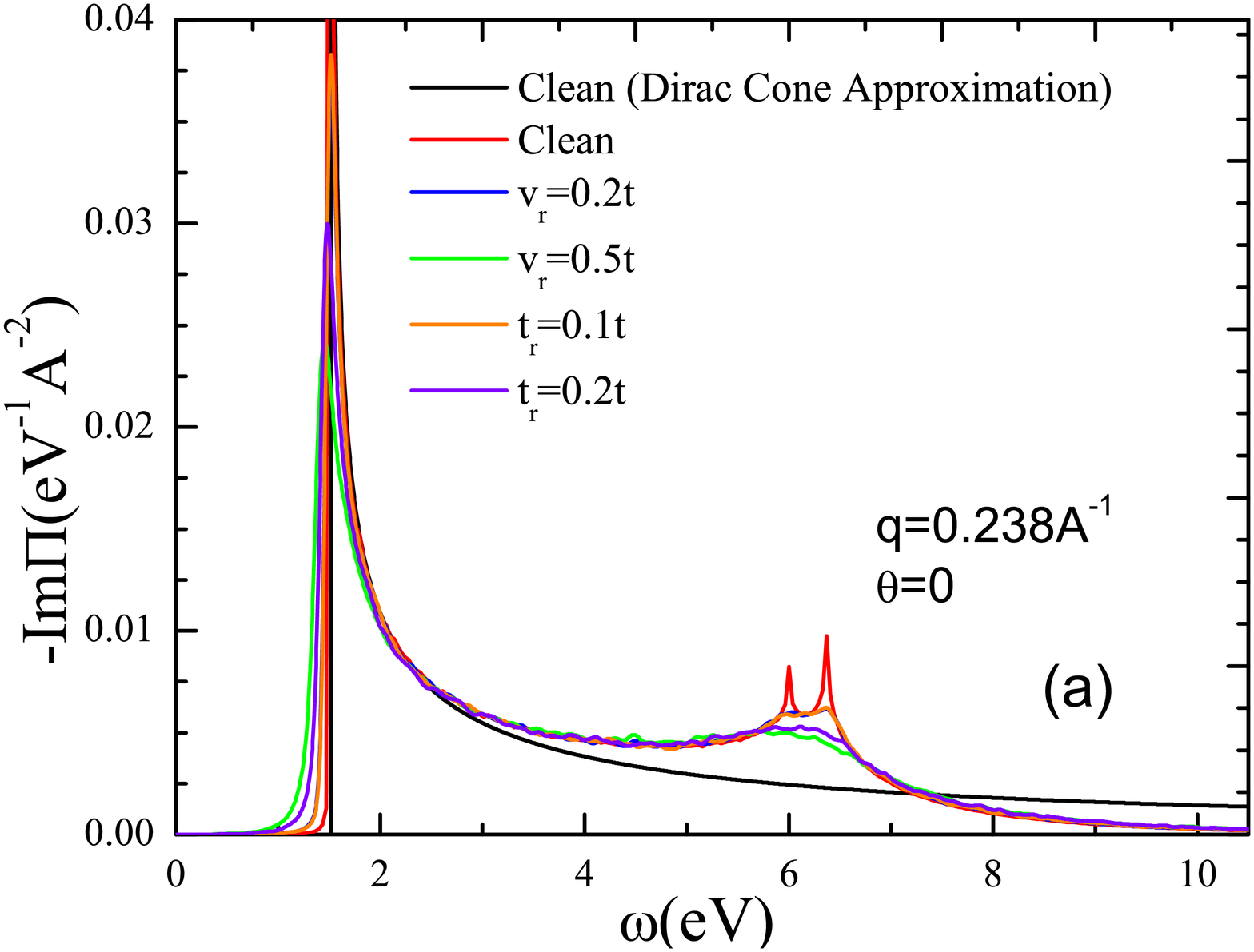}
\includegraphics[width=7cm]{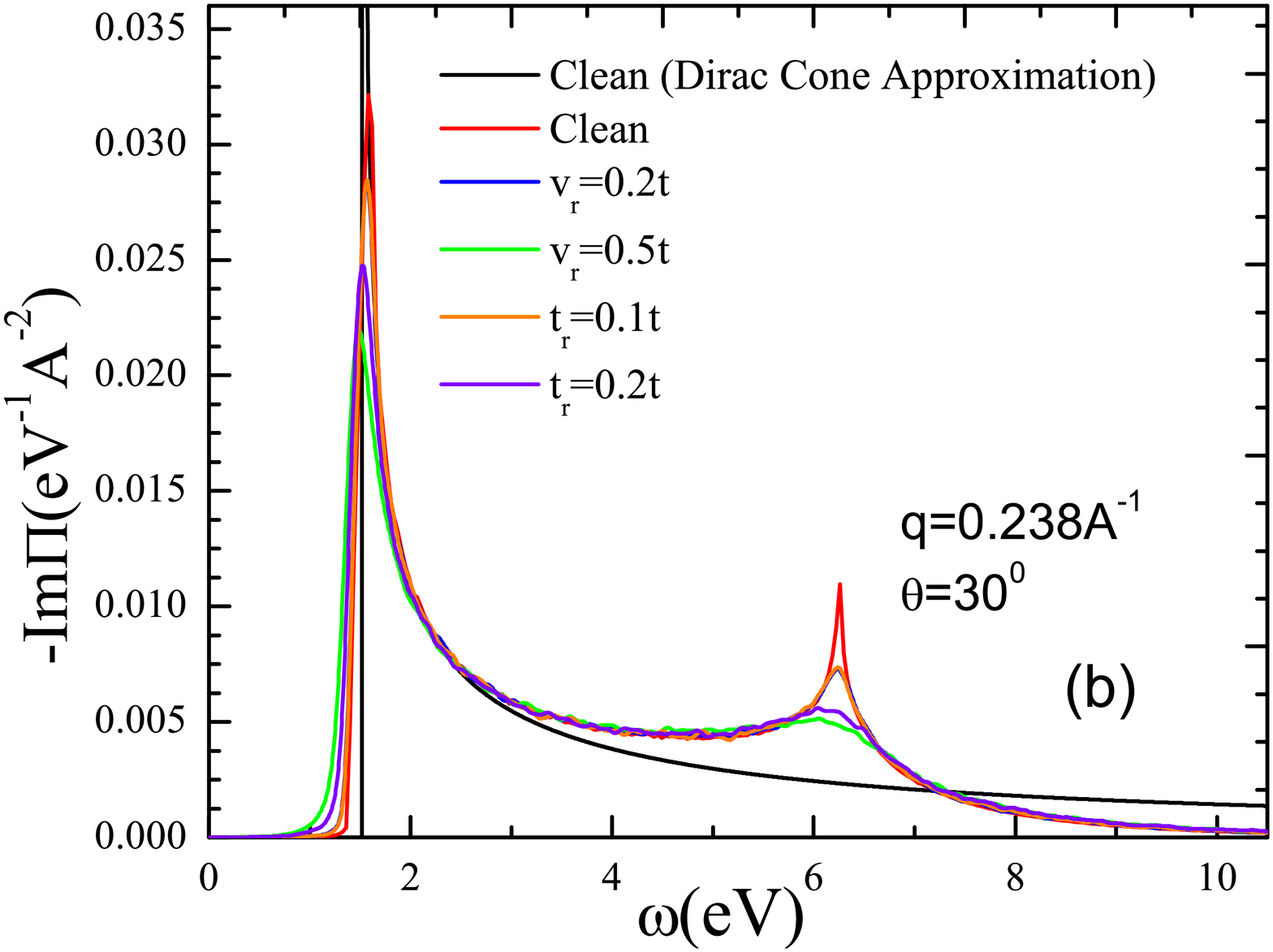}
} 
\mbox{
\includegraphics[width=7cm]{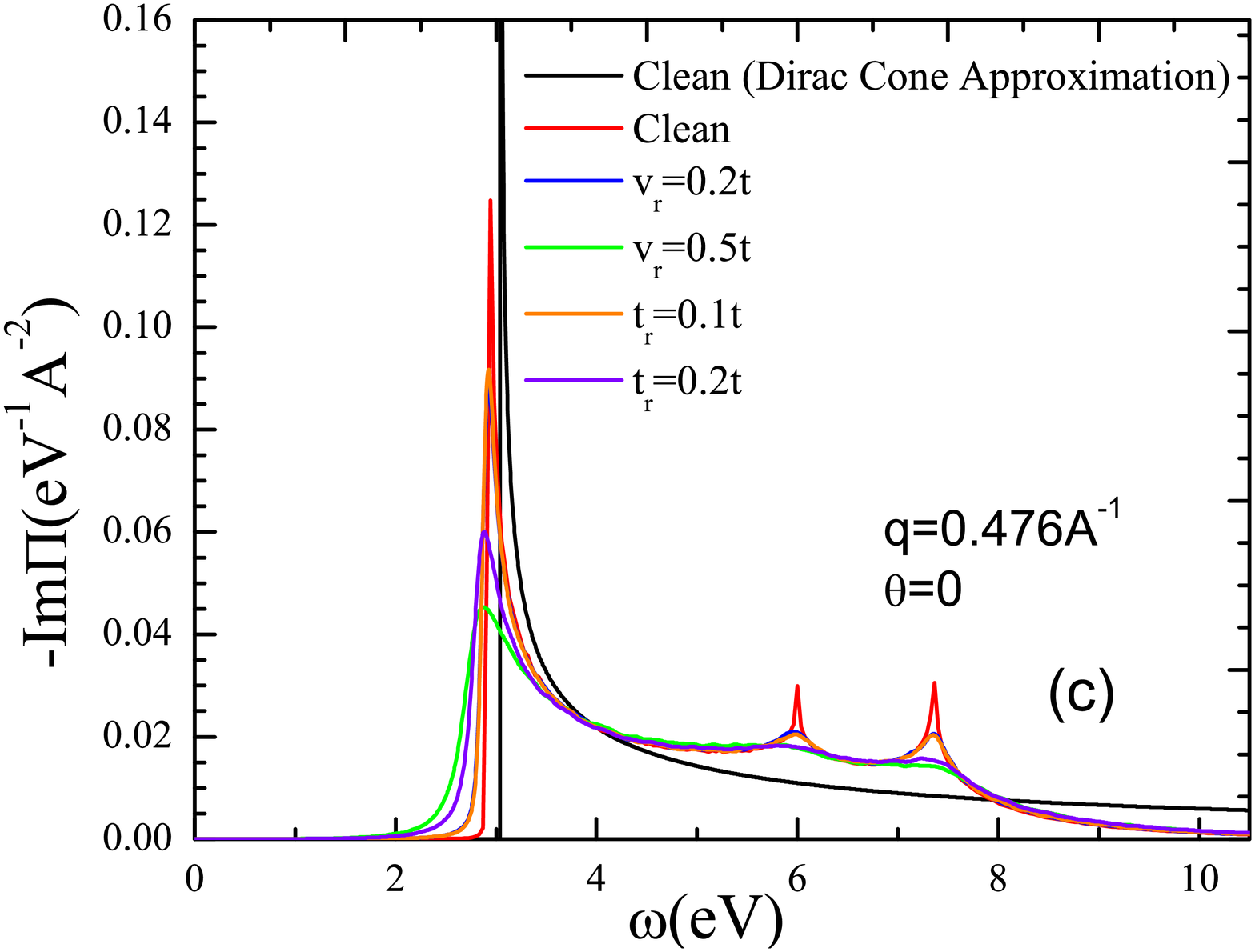}
\includegraphics[width=7cm]{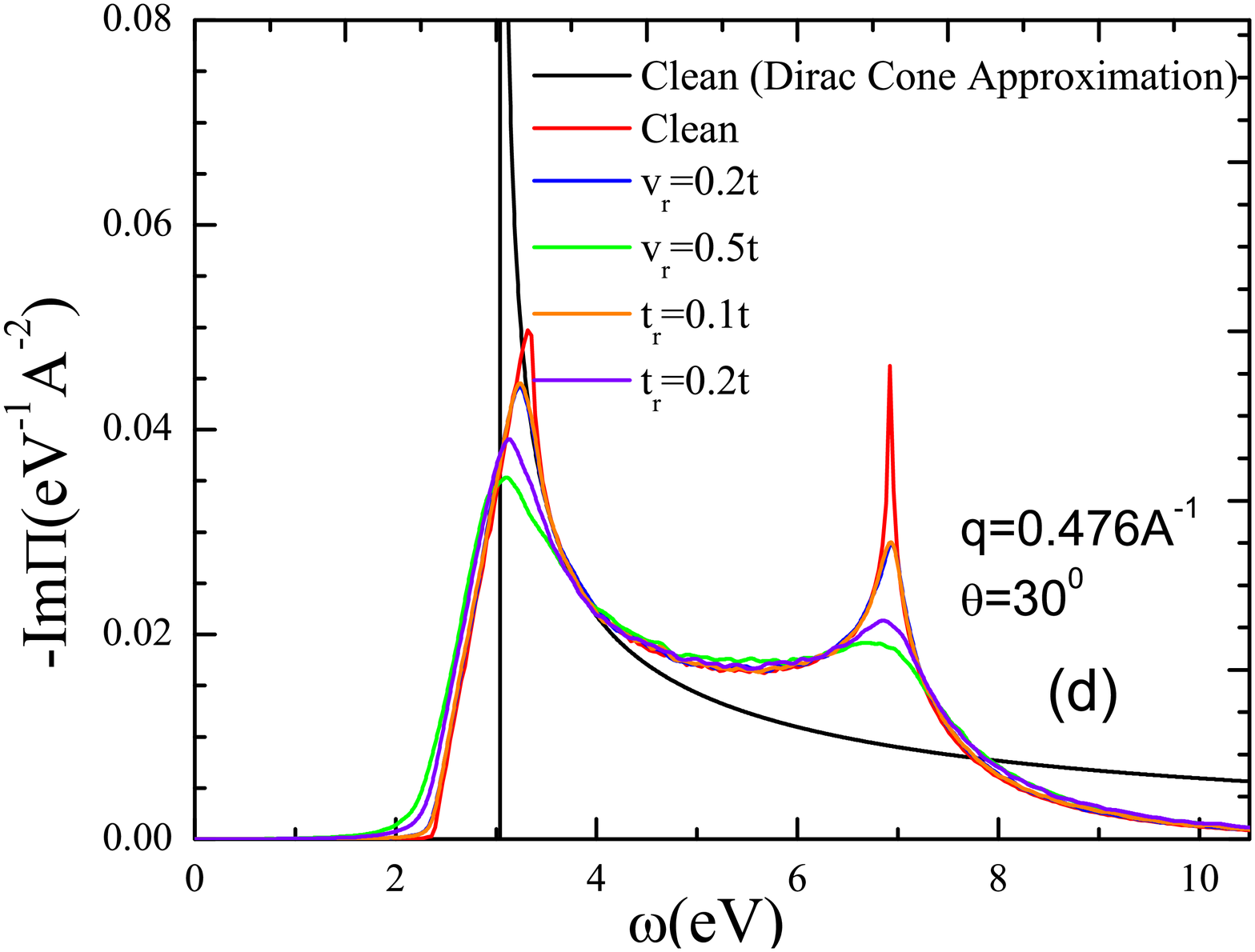}
}
\end{center}
\caption{(Color online) $-\mathrm{Im}~\Pi (\mathbf{q},\protect\omega )$ for
different kinds of disorder and for different values and orientation of the
wave-vector $\mathbf{q}$. In all the plots, the results using the Kubo
formula Eq. (\protect\ref{Eq:Kubo}) are compared to the Dirac cone
approximation. The sample size of SLG is $4096\times 4096.$}
\label{Fig:ImPiSLG-disorder}
\end{figure*}

\begin{figure*}[t]
\begin{center}
\mbox{
\includegraphics[width=7cm]{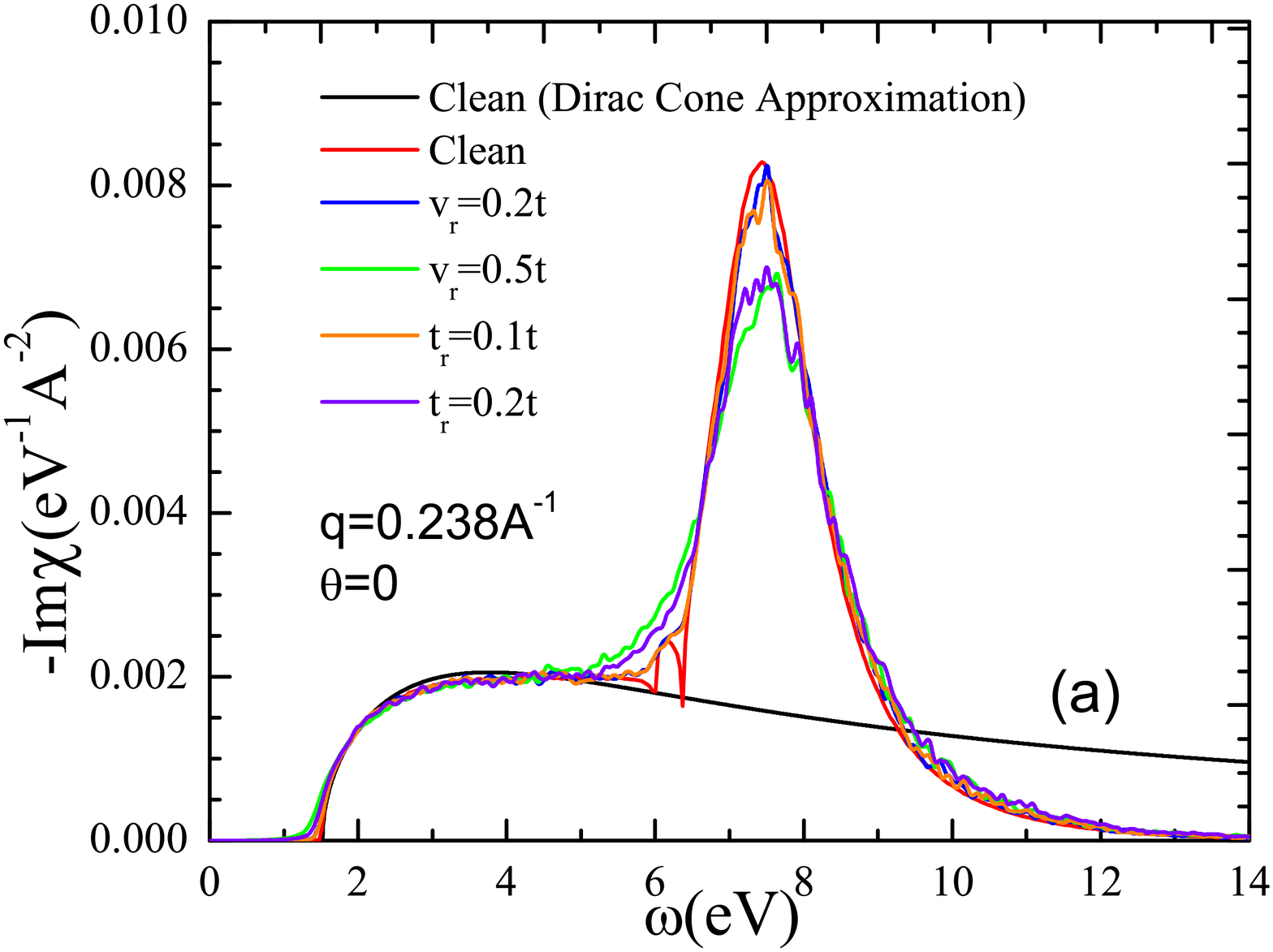}
\includegraphics[width=7cm]{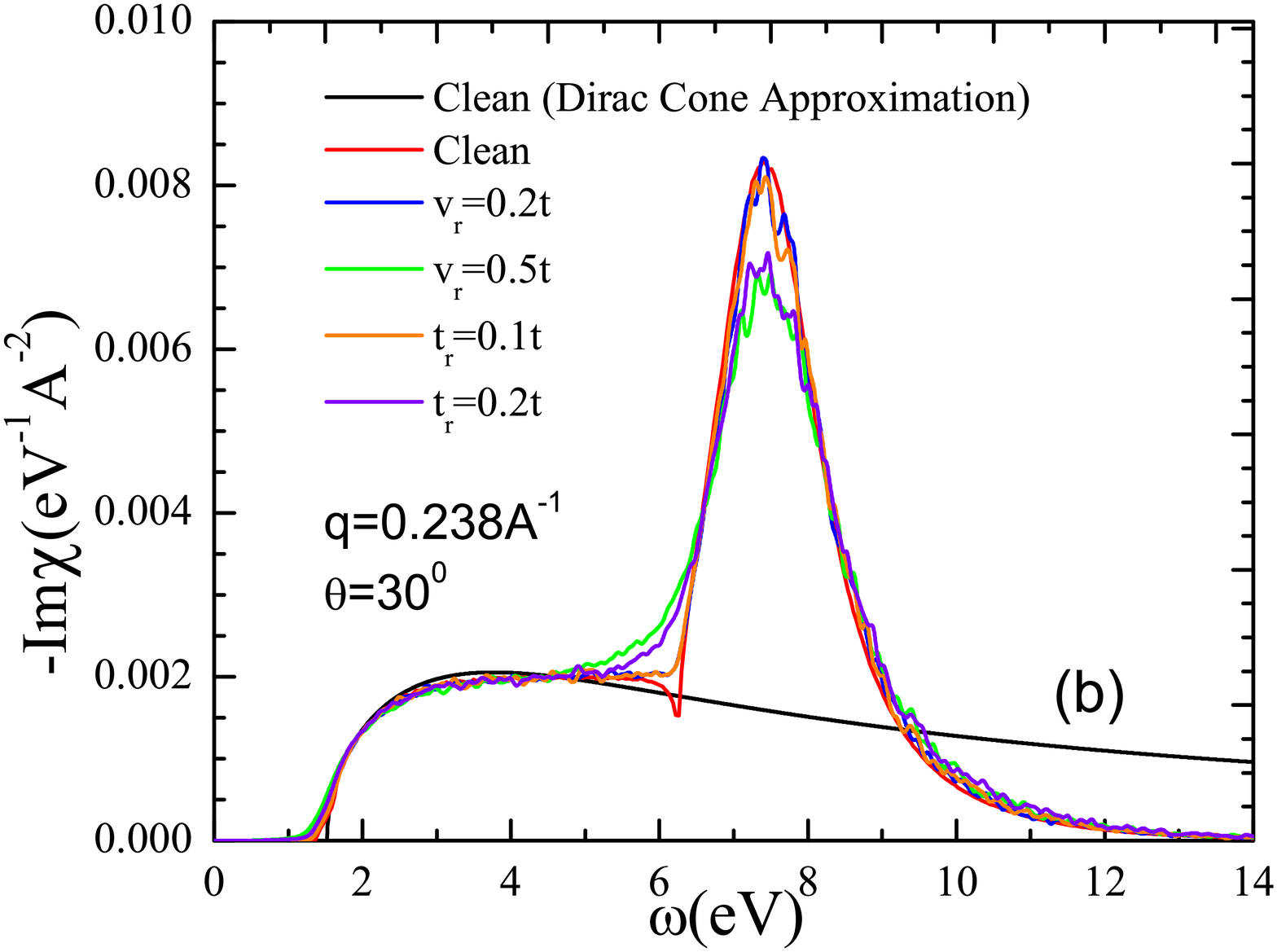}
} 
\mbox{
\includegraphics[width=7cm]{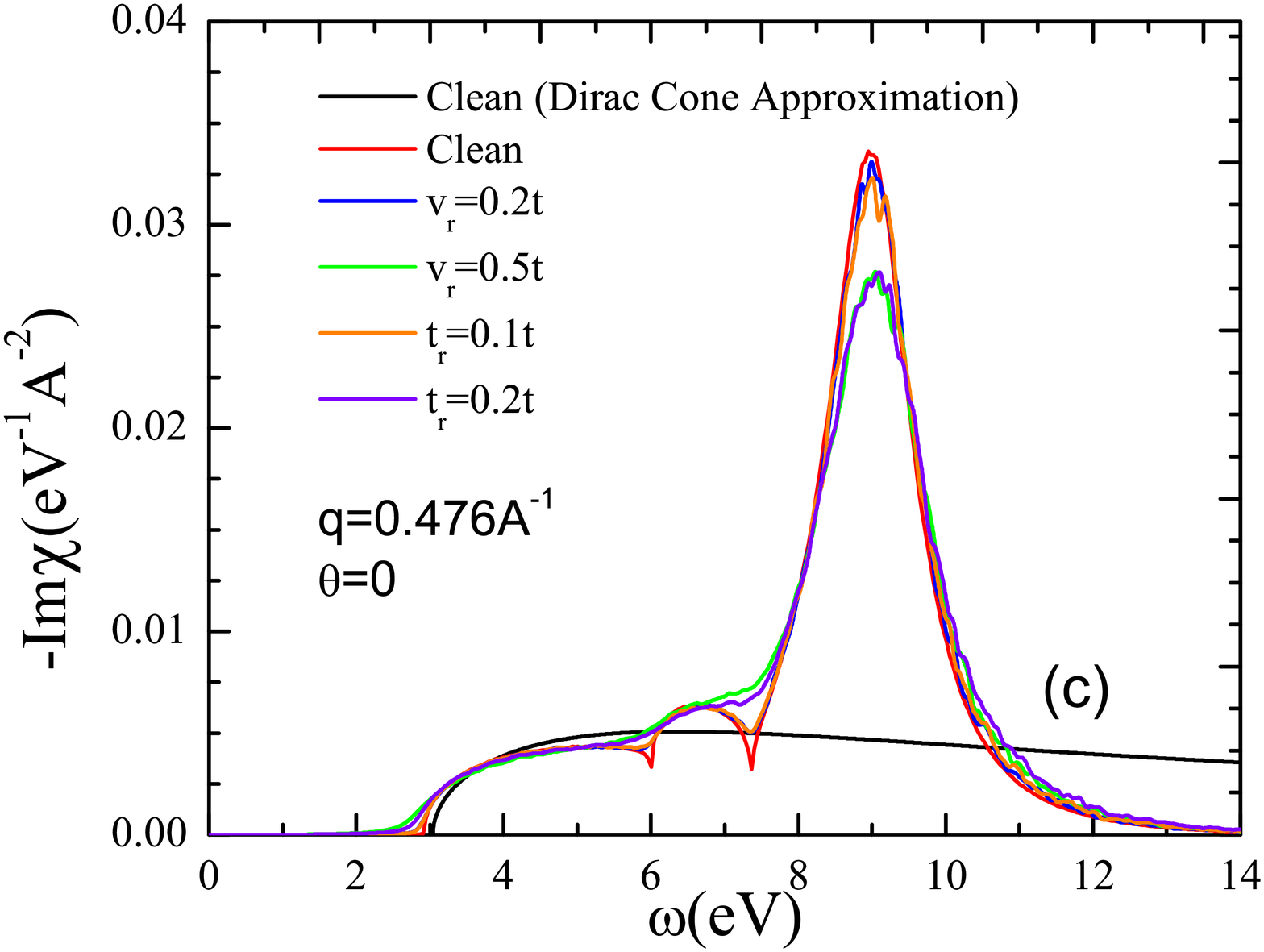}
\includegraphics[width=7cm]{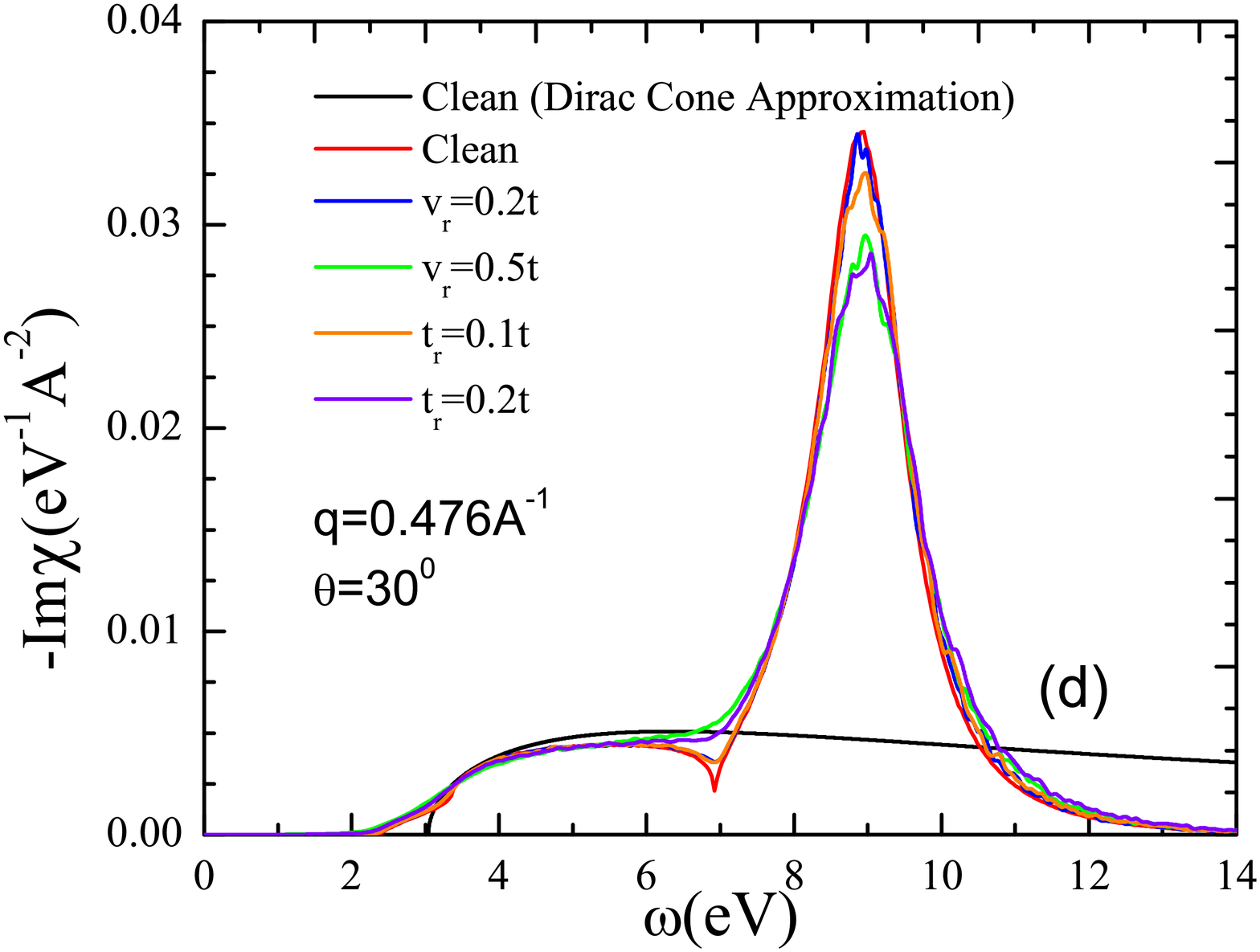}
}
\end{center}
\caption{(Color online) $-\mathrm{Im}~\protect\chi(\mathbf{q},\protect\omega%
) $ for the same values of $\mathbf{q}$ and disorder as in Fig. \protect\ref%
{Fig:ImPiSLG-disorder}.}
\label{Fig:ImchiSLG}
\end{figure*}

Second, for larger wave-vectors [Figs. \ref{Fig:ImPiSLG}(c)-(f)] one
observes strong differences in the spectrum depending on the orientation of $%
\mathbf{q}$, effect which has been discussed previously.\cite{SSP10,SG10} If 
$\mathbf{q}$ is along the $\Gamma $-K direction, there is a splitting of the
peak associated to the Van Hove singularity at $\omega \sim 2t$. At low
energies, we also observe a finite contribution to the spectral weight to
the left of the $\omega \approx v_{\mathrm{F}}q$ peak for momenta along the $%
\Gamma $-M direction [plots Figs. \ref{Fig:ImPiSLG}(d) and (f)]. Finally,
trigonal warping effects are important as we increase the magnitude of $|%
\mathbf{q}|$, due to the deviation of the band dispersion with respect to the linear
cone approximation. As a consequence, the constant energy contours are not 
any more circles around the Dirac points, but present a triangular shape. 
The consideration of this effect leads to a redshift of the $\omega \approx v_{\mathrm{F}}q$
peak with respect to the Dirac cone approximation, as seen clearly in Fig. %
\ref{Fig:ImPiSLG}(e).

\begin{figure}[t]
\begin{center}
\mbox{
\includegraphics[width=4cm]{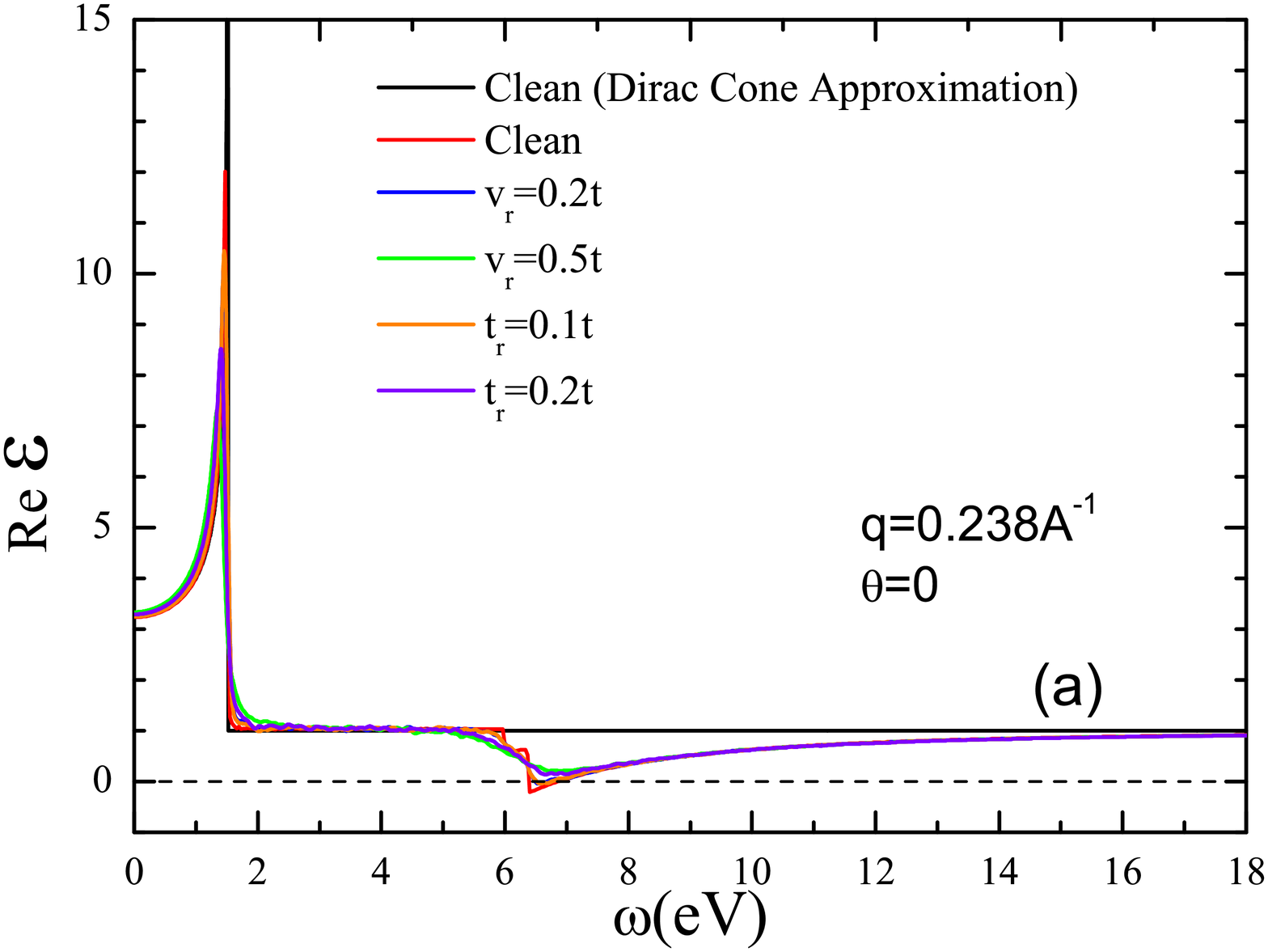}
\includegraphics[width=4cm]{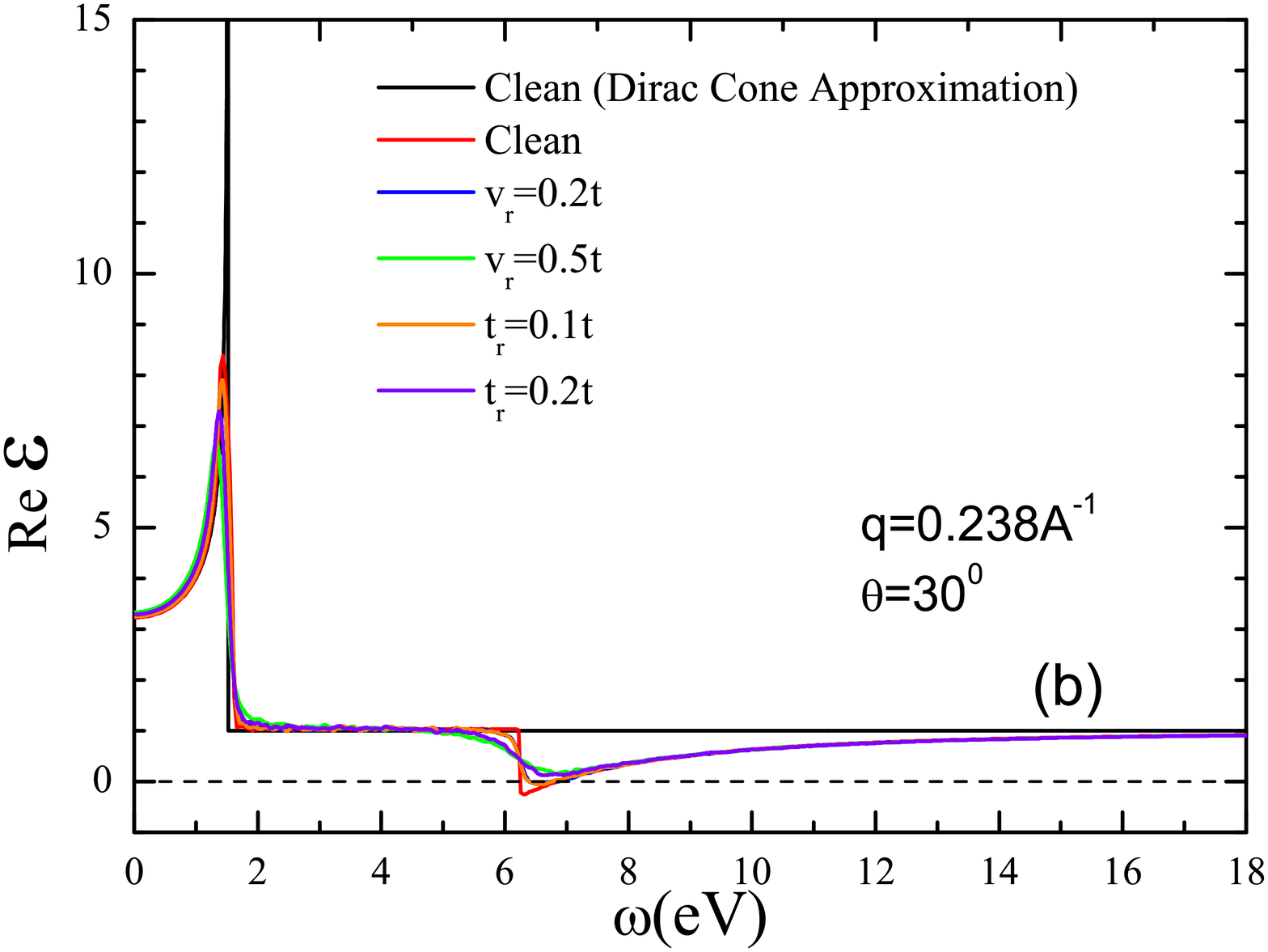}
} 
\mbox{
\includegraphics[width=4cm]{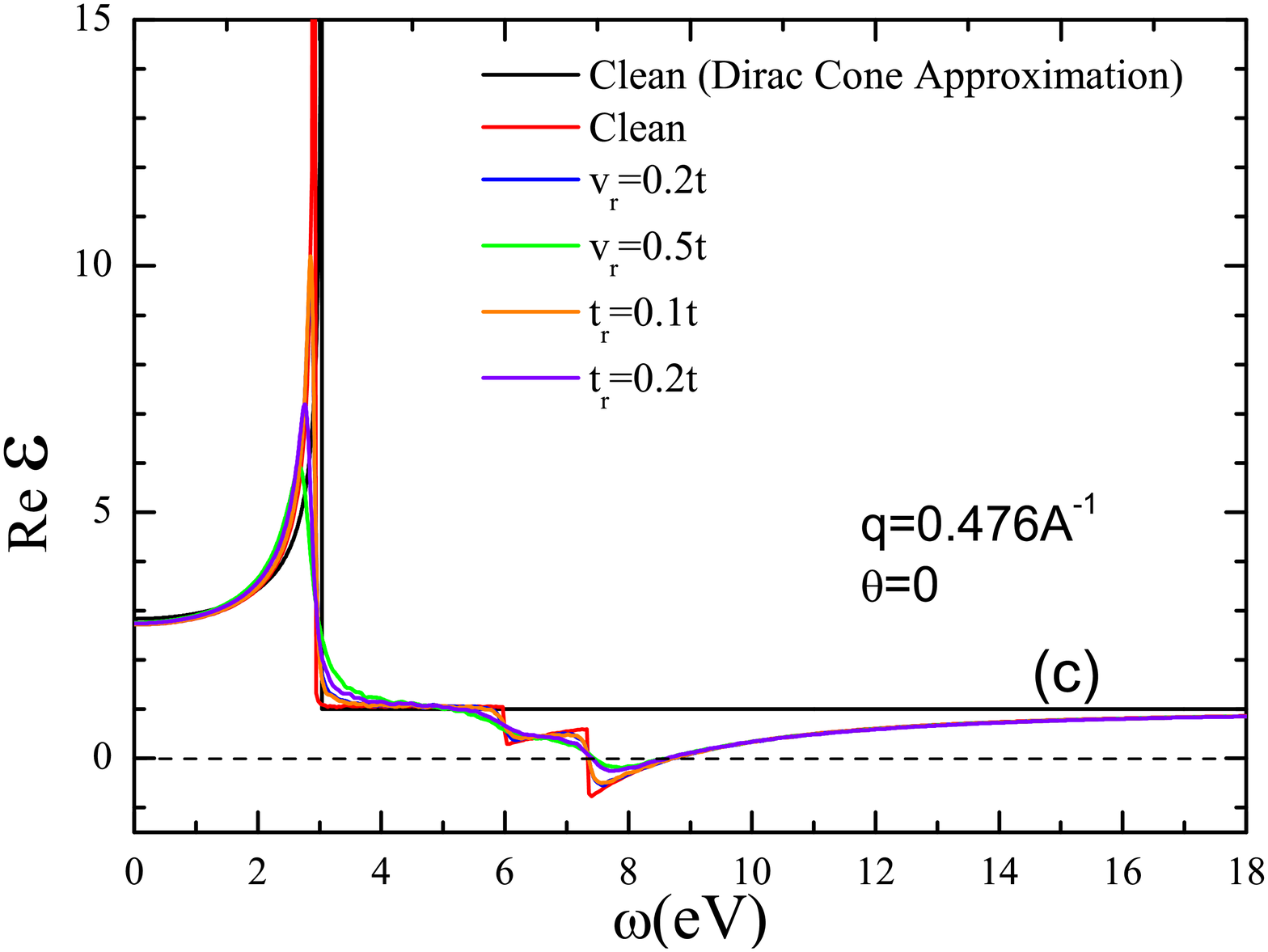}
\includegraphics[width=4cm]{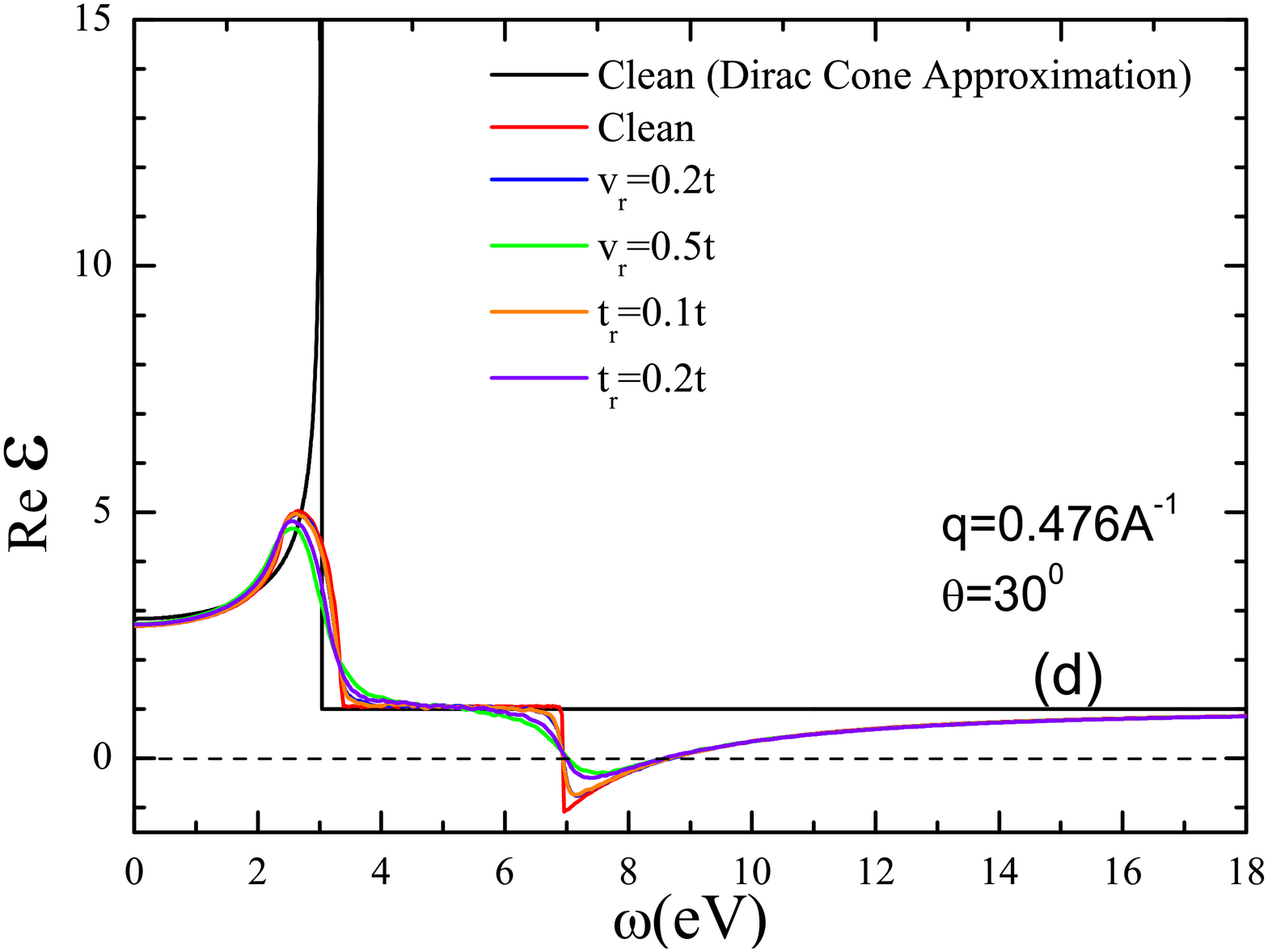}
}
\end{center}
\caption{(Color online) $\mathrm{Re}~\protect\varepsilon (\mathbf{q},\protect%
\omega )$ for the same values of $\mathbf{q}$ and disorder as in Fig. 
\protect\ref{Fig:ImPiSLG-disorder} and \protect\ref{Fig:ImchiSLG}.}
\label{Fig:ReepsSLG}
\end{figure}

Once we have discussed the clean case, we consider the effect of disorder on
the excitation spectrum as explained in Sec. \ref{Sec:Method}. Two different
kinds of disorder are considered: random local change of on-site potentials
and random renormalization of the hopping, which correspond to the diagonal
and off-digonal disorders in the single-layer Hamiltonian Eq.~(\ref%
{Hamiltonian_SLG}), respectively. The former acts as a chemical potential
shift for the Dirac fermions, i.e., shifts locally the Dirac point, and the
later rises from the changes of distance or angles between the $p_{z}$
orbitals. In Fig. \ref{Fig:DOS-disorder} we show the DOS of SLG for
different kinds and magnitudes of disorder. The DOS for clean graphene has
been plotted by using the analytical expression given in Ref.~%
\onlinecite{HN53}. The DOS of the disordered systems are calculated by
Fourier transform of the time-dependent correlation functions \cite{YRK10} 
\begin{equation}
\rho \left( \varepsilon \right) =\frac{1}{2\pi }\int_{-\infty }^{\infty
}e^{i\varepsilon t}\left\langle \varphi \right\vert e^{-iHt}\left\vert
\varphi \right\rangle dt,  \label{Eq:DOS}
\end{equation}%
with the same initial state $\left\vert \varphi \right\rangle $ defined in
Eq.~(\ref{Eq:phi0}). As shown in Ref.~\onlinecite{YRK10}, the result
calculated from a SLG with $4096\times 4096$ lattice sites matches very well
with the analytical expression, and here we use the same sample size in the
disordered systems. We consider that the on-site potential $v_{i}$ is random
and uniformly distributed (independently on each site $i$ ) between $-v_{r}$
and $+v_{r}$. Similarly, the in-plane nearest-neighbor hopping $t_{ij}$ is
random and uniformly distributed (independently on sites $i,j$) between $%
t-t_{r}$ and $t+t_{r}$. The main effect is a smearing of the Van Hove
singularity at $E=t$, as observed in the right hand side inset of Fig. \ref%
{Fig:DOS-disorder}.

The effect of disorder is also appreciable in the non-interacting excitation
spectrum of the system, as shown by Fig. \ref{Fig:ImPiSLG-disorder}. A
broadening of the $\omega\approx v_{\mathrm{F}} q$ and $\omega\approx 2t$
peaks is observed in all the cases. Furthermore, disorder leads to a slight
but appreciable redshift of the peaks with respect to the clean limit. This
effect is more important for higher wave-vectors, as it can be seen in Fig. %
\ref{Fig:ImPiSLG-disorder}(c)-(d). Finally, the disorder broadening of the
peaks leads in all the cases to a transfer of spectral weight to low
energies (below $\omega=v_{\mathrm{F}} q$), as it is appreciable in Fig. \ref%
{Fig:ImPiSLG-disorder}(a)-(d).

The next step is to consider both, disorder and electron-electron
interaction in the system. In the RPA, the response function is calculated
as in Eq. (\ref{Eq:chi}). The results are shown in Fig. \ref{Fig:ImchiSLG},
where $-\mathrm{Im}~\chi(\mathbf{q},\omega)$ is plotted for the same
wave-vectors and disorder used in Fig. \ref{Fig:ImPiSLG-disorder}. We
observe that the Dirac cone approximation (black line) captures well the low
energy region of the spectrum. However, the large peak at $\omega\sim 2t$
cannot be captured by the continuum approximation. They are due to a plasmon
mode associated to transitions between electrons in the saddle points of the 
$\pi$-bands. Strictly speaking, those modes cannot be considered as \textit{%
fully coherent} collective modes, as for example, the low energy $\sqrt{q}$%
-plasmon which is present in doped graphene.\cite{S86} For doped graphene,
the acoustic $\sqrt{q}$-plasmon is undamped above the threshold $\omega=v_{%
\mathrm{F}} q$ until it enters the inter-band particle-hole continuum, when
it starts to be damped and decays into electron-hole pairs. However, the $%
\pi $-plasmon, although it corresponds to a zero of the dielectric function
as it can be seen in Fig. \ref{Fig:ReepsSLG}, it is a mode which lies 
\textit{inside} the continuum of particle-hole excitations: $-\mathrm{Im}%
~\Pi(\mathbf{q},\omega_{pl})>0$ at the $\pi$-plasmon energy $\omega_{pl}$,
and the mode will be damped even at $q\rightarrow 0$. In any case, it is a
well defined mode which has been measured experimentally for SLG and MLG.%
\cite{KP08,EB08,GG08,RA10,MSH11} Coming back to our results, notice that the
height of the peaks is reduced when the effect of disorder is considered,
although the position is unaffected by it. For small wave-vectors, this mode
is highly damped due to the strong spectral weight of the particle-hole
excitation spectrum at this energy, as seen e.g. by the peak of $-\mathrm{Im}%
~\Pi(\mathbf{q},\omega)$ at $\omega=2t$ in Fig. \ref{Fig:ImPiSLG}(a)-(b).
The position of the collective modes can be alternatively seen by the zeroes
of the dielectric function Eq. (\ref{Eq:Plasmons}), which is shown in Fig. %
\ref{Fig:ReepsSLG}. Notice that the Dirac cone approximation (solid black
lines in Fig. \ref{Fig:ReepsSLG}) is completely insufficient to capture this
high energy $\pi$-plasmon, which predicts always a finite $\mathrm{Re}%
~\varepsilon(\mathbf{q},\omega)$. As for the polarization, we see
that disorder lead to an important smearing of the singularities of the
dielectric function, as seen in Fig. \ref{Fig:ReepsSLG}. 
Finally, we mention that the application of our method to even higher wave-vectors 
and energies as the ones considered in the present work, should be accompanied by 
the inclusion of local field effects (LFE) in the dielectric function, which are related
to the periodicity of the crystalline lattice.\cite{A62} In fact, for SLG and for wave-vectors 
along the zone boundary between the M and the K points of the Brillouin zone (see Fig. \ref{Fig:BZ}),
the inclusion of LFEs leads to a new optical plasmon mode at an energy of 20-25 eV.\cite{PAP10}

\section{Excitation spectrum for multi-layer graphene}

\label{Sec:MLG}

\begin{figure}[t]
\begin{center}
\mbox{
\includegraphics[width=4.0cm]{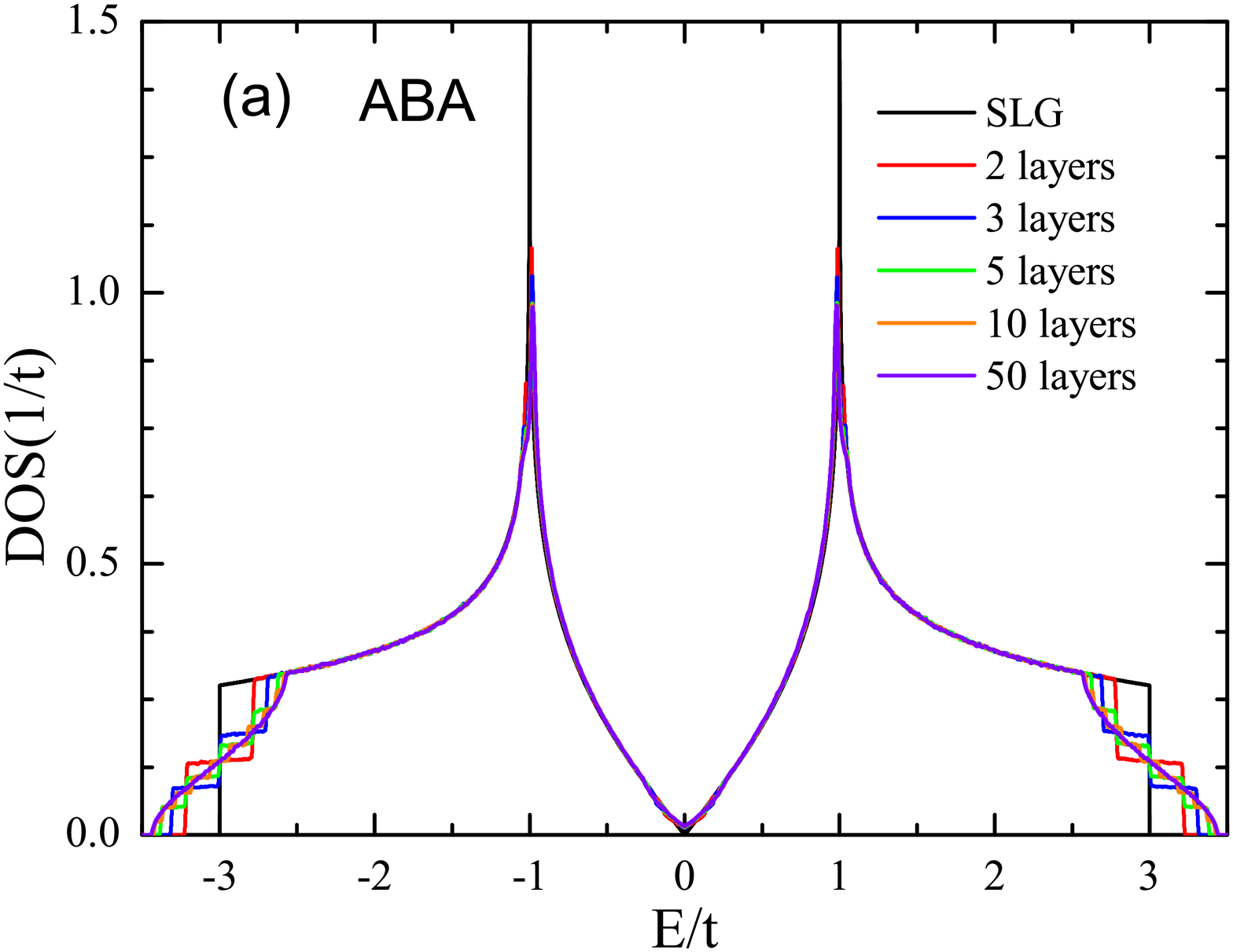}
\includegraphics[width=4.0cm]{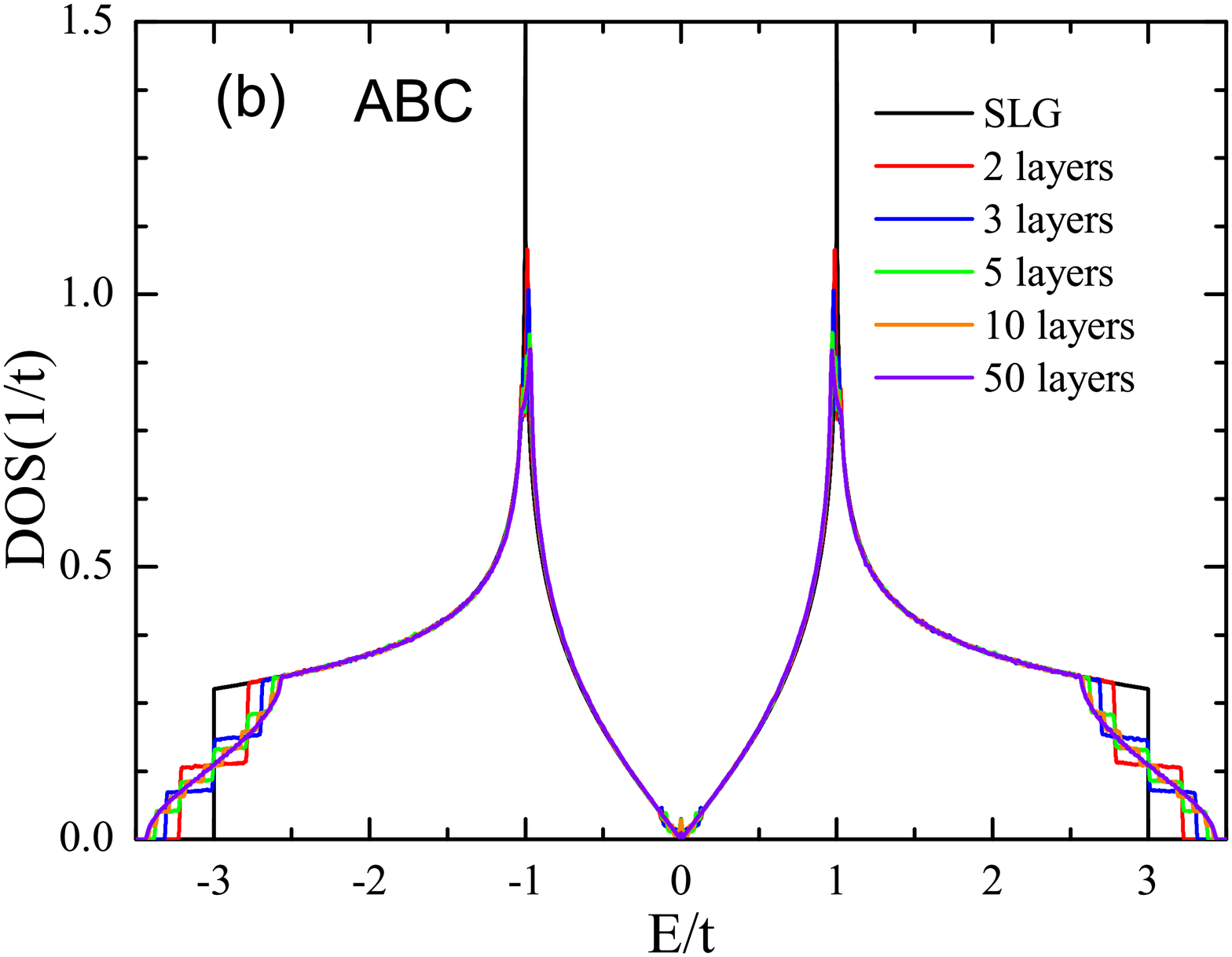}
} 
\mbox{
\includegraphics[width=4.0cm]{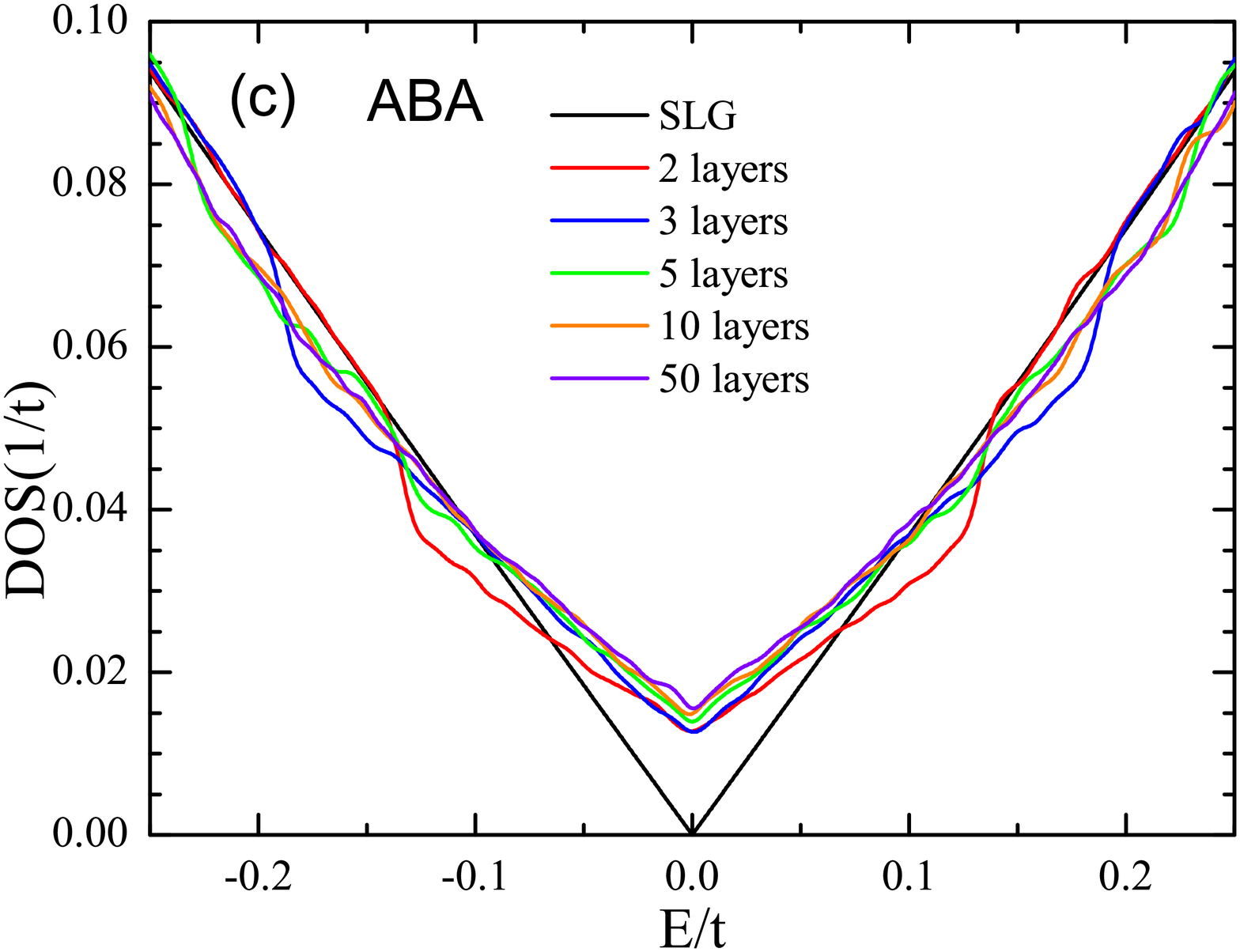}
\includegraphics[width=4.0cm]{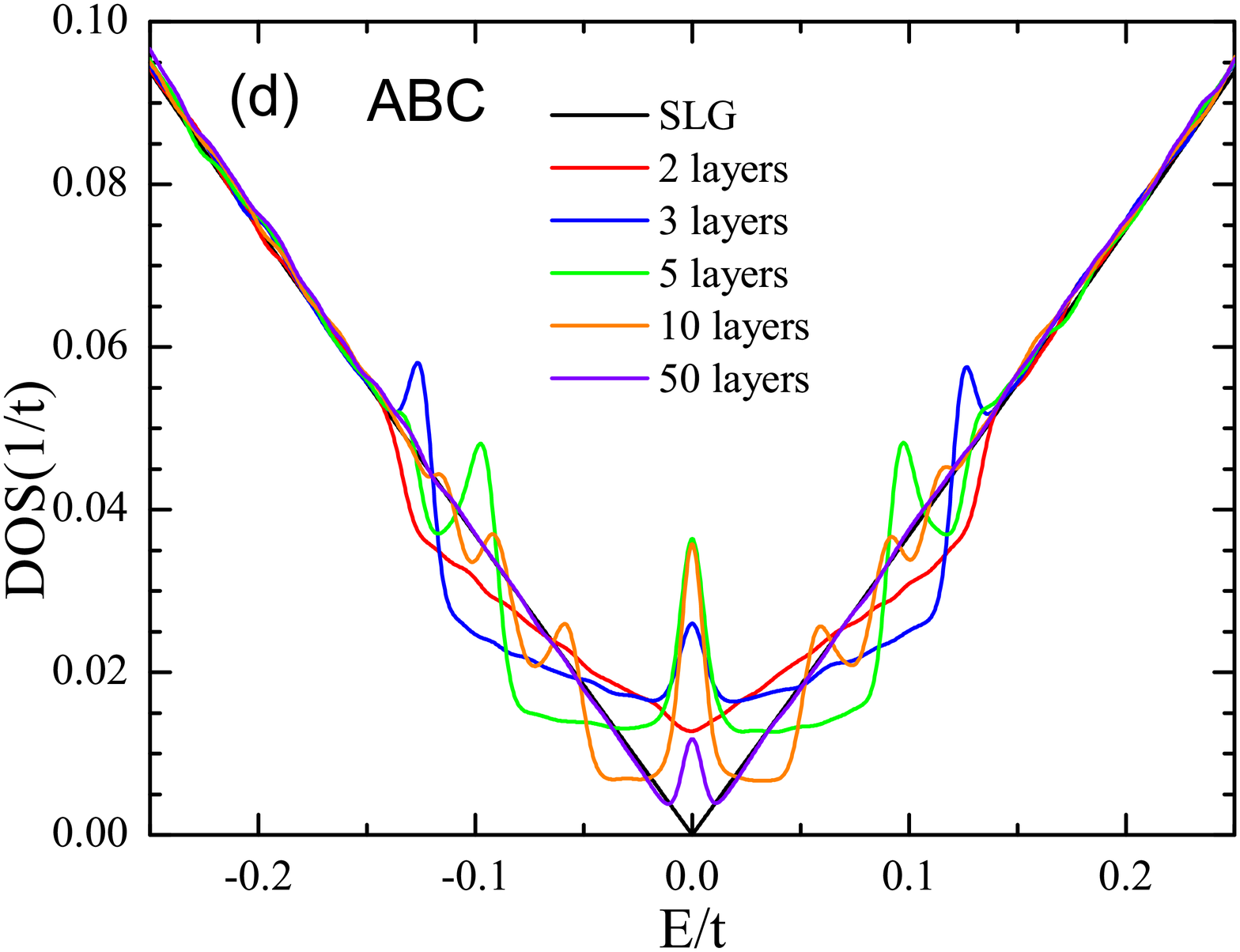}
} 
\mbox{
\includegraphics[width=4.0cm]{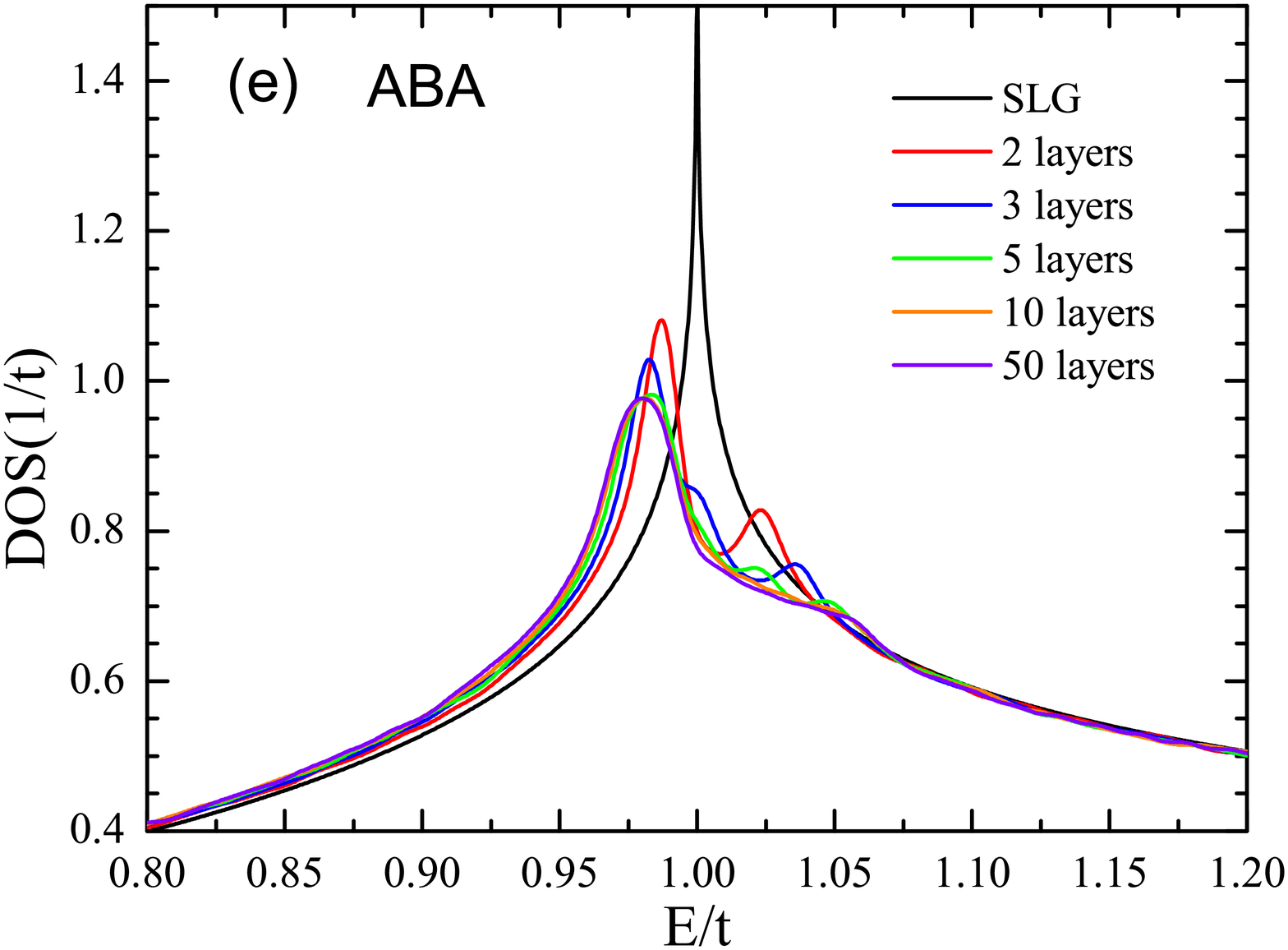}
\includegraphics[width=4.0cm]{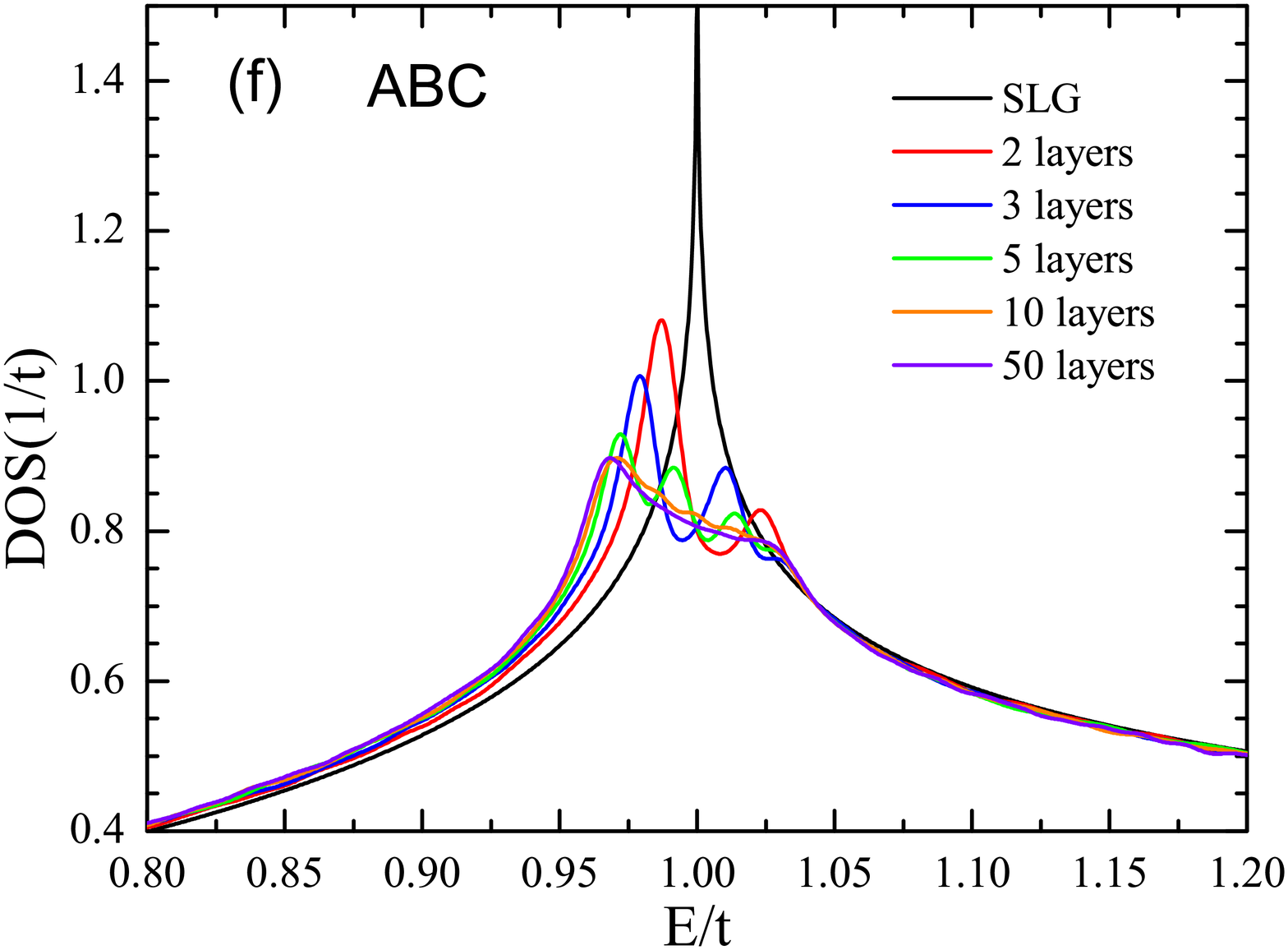}
}
\end{center}
\caption{(Color online) (a,b) Density of states of ABA- (left panels) and
ABC-stacked (right panels) multilayer graphene. A zoom of the DOS around the
Dirac point ($E=0$) is shown in (c,d), and around the Van Hove singularity ($%
E=t$) is shown in (e,f). The sample sizes of each layer in MLG are: $%
4096\times 4096$ atoms in bilayer; $3200\times 3200$ in trilayer; $%
2048\times 2048$ in 5 layers; $1600\times 1600$ in 10 layers and $800\times
800$ in 50 layers.}
\label{Fig:DOS-MLG}
\end{figure}

\begin{figure}[t]
\begin{center}
\mbox{
\includegraphics[width=4.2cm]{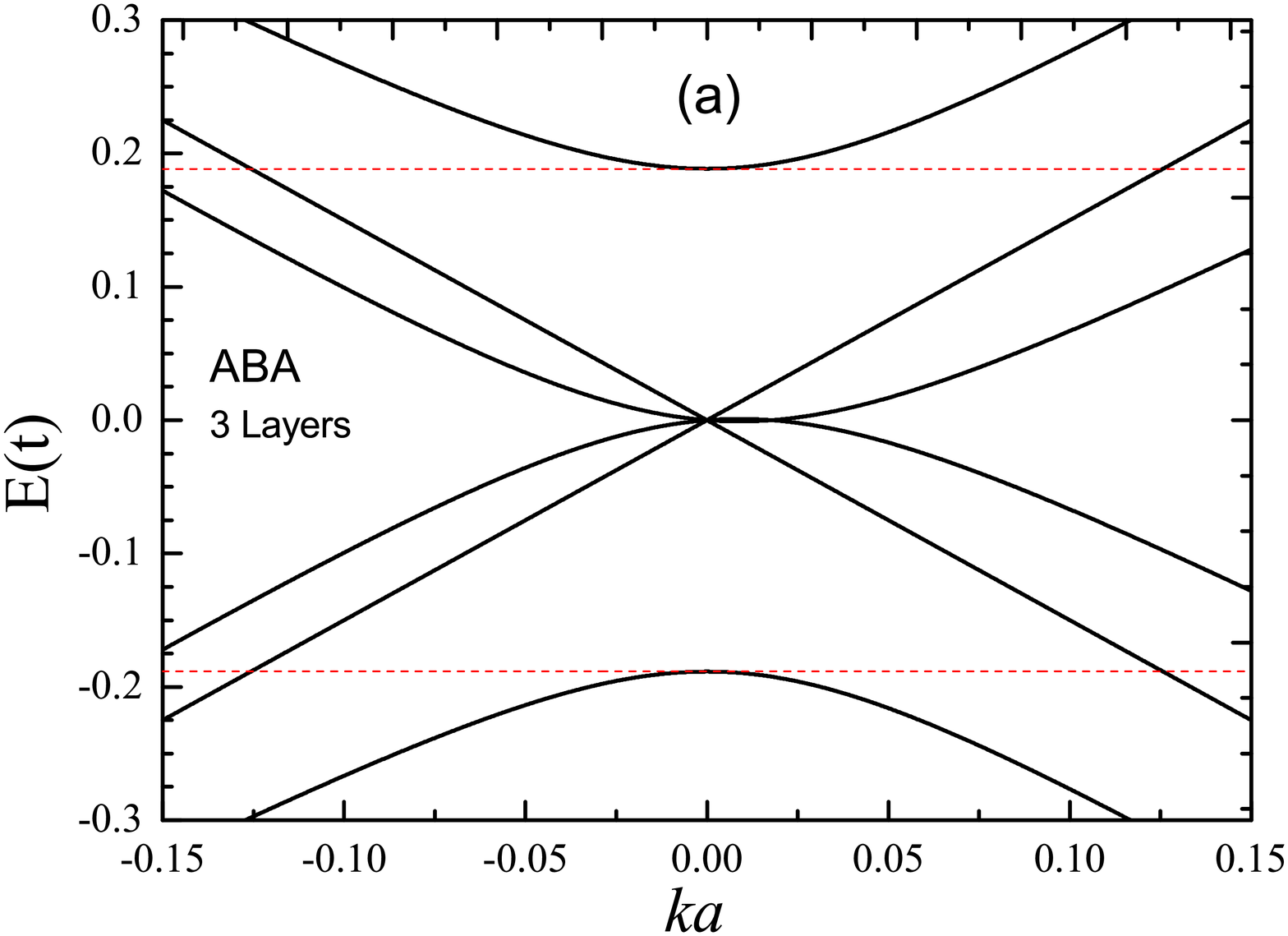}
\includegraphics[width=4.2cm]{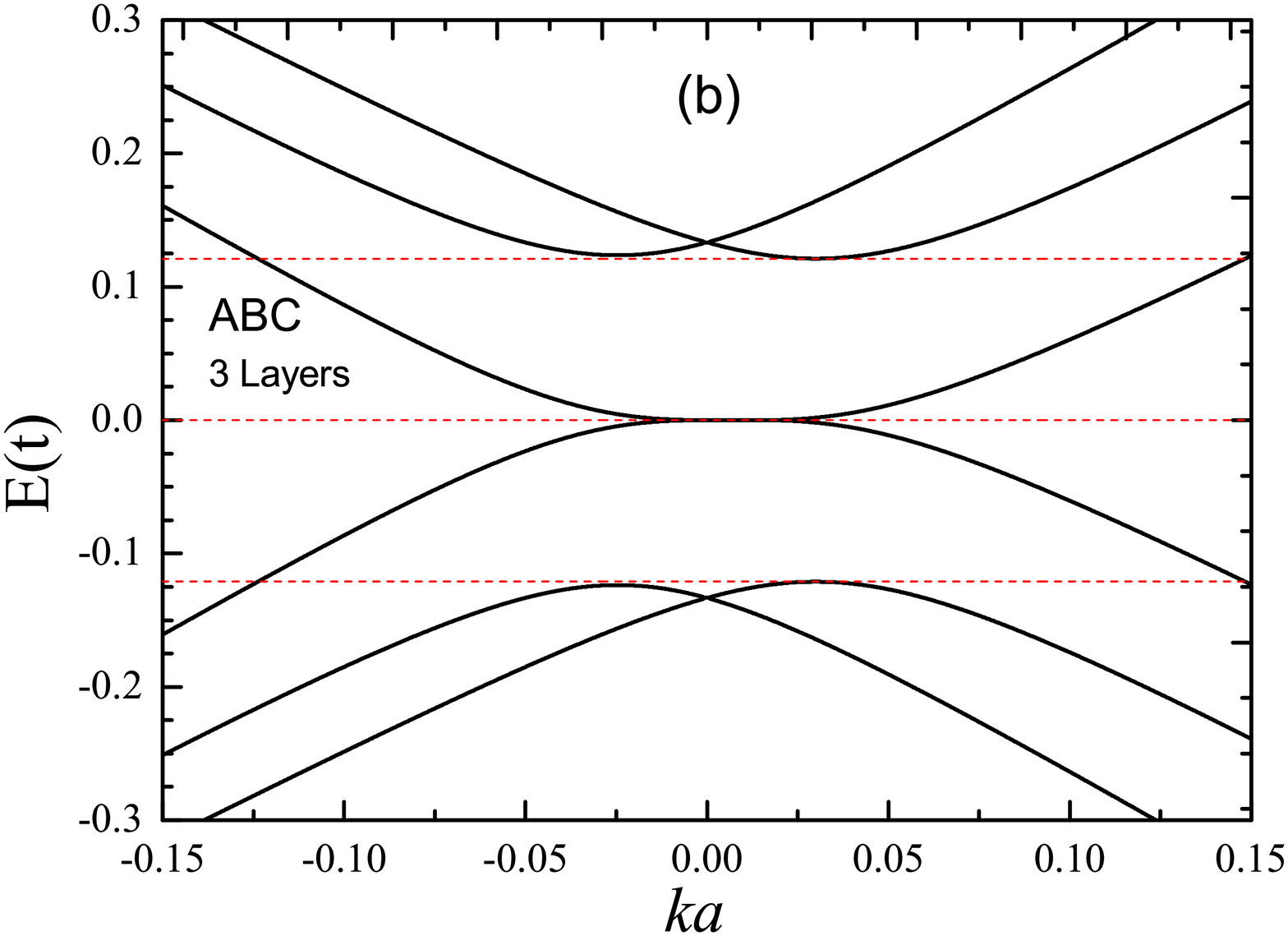}
}
\end{center}
\caption{(Color online) Band structure of ABA- and ABC-stacked trilayer graphene. Left
panel: the red dashed lines indicate the position of the jump in DOS of
ABA-stacked trilayer graphene at $\left\vert E\right\vert \approx 0.19t$. Right panel: the
red dashed lines indicate position of the peaks in DOS of ABC-stacked trilayer graphene at 
$E=0$ and $\left\vert E\right\vert \approx 0.12t$.}
\label{Fig:BandStructure}
\end{figure}

In the following, we study the excitation spectrum and collective modes of
MLG. For this, we consider not only the Coulomb interaction between
electrons on different layers, but also the possibility for the carriers to
tunnel between neighboring layers, as described in Sec. \ref{Sec:Method}.
The importance of considering inter-layer hopping has been already shown in
the study of screening properties of MLG.\cite{G07} First, we see that the
results are sensitive to the relative orientation between layers. In Fig. %
\ref{Fig:DOS-MLG} we show the density of states for ABA- and ABC- stacked
MLG (see Fig. \ref{Fig:Stacking} for details on the difference between those
two orientations). As seen in Fig. \ref{Fig:DOS-MLG}(c)-(d), all the MLGs
present a finite DOS at $E=0$, contrary to SLG which has a vanishing DOS at
the Dirac point. The main difference between the two kinds of stacking is
that for ABC there is a central peak together with a series of satellite
peaks around $E=0$ [Fig. \ref{Fig:DOS-MLG}(d)], whereas for ABA the DOS
follows closer the behavior of the SLG [Fig.\ref{Fig:DOS-MLG}(c)]. 
The different structure in the DOS can be understood by looking at Fig. \ref{Fig:BandStructure}, 
where we show the low energy band structure of a trilayer graphene with ABA 
[Fig. \ref{Fig:BandStructure}(a)] and ABC [Fig. \ref{Fig:BandStructure}(b)] orientations. 
The different jumps and peaks in the DOS of Fig. \ref{Fig:DOS-MLG}(c)-(d) are associated to 
the regions of the band dispersion marked by the horizontal red lines of Fig. \ref{Fig:BandStructure}, 
the energy of which depends on the values of the tight-binding parameters associated 
to inter-layer tunneling ($\gamma_1$ and $\gamma_3$ in our case). In the
two cases, we observe a splitting of the Van Hove peak, as seen in Fig. \ref%
{Fig:DOS-MLG}(e)-(f). Notice that when we have a high number of layers (e.g.
above 10 layers), there is a weak effect on adding a new graphene sheet to
the system, as it can be seen from the similar DOS between the 10- and the
50-layers cases of Fig. \ref{Fig:DOS-MLG}.

\begin{figure}[t]
\begin{center}
\mbox{
\includegraphics[width=4.0cm]{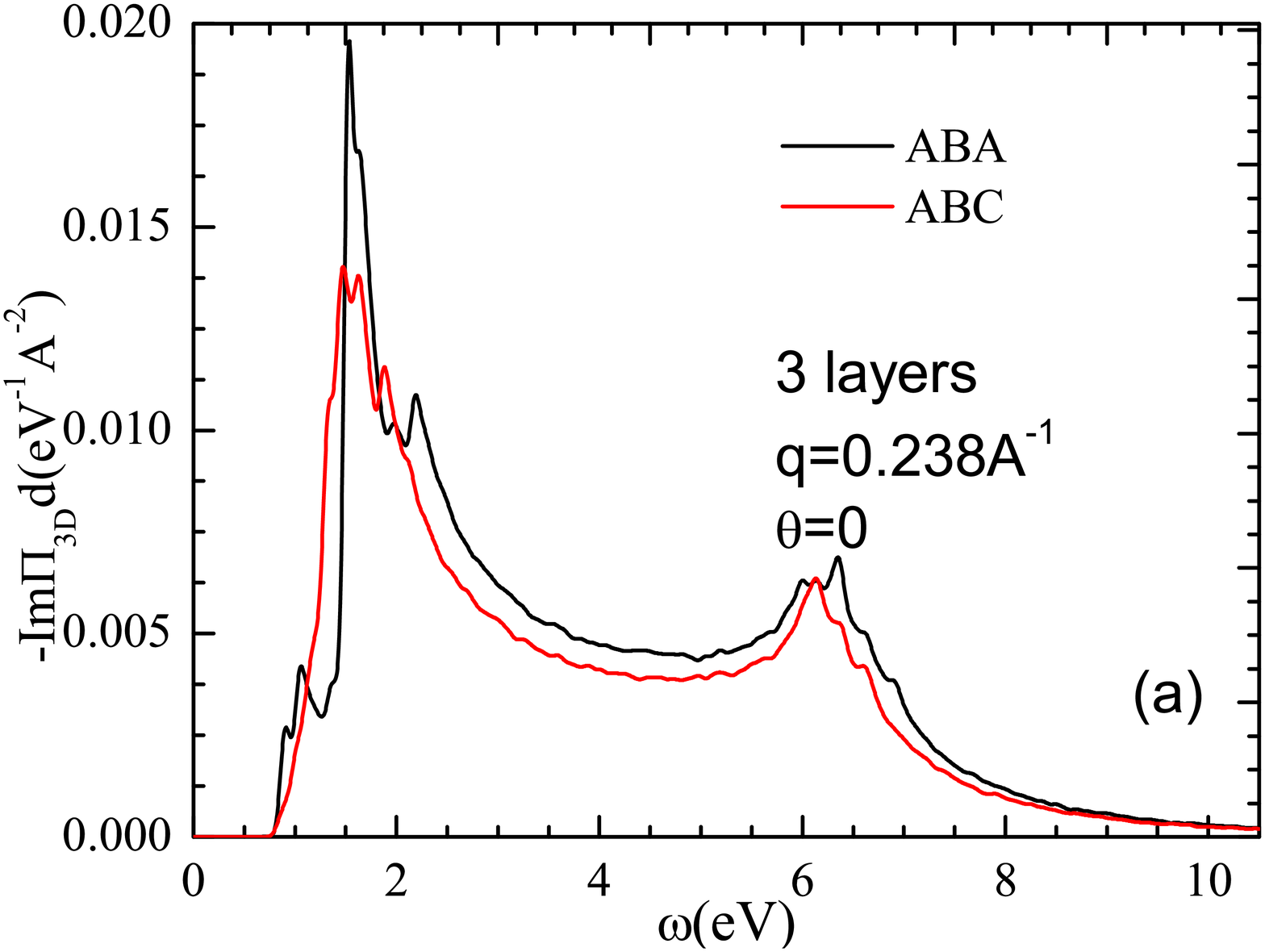}
\includegraphics[width=4.0cm]{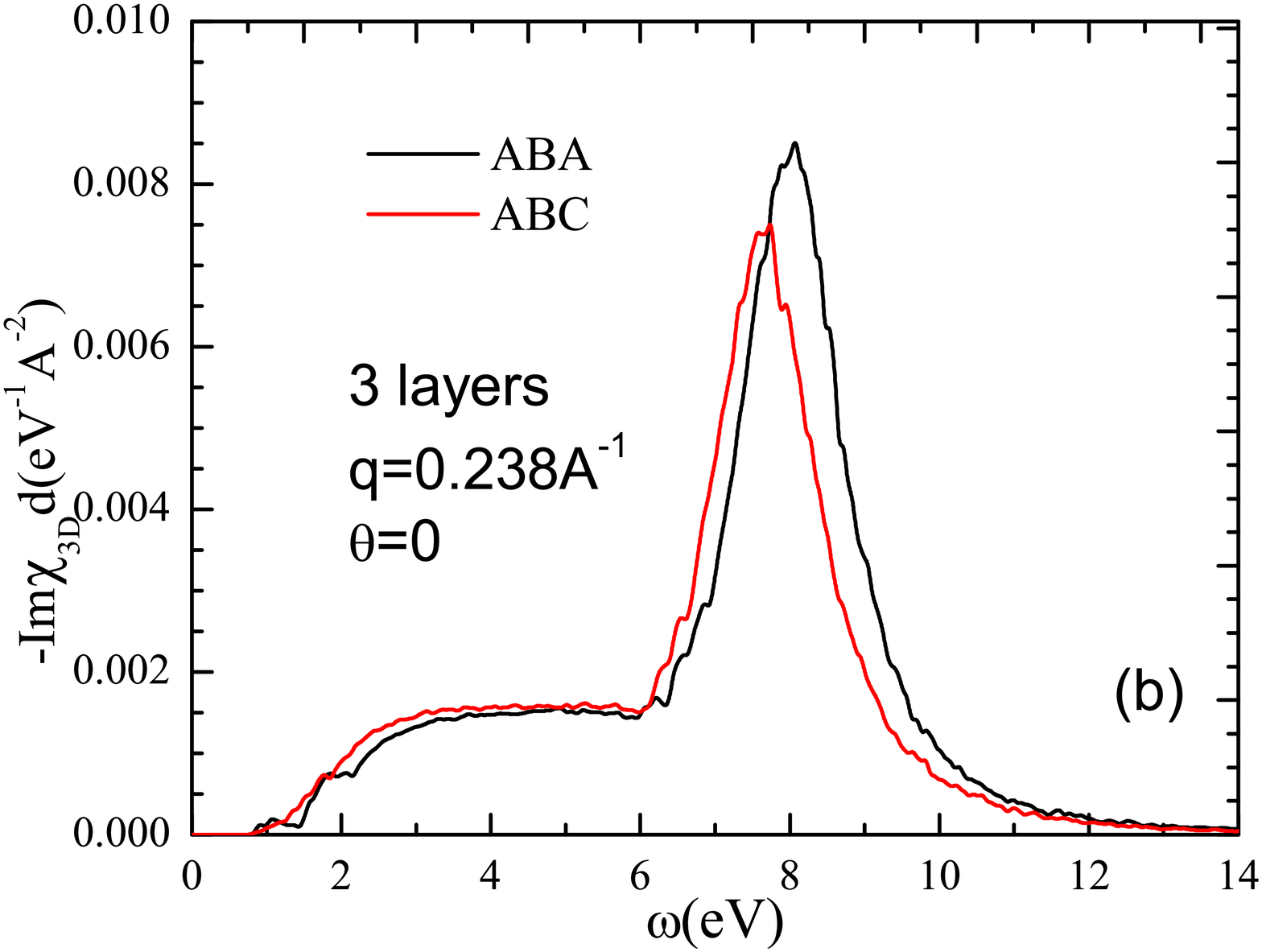}
} 
\mbox{
\includegraphics[width=4.0cm]{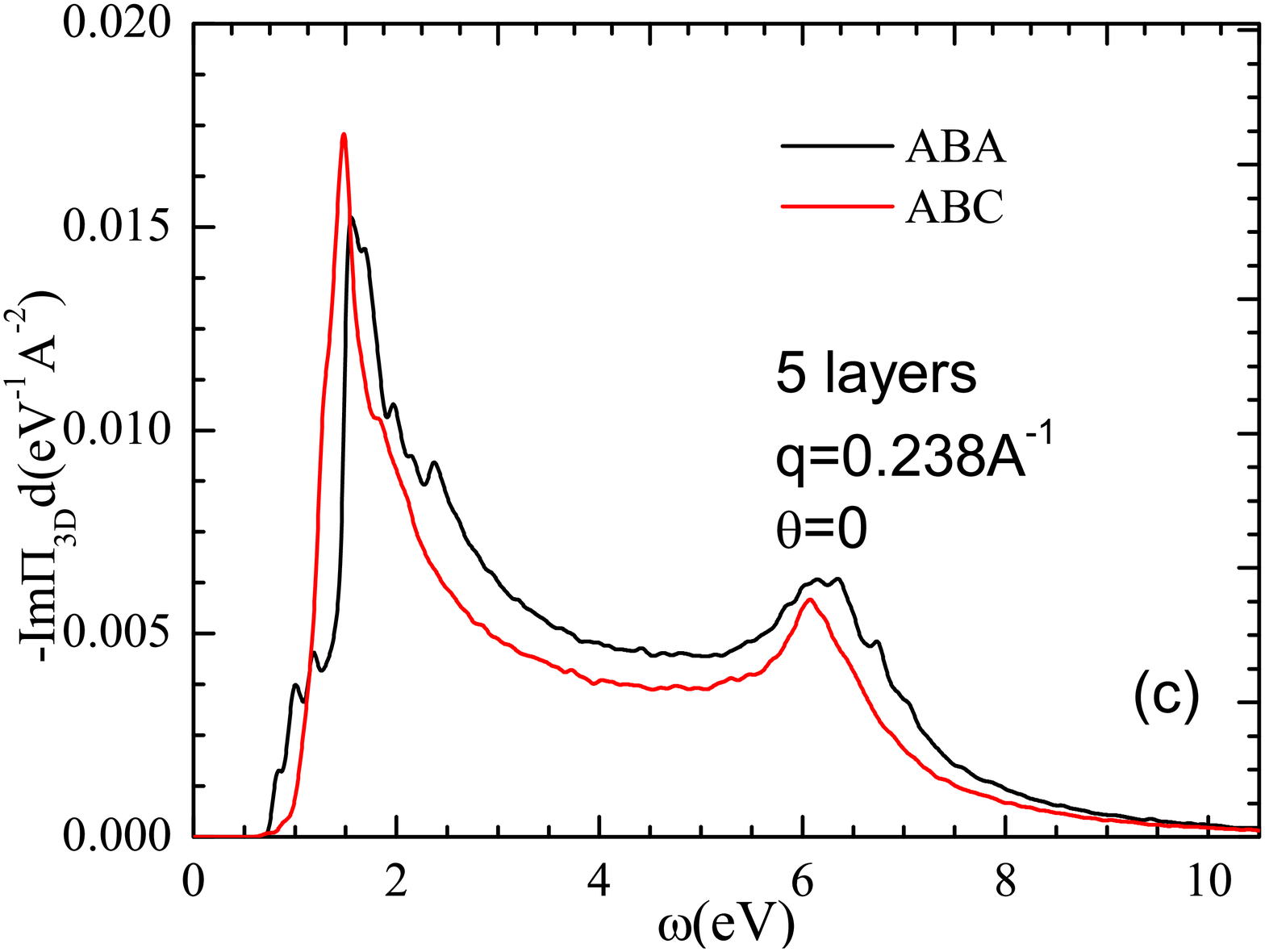}
\includegraphics[width=4.0cm]{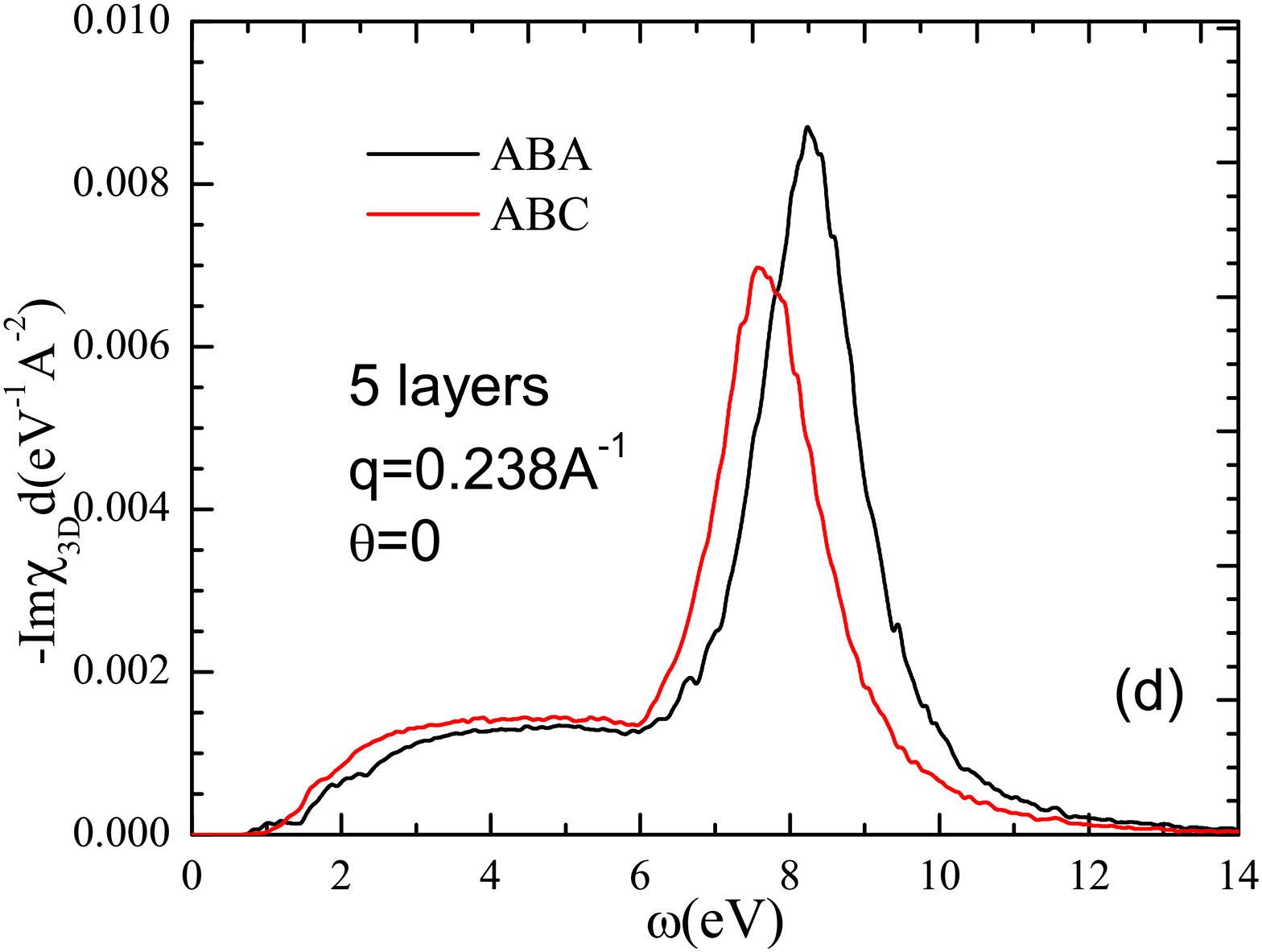}
} 
\mbox{
\includegraphics[width=4.0cm]{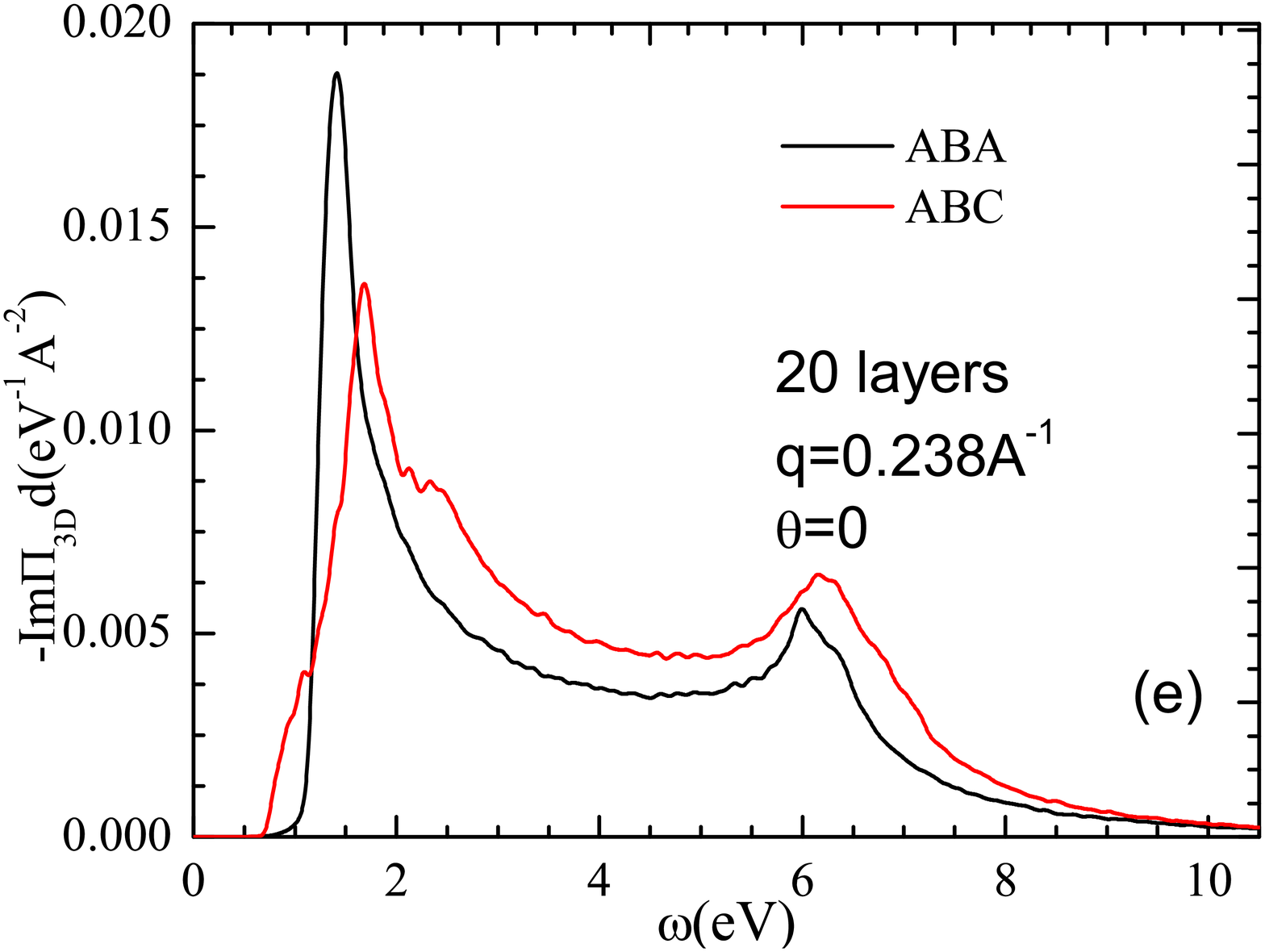}
\includegraphics[width=4.0cm]{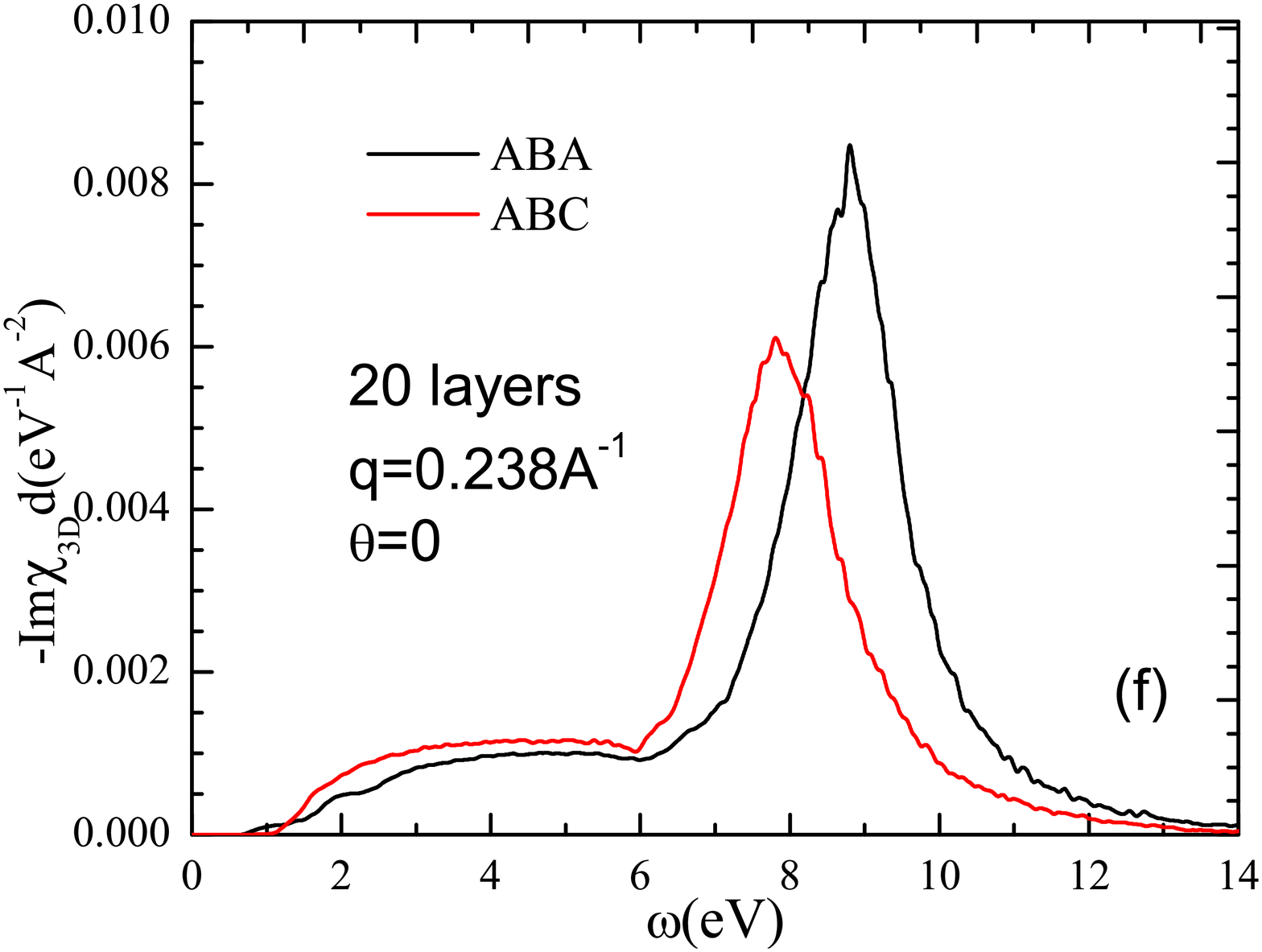}
}
\end{center}
\caption{(Color online) Dynamical polarization and response function of ABA-
and ABC-stacked multilayer graphene. The number of layers are 3 layers in
(a,b), 5 layers in (c,d) and 20 layers in (e,f). The size of each layer is
3600$\times$3600 atoms for tri-layer (a,b); 3200$\times$3200 atoms for
5-layer samples (c,d), and 1600$\times$1600 atoms for 20-layer (e,f).}
\label{Fig:Pi-ABA-ABC}
\end{figure}

In Fig. \ref{Fig:Pi-ABA-ABC} we show the non-interacting (left panels) and
the RPA (right panels) polarization function of MLG, for systems made of 3,
5 and 20 layers, and for ABA- and ABC-stacking. For the spectrum in the
absence of electron-electron interaction, as shown in Fig. \ref%
{Fig:Pi-ABA-ABC}(a), (c) and (e), one does not observe any specific
difference in the two spectra apart from different intensities depending on
the kind of stacking and on the number of layers considered for the
calculation. On the other hand, the energy of the $\pi$-plasmon of ABC
samples is redshifted with respect to the ABA-stacking. This can be see from
the relative position of the peaks of $-\mathrm{Im}~\chi(\mathbf{q},\omega)$
in Fig. \ref{Fig:Pi-ABA-ABC}(b), (d) and (e). Also, notice that the
separation between the two peaks grows with the number of layers, and for a
20-layers system, the difference can be of the order of 1~eV, as it can be
seen in Fig. \ref{Fig:Pi-ABA-ABC}(f). In the following and unless we say the
opposite, all the results will be calculated for the more commonly found
ABA-stacking.

\begin{figure}[t]
\begin{center}
\includegraphics[width=6.5cm]{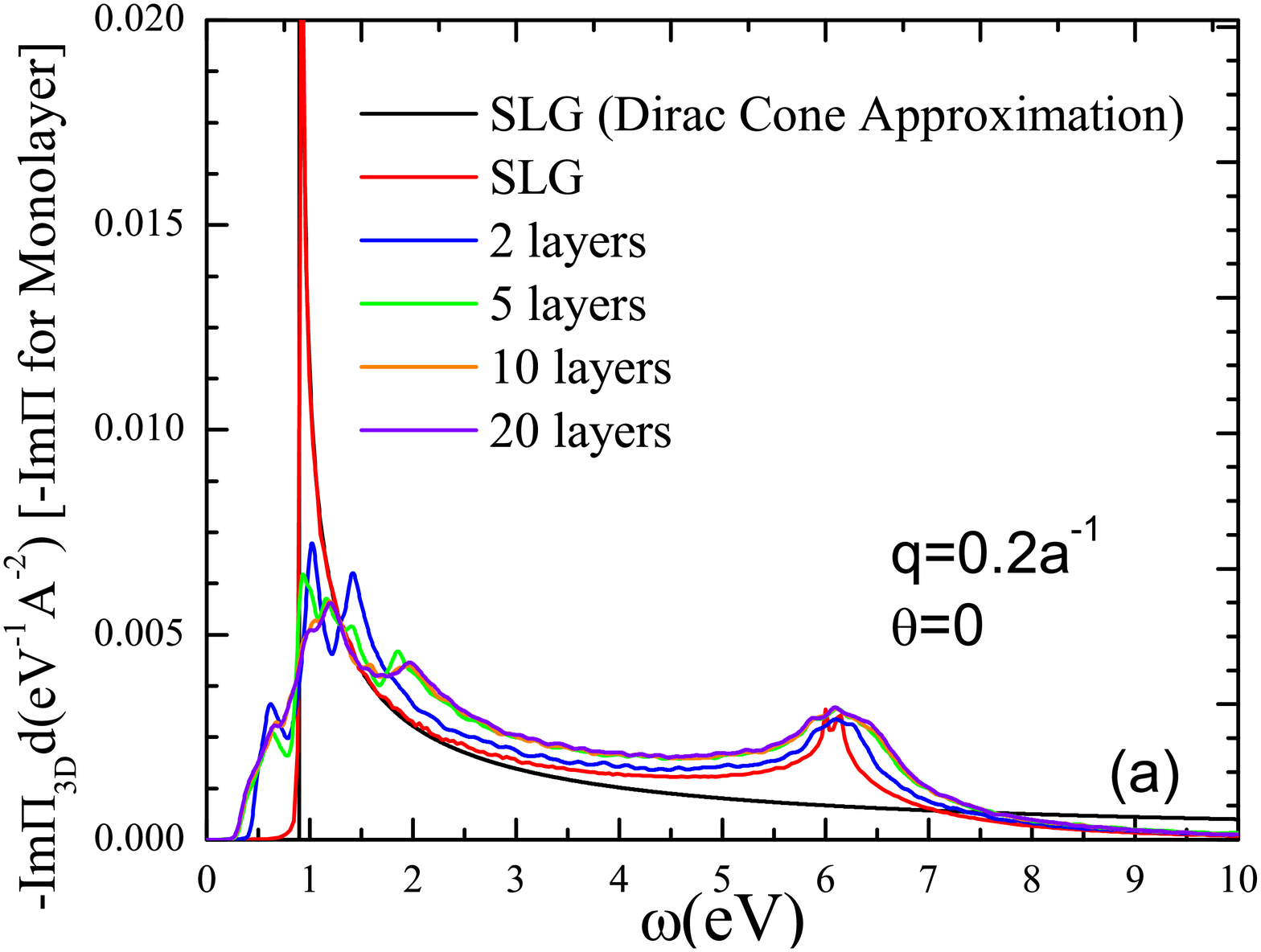} %
\includegraphics[width=6.5cm]{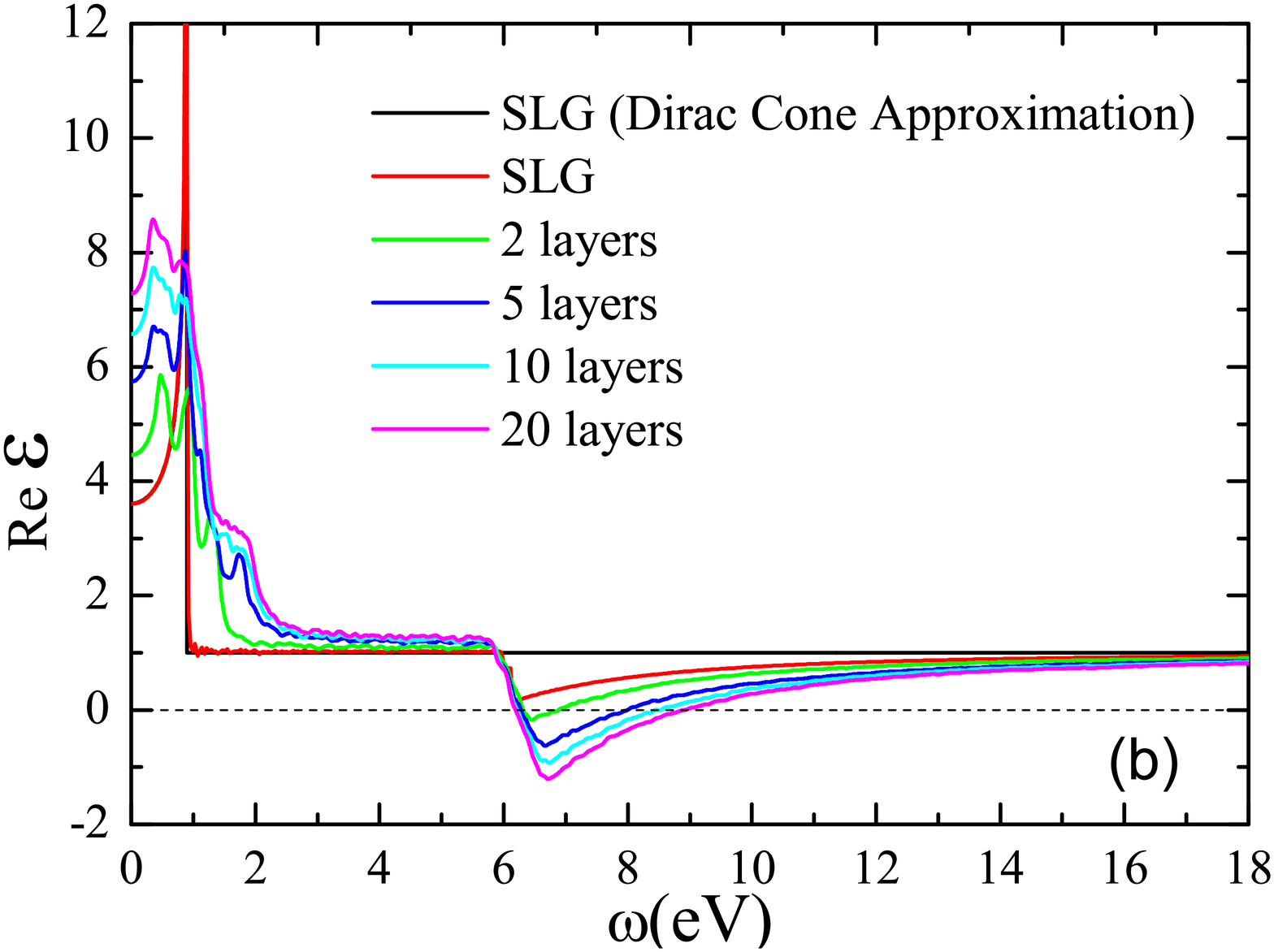} 
\end{center}
\caption{(Color online) (a) $\mathrm{Im}\Pi (\mathbf{q},\protect\omega )$
for SLG and MLG. The results for SLG obtained from the Kubo formula Eq. (%
\protect\ref{Eq:Kubo}) are compared to those obtained from the Lindhard
function Eq. (\protect\ref{Eq:Lindhard}), and to the Dirac cone
approximation. (b) $\mathrm{Re}~\protect\varepsilon (\mathbf{q},\protect%
\omega )$ for SLG and MLG, and comparison to the Dirac cone approximation,
for the same value of $q$ as in (a).}
\label{Fig:SpectrumMLG}
\end{figure}

For a more clear understanding about the evolution of the particle-hole
excitation spectrum with the number of layers, we plot in Fig. \ref%
{Fig:SpectrumMLG}(a) the imaginary part of $\Pi(\mathbf{q},\omega)$ for SLG
and MLG of several number of layers, and compare the results to the
polarization obtained using the Dirac cone approximation. It is very
important to notice that multi-layer graphene presents some spectral weight
at low energies as compared to graphene, which can be seen from the finite
contribution of $\mathrm{Im}\Pi(\mathbf{q},\omega)$ that appears to the left
of the big peak of the graphene polarizability at $\omega=v_{\mathrm{F}} q$ (%
$\sim 1$eV for the used parameters), in terms of the Fermi velocity near the
Dirac point, $v_{\mathrm{F}}=3at/2$. This is due to the low energy
parabolic-like dispersion of bilayer and multilayer graphene, as compared to
the linear dispersion of single layer graphene, and it can only be captured
by considering the inter-layer hopping contribution to the kinetic
Hamiltonian Eq. (\ref{Eq:H-interlayer}). Furthermore, the spectrum presents
a series of peaks for $\omega\approx v_{\mathrm{F}} q$, the number of which
depends on the number of layers. This is due to the fact that as we increase
the number of coupled graphene planes, the number of bands available for
particle-hole excitations also grows leading to peaks at different energies
for a given wave-vector.\cite{ZM08}

The difference between SLG and MLG is also relevant in the low energy region
of the dielectric function, as it can be seen in Fig. \ref{Fig:SpectrumMLG}%
(b). In fact, the $\omega\rightarrow 0$ limit of $\mathrm{Re}~\varepsilon(%
\mathbf{q},\omega)$ calculated within the RPA grows with the number of
layers. Moreover, as we have discussed above, the zeroes of $\mathrm{Re}%
~\varepsilon(\mathbf{q},\omega)$ signal the position of collective
excitations in the system (plasmons). In Fig. \ref{Fig:SpectrumMLG}(b) we
see that $\mathrm{Re}~\varepsilon(\mathbf{q},\omega)$, for the small
wave-vector used, crosses 0 for MLG, revealing the existence of a solution
of Eq. (\ref{Eq:Plasmons}), but not so for SLG, as it was pointed out in
Ref. \onlinecite{SSP10}. However, we emphasize that the very existence of
solutions for the $\mathrm{Re}~\varepsilon(\mathbf{q},\omega)=0$ equation
for MLG does not imply the existence of long-lived plasmon modes. In fact,
as we have already discussed in Sec. \ref{Sec:SLG}, these modes disperse
within the continuum of particle-hole excitations [$\mathrm{Im}~\Pi(\mathbf{q%
},\omega_{pl})\ne 0$, where $\omega_{pl}$ is the solution of Eq. (\ref%
{Eq:Plasmons})], so they will be Landau damped and will decay into
electron-hole pairs with a damping given by Eq. (\ref{Eq:Damping}).
Furthermore, we remember that for a given wave-vector, the energy of the
mode is controlled by the background dielectric constant $\kappa$, as given
by Eq. (\ref{Eq:kappa}). For the systems under consideration, $\kappa$
changes between 1 (for SLG) and 2.4 (for graphite). The value of $\kappa$,
together with the form factor Eq. (\ref{Eq:FormFactor}) that takes into
account inter-layer Coulomb interaction, fix the position of the modes in
each case.

\begin{figure}[t]
\begin{center}
\mbox{
\includegraphics[width=4.0cm]{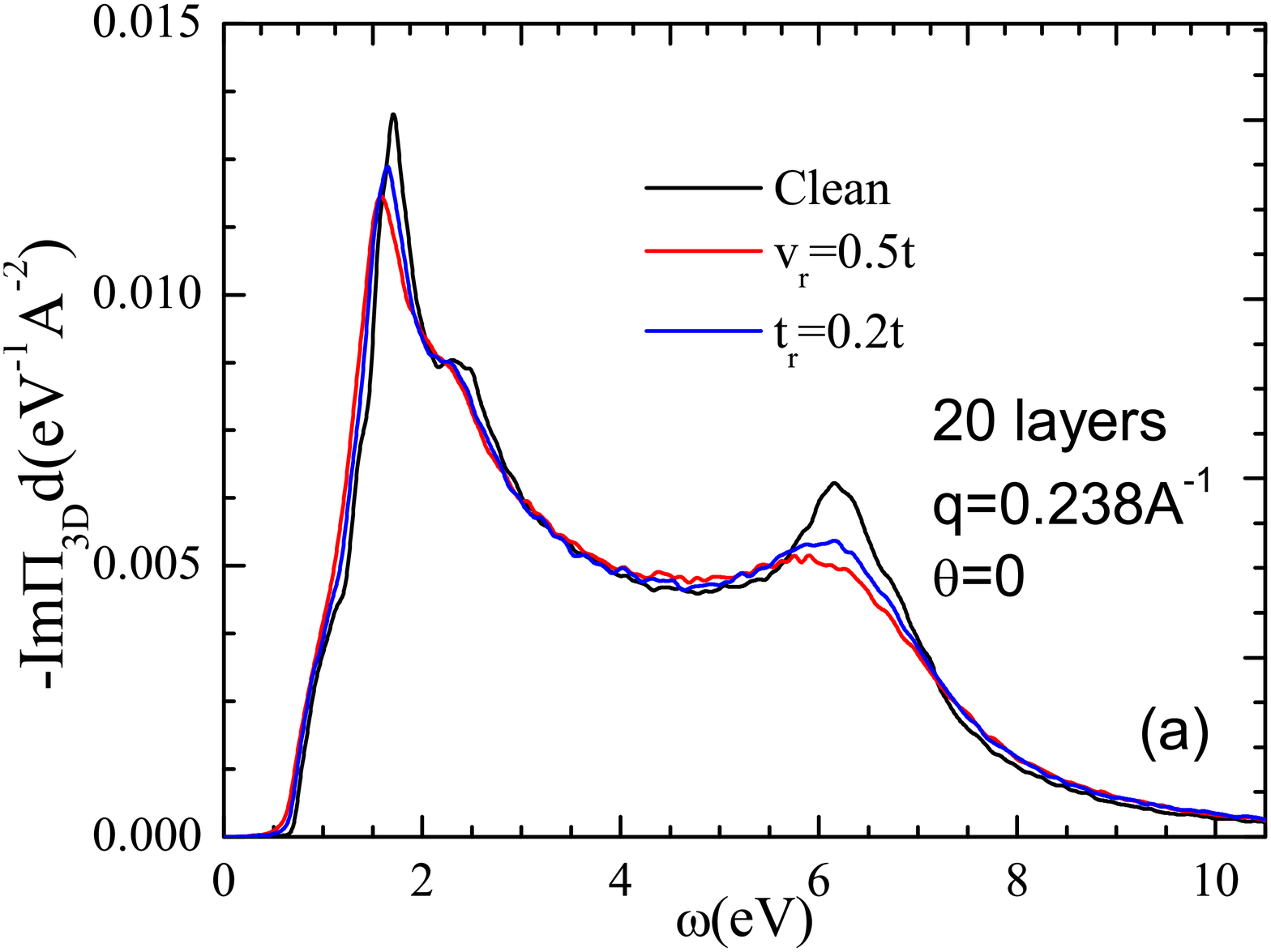}
\includegraphics[width=4.0cm]{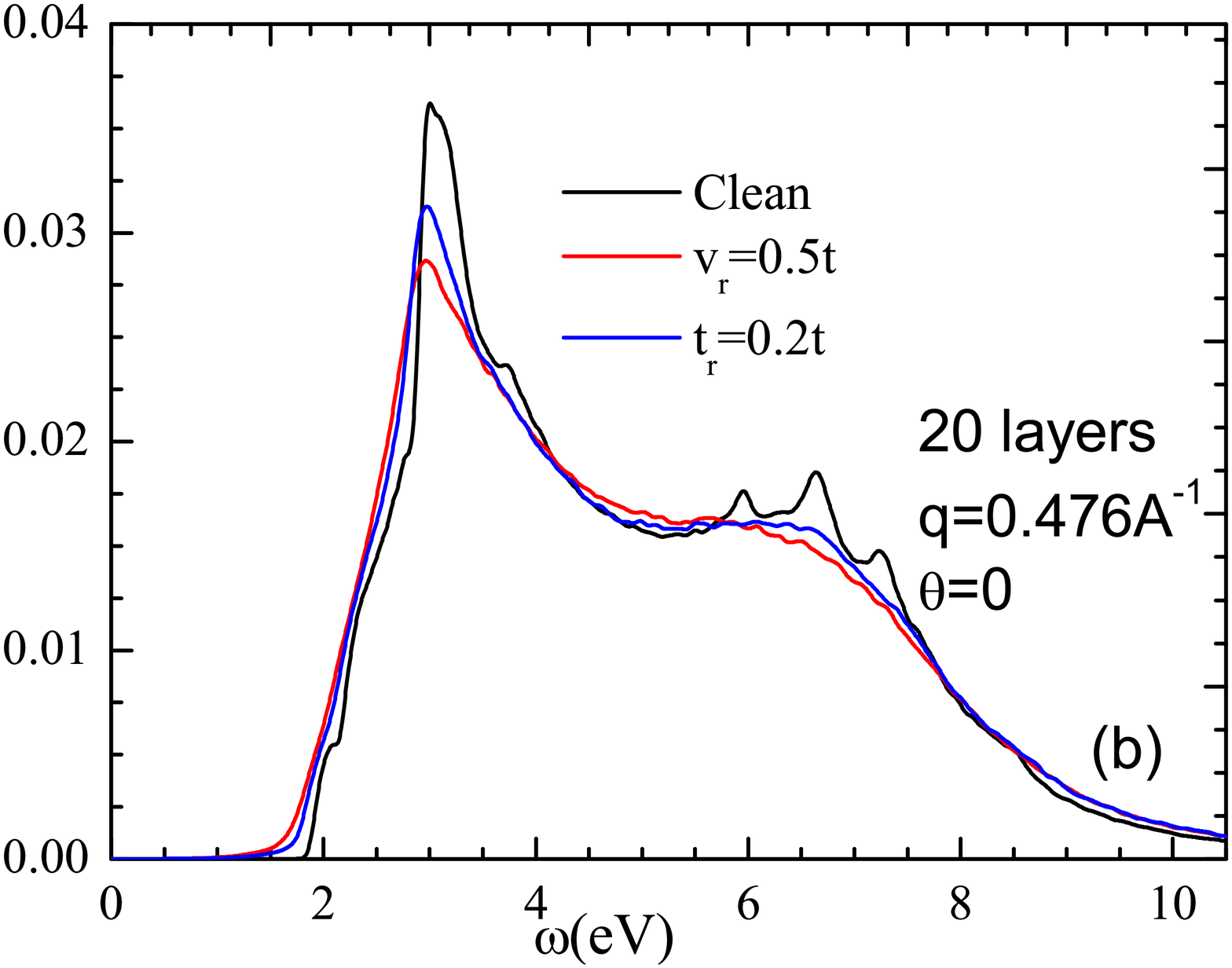}
} 
\mbox{
\includegraphics[width=4.0cm]{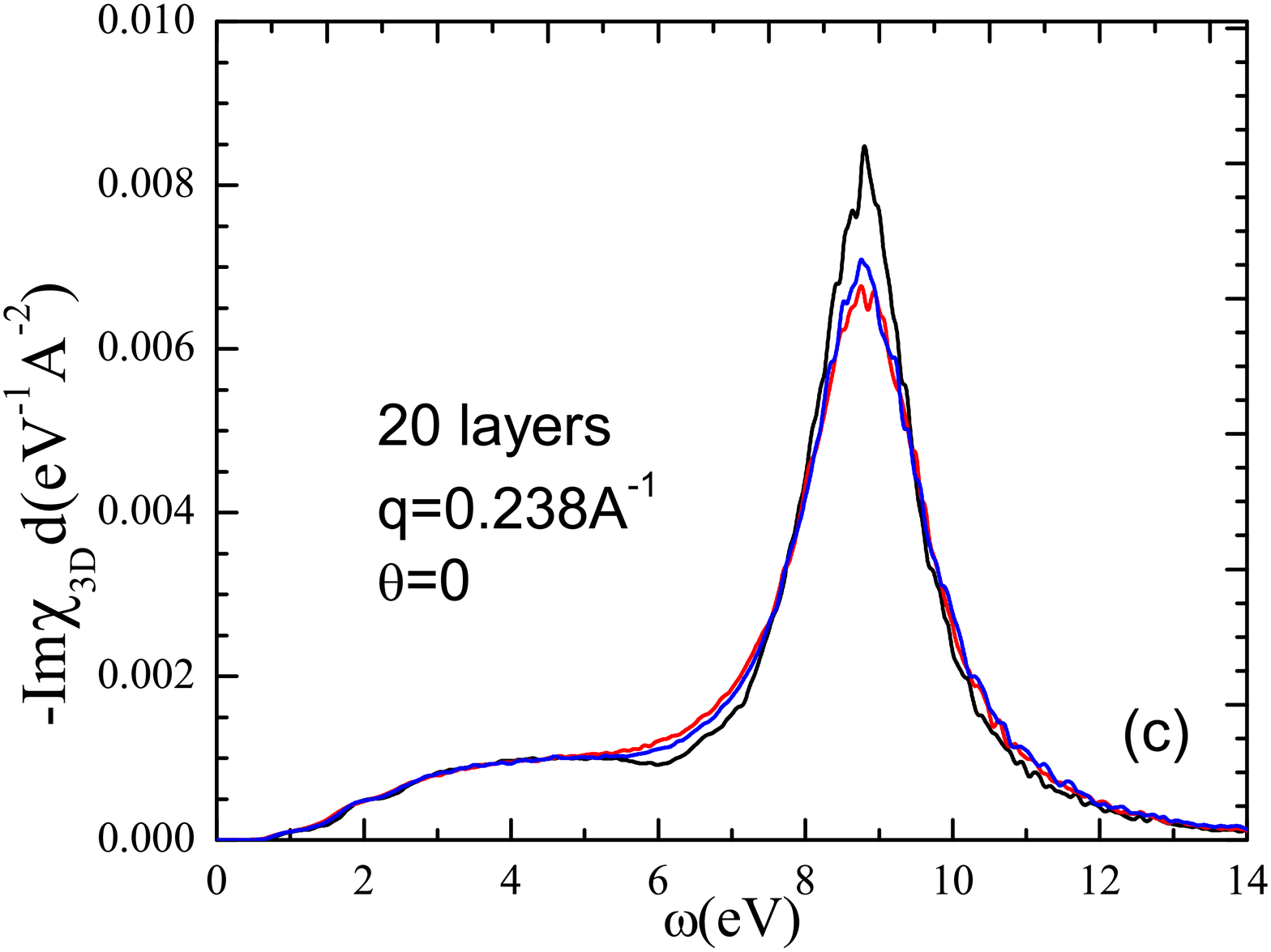}
\includegraphics[width=4.0cm]{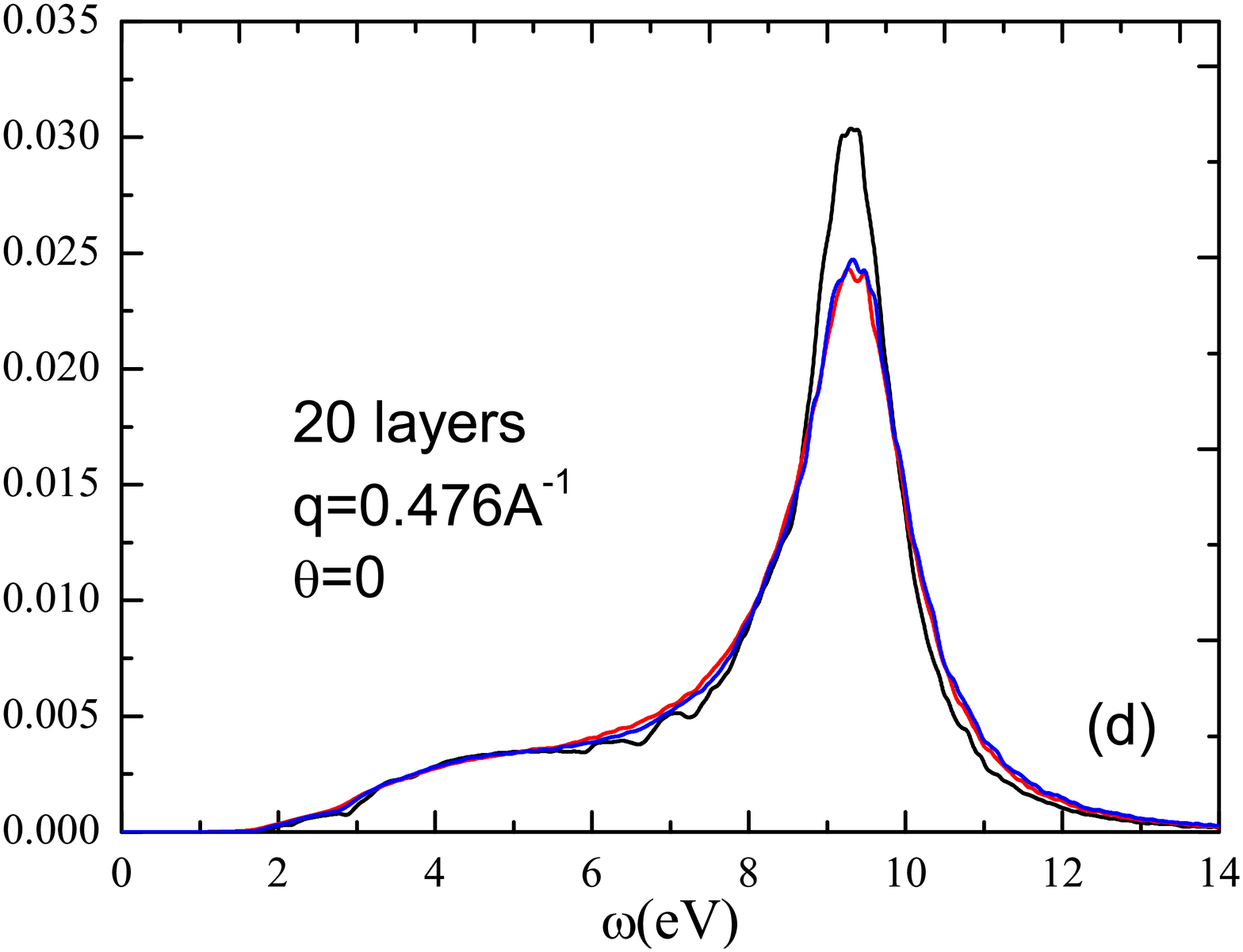}
}
\end{center}
\caption{{}(Color online) Dynamical polarization and response function of
ABA-stacked 20-layer graphene with disorders. The sample size of each layer
is $1600\times 1600$.}
\label{Fig:20layer}
\end{figure}

The effect of disorder in MLG is considered in Fig. \ref{Fig:20layer}, where
we show the polarization function of a 20-layer graphene system for
different kinds of disorder. As in the SLG, we find that disorder leads to a
slight redshift of the peaks of the non-interacting spectrum [Fig. \ref%
{Fig:20layer}(a) and (b)], together with a smearing of the peaks at $%
\omega\sim v_{\mathrm{F}} q$ and $\omega\sim 2t$. On the other hand, the
interacting polarization function presents a reduction of the intensity of
the plasmon peak due to disorder, as seen in Fig. \ref{Fig:20layer}(c) and
(d), also in analogy with the SLG case.

\section{Discussion and comparison to experimental results}

\label{Sec:ComparExp}

\begin{figure}[t]
\begin{center}
\includegraphics[width=6.5cm]{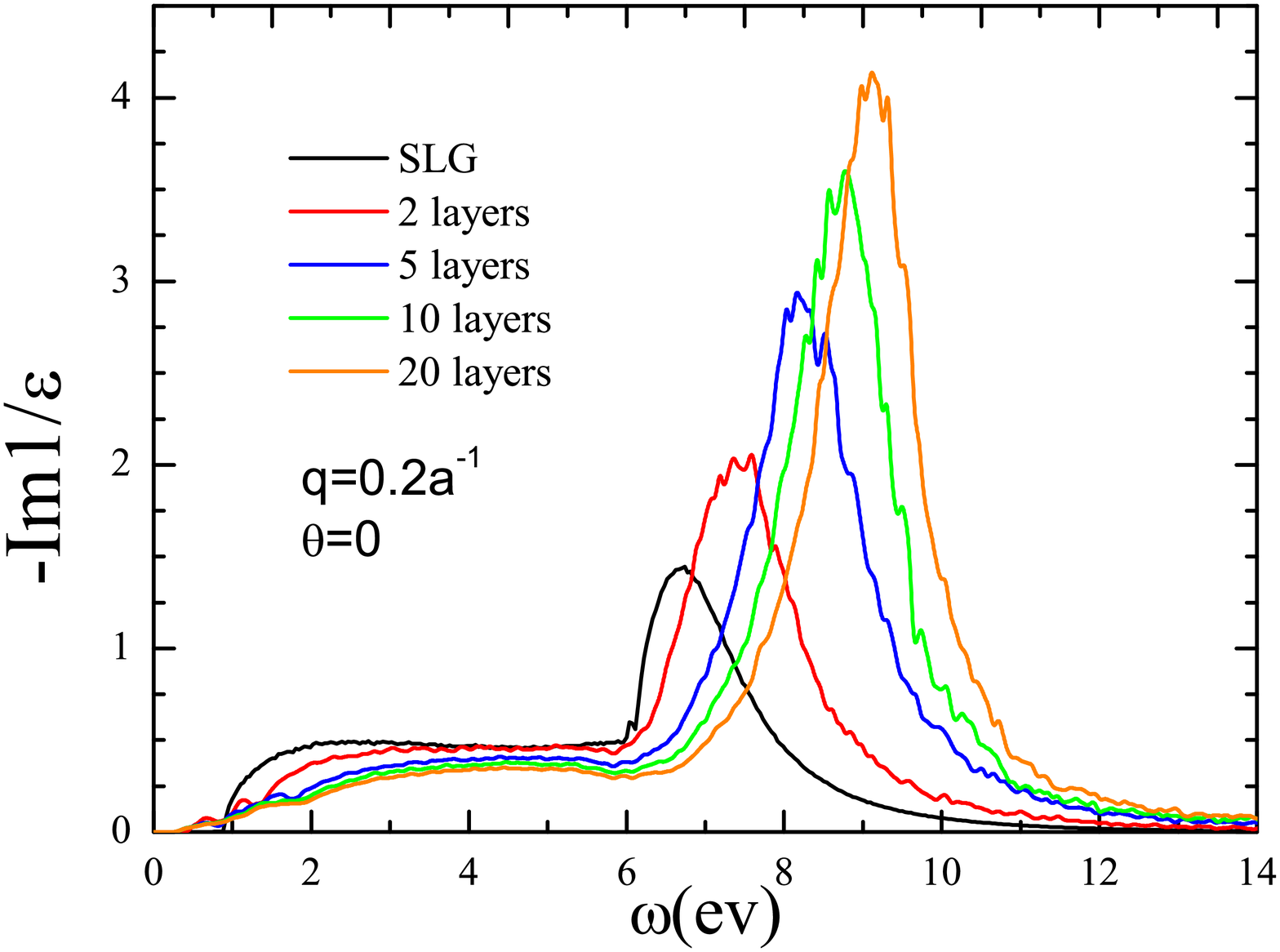}
\end{center}
\caption{(Color online) Loss function $-\mathrm{Im}~1/\protect\varepsilon (%
\mathbf{q},\protect\omega )$ for SLG and MLG, which is proportional to the
spectrum obtained by EEL experiments.\protect\cite{EB08} We have used the
same $q$ as in Fig. \protect\ref{Fig:SpectrumMLG}.}
\label{Fig:EELS}
\end{figure}

\begin{figure*}[t]
\begin{center}
\mbox{
\includegraphics[width=6.4cm]{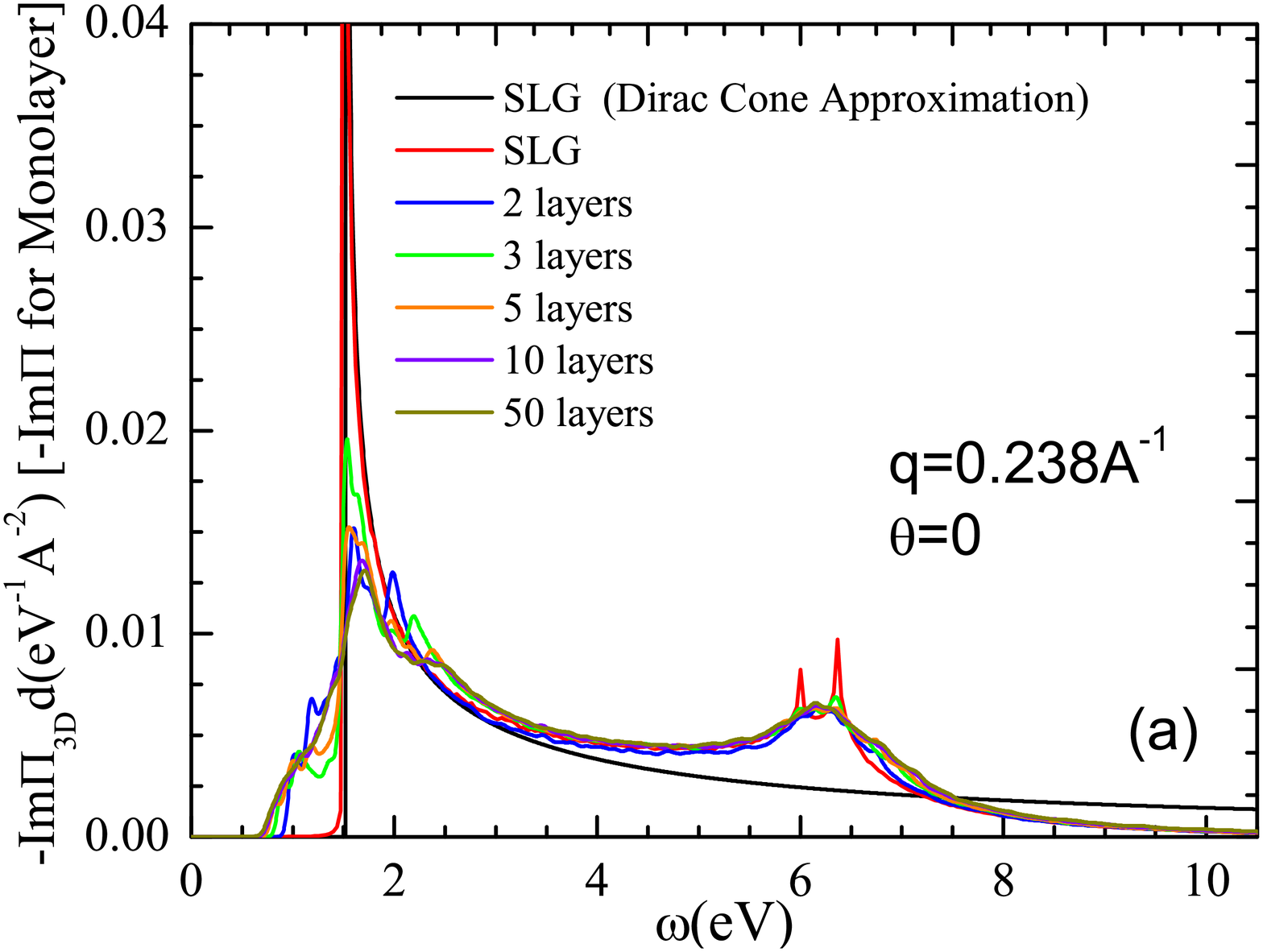}
\includegraphics[width=6.4cm]{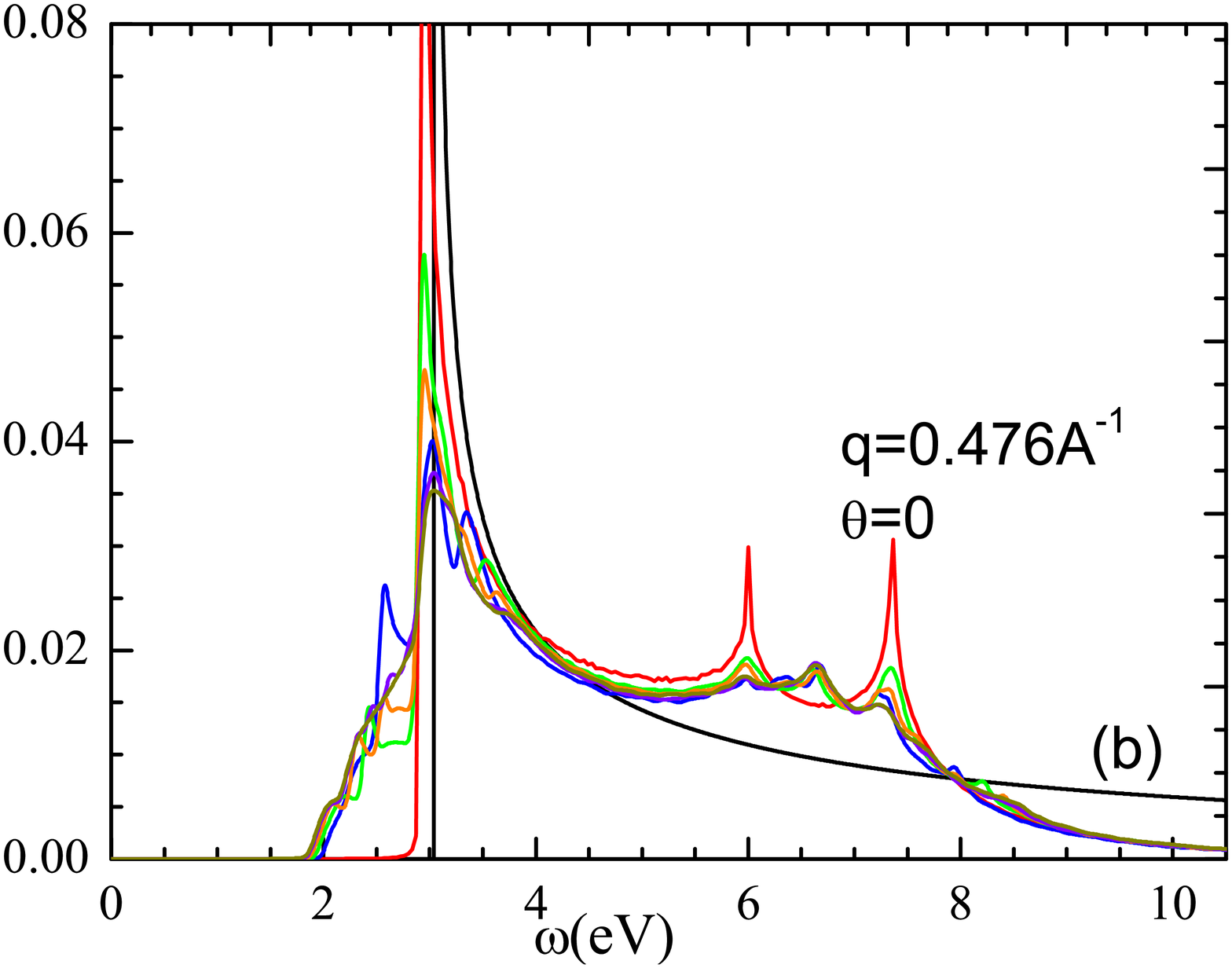}
} 
\mbox{
\includegraphics[width=6.4cm]{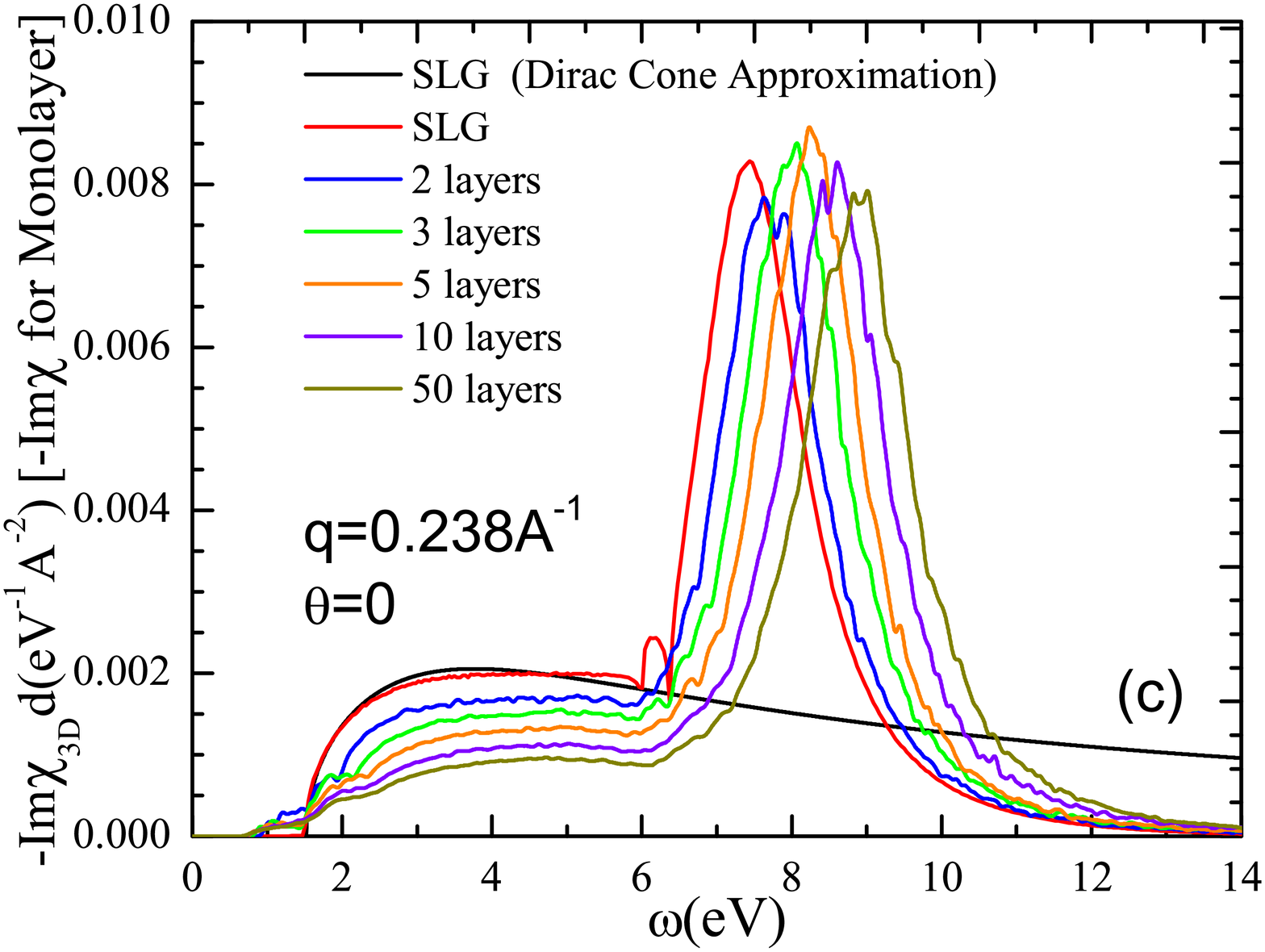}
\includegraphics[width=6.4cm]{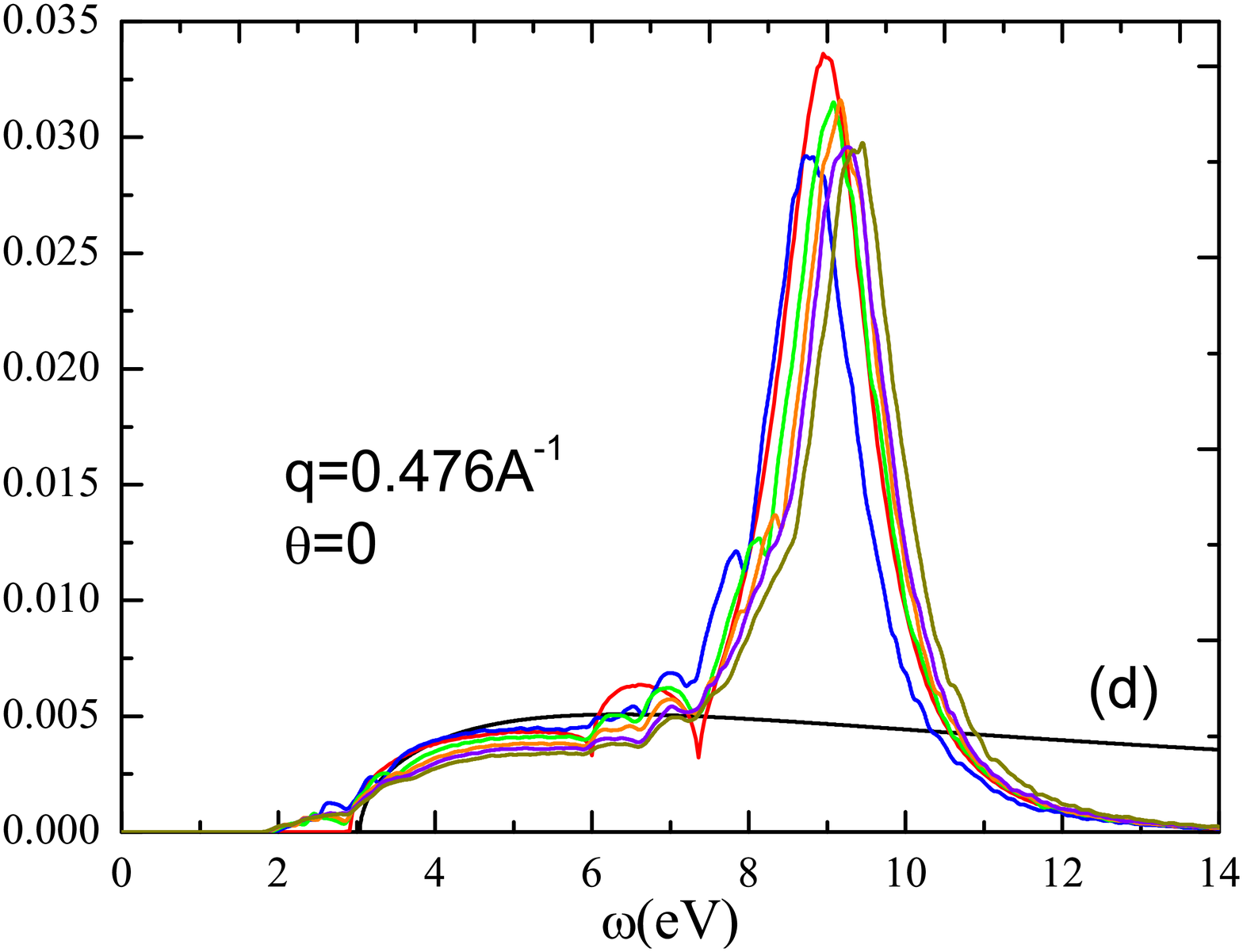}
}
\end{center}
\caption{(Color online) Non-interacting polarization function $\mathrm{Im}%
\Pi(\mathbf{q},\protect\omega)$ [plots (a) and (b)] and RPA response
function $\mathrm{Im}\protect\chi(\mathbf{q},\protect\omega)$ [plots (c) and
(d)] for SLG and MLG, for two different wave-vectors. The wave-vectors are
chosen as in Ref. \onlinecite{RA10}. }
\label{Fig:Abb}
\end{figure*}

In this section we compute quantities which are directly comparable to
recent experimental results on SLG and MLG. We start by calculating the loss
function $-\mathrm{Im}~1/\varepsilon(\mathbf{q},\omega)$, which is
proportional to the spectrum measured by EELS. Our results, shown in Fig. %
\ref{Fig:EELS}, are in good agreement with the experimental data of Ref. %
\onlinecite{EB08}: as in the experiments, we observe a redshift of the
plasmon peak as one decrease the number of layers, as well as an increase of
the intensity with the number of layers. Notice that, due to finite size
effects, there is an infra-red cutoff for the wave-vectors used in our
calculations which prevents to reach the long wavelength limit. In Fig. \ref%
{Fig:EELS} we show the results for the smallest wave-vector available, and
we emphasize that the peaks will be further shifted to the left for smaller
values of $q$. A further redshift of the peaks would be obtained beyond RPA,
as it has been reported for single- and bilayer graphene, where excitonic
effects have been included.\cite{YL09}

We have also used our method to study the IXS experiments of Reed \textit{et
al}.\cite{RA10} In Fig. \ref{Fig:Abb}(a)-(b) we plot the imaginary part of
the non-interacting polarization function for SLG and MLG, for two values of 
$q$ similar to the ones used in Ref. \onlinecite{RA10}. As we have discussed
in Sec. \ref{Sec:MLG}, inter-layer hopping leads to a finite contribution to
the spectral weight in the low energy region of MLG as compared to the SLG
spectrum. Notice that the number of peaks at this energy $\omega \approx v_{%
\mathrm{F}} q$ scales with the number of accessible bands and therefore,
with the number of layers. We emphasize that this effect is not included by
the usually employed approximation of considering MLG as a series of
single-layers of graphene, only coupled via direct Coulomb interaction.\cite%
{RA10} Without the possibility of inter-layer hopping, the polarization
function of graphene and graphite are, apart from some multiplicative
factor, the same. As we have seen in Sec. \ref{Sec:MLG}, this simplification
does not capture the low energy part of the spectrum, with some finite
spectral weight due to low energy inter-band transitions between
parabolic-like bands.

At an energy of the order of $\omega\approx 2t$ one observes the peak due to
transitions between electrons from the Van Hove singularity of the occupied
band to the singularity of the empty band. For SLG, the peak is split into
two peaks if the wave-vector points in the $\Gamma$-K direction (as it is
the case here), the separation of which increases with the modulus $q=q_x$.
However the amplitude of these peaks is highly suppressed from SLG to MLG.
Finally, one observes in Fig. \ref{Fig:Abb}(b) that for higher values of $q$%
, as the one used here, there is a redshift of the peak of $\mathrm{Im}~\Pi(%
\mathbf{q},\omega)$ at the energy $\omega\approx v_{\mathrm{F}} q$ with
respect to the Dirac cone approximation. This is due to trigonal warping
effects, which are beyond the continuum approximation. Summarizing, we find
two effects that lead to a global contribution to the polarization at low
energies: one is the contribution to the spectral weight due to inter-layer
hopping in MLG, and the other is the redshift of the peaks at $\omega\approx
v_{\mathrm{F}} q$ due to trigonal warping effects. Notice that, because we
are studying the \textit{non-interacting} polarization function, no
excitonic effects are present in the results of Fig. \ref{Fig:Abb}(a)-(b).

Once the polarization function $\Pi(\mathbf{q},\omega)$ is known, we compute
the response function $\chi(\mathbf{q},\omega)$ at the RPA level, as shown
in Fig. \ref{Fig:Abb}(c)-(d). Again, we find a redshift of the position of
the peaks as we decrease the number of layers. The different position of the
peaks is due to the different contribution of inter-layer electron-electron
interaction for each case, as well as to the different value of $\kappa$ as
given by Eq. (\ref{Eq:kappa}). Our results agree reasonably well with those
of Ref. \onlinecite{RA10}.

\begin{figure}[t]
\begin{center}
\includegraphics[width=6.5cm]{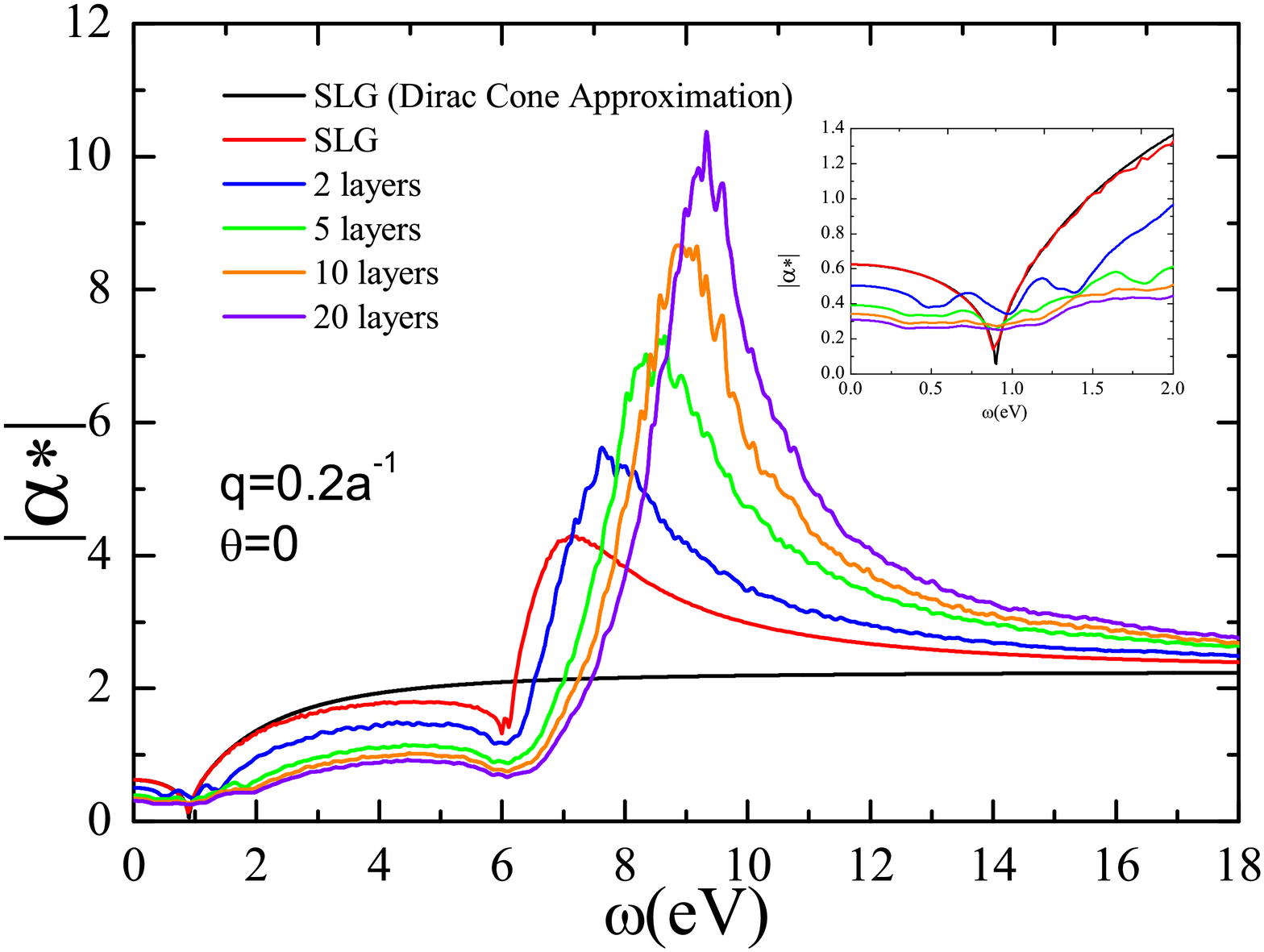}
\end{center}
\caption{(Color online) Modulus of the screened fine structure constant $|%
\protect\alpha^{*}|$, calculated from Eq. (\protect\ref{Eq:alpha}), for SLG
and MLG of a different number of layers. The inset is a zoom for the more
experimentally relevant $\protect\omega \rightarrow 0$ region of the
spectrum (see text). $|\protect\alpha^*|\approx 0.6$ for SLG, whereas this
value is highly reduced for MLG: $|\protect\alpha^*|\approx 0.3$ for a
20-layers sample in our numerical calculation, which has a behavior very
similar to graphite.}
\label{Fig:alpha}
\end{figure}

Finally, we calculate the renormalization of the fine structure constant $%
\alpha=e^2/v_{\mathrm{F}}$ due to dynamic screening associated to the
inter-band transitions from the valence band. For this, and in analogy with
Ref. \onlinecite{RA10}, we define 
\begin{equation}  \label{Eq:alpha}
\alpha^*(\mathbf{q},\omega)=\frac{\alpha}{\varepsilon(\mathbf{q},\omega)}
\end{equation}
The results for the modulus $|\alpha^*|$ for SLG and MLG are shown in Fig. %
\ref{Fig:alpha}. In this plot we have used the value of the Fermi velocity
valid near the Dirac point, i.e., $v_{\mathrm{F}}=3at/2$. Therefore, we
emphasize that these results should be reliable only at low energies. At $%
\omega \rightarrow 0$ and for the smallest wave-vector we can access ($%
q=0.2a^{-1}$), RPA predicts $|\alpha^*|\approx 0.6$ for SLG, which is
considerably higher than the value estimated in Ref. \onlinecite{RA10}: $%
|\alpha^*|\approx 0.15$. However, the results that we obtain for MLG are
much closer to this value: we find that $|\alpha^*|\approx 0.3$ for
graphite, only slightly higher (a factor of 2) than the experimental results
of Ref. \onlinecite{RA10}, which are actually obtained from graphite.

\section{Conclusions}

\label{Sec:Conclusions}

In conclusion, we have studied the excitation spectrum of single- and
multi-layer graphene using a full $\pi$-band tight-binding model in the
random phase approximation. We have found that, for MLG, the consideration
of inter-layer hopping is very important to properly capture the low energy
region ($\omega\sim v_{\mathrm{F}} q$) of the spectrum. This, together with
trigonal warping effects, lead to a finite contribution to the spectral
weight at low energies as well as a redshift of the peaks with respect to
the Dirac cone approximation. We have also studied the high energy plasmons
which are present in the spectrum of SLG and MLG at an energy of the order
of $\omega\sim 2t\approx 6$eV and which are associated to the enhanced DOS
at the Van Hove singularities of the $\pi$-bands. The energy of the $\pi$%
-plasmon depend also on the orientation between adjacent layers, and we find
that, for a given wave-vector, the energy of the mode for ABC-stacked MLG is
redshifted with respect to the corresponding energy of ABA ordering. This
difference is higher as we increase the number of graphene layers of the
system.

The effect of disorder has been considered by the inclusion of a random
on-site potential and by a renormalization of the nearest neighbor hopping.
Both kinds of disorder lead to a redshift of the $\omega\approx v_{\mathrm{F}%
} q$ and $\omega\approx 2t$ peaks of the non-interacting excitation spectrum
and to a smearing of the Van Hove singularities. The position of the $\pi$%
-plasmons is unaffected by disorder, although the height of the absorption
peaks is reduced as compared to the clean limit.

Finally, we have compared our results to some recent experiments. Our
calculations for the loss function $\mathrm{Im}~1/\varepsilon (\mathbf{q}%
,\omega )$ show a redshift of the SLG mode with respect to graphite, and
compare reasonably well with experimental EELS data.\cite{EB08,GG08}
Furthermore, we also obtain good agreement with the IXS results for the
response function obtained in Ref. \onlinecite{RA10}. We obtain a static
dielectric function which grows with the number of layers of the system. In
the long wavelength and $\omega \rightarrow 0$ limit, the dynamically
screened fine structure constant is found to be highly reduced from graphene
to graphite. The value that we find for a MLG in the RPA, without
considering any excitonic effects, is about two times larger than the one
estimated in Ref. \onlinecite{RA10} for graphene. More accurate results
could be obtained going beyond single-band RPA,\cite{SK11} which is beyond
the scope of this work.

\section{Acknowledgement}

The support by the Stichting Fundamenteel Onderzoek der Materie (FOM) and
the Netherlands National Computing Facilities foundation (NCF) are
acknowledged. We thank the EU-India FP-7 collaboration under MONAMI.

\bibliographystyle{apsrev4-1}
\bibliography{BibliogrGrafeno}

\newcommand{\npb}{Nucl. Phys.}\newcommand{\adv}{Adv.
  Phys.}\newcommand{\epl}{Europhys. Lett.}
\begin{thebibliography}{45}%
\makeatletter
\providecommand \@ifxundefined [1]{%
 \@ifx{#1\undefined}
}%
\providecommand \@ifnum [1]{%
 \ifnum #1\expandafter \@firstoftwo
 \else \expandafter \@secondoftwo
 \fi
}%
\providecommand \@ifx [1]{%
 \ifx #1\expandafter \@firstoftwo
 \else \expandafter \@secondoftwo
 \fi
}%
\providecommand \natexlab [1]{#1}%
\providecommand \enquote  [1]{``#1''}%
\providecommand \bibnamefont  [1]{#1}%
\providecommand \bibfnamefont [1]{#1}%
\providecommand \citenamefont [1]{#1}%
\providecommand \href@noop [0]{\@secondoftwo}%
\providecommand \href [0]{\begingroup \@sanitize@url \@href}%
\providecommand \@href[1]{\@@startlink{#1}\@@href}%
\providecommand \@@href[1]{\endgroup#1\@@endlink}%
\providecommand \@sanitize@url [0]{\catcode `\\12\catcode `\$12\catcode
  `\&12\catcode `\#12\catcode `\^12\catcode `\_12\catcode `\%12\relax}%
\providecommand \@@startlink[1]{}%
\providecommand \@@endlink[0]{}%
\providecommand \url  [0]{\begingroup\@sanitize@url \@url }%
\providecommand \@url [1]{\endgroup\@href {#1}{\urlprefix }}%
\providecommand \urlprefix  [0]{URL }%
\providecommand \Eprint [0]{\href }%
\providecommand \doibase [0]{http://dx.doi.org/}%
\providecommand \selectlanguage [0]{\@gobble}%
\providecommand \bibinfo  [0]{\@secondoftwo}%
\providecommand \bibfield  [0]{\@secondoftwo}%
\providecommand \translation [1]{[#1]}%
\providecommand \BibitemOpen [0]{}%
\providecommand \bibitemStop [0]{}%
\providecommand \bibitemNoStop [0]{.\EOS\space}%
\providecommand \EOS [0]{\spacefactor3000\relax}%
\providecommand \BibitemShut  [1]{\csname bibitem#1\endcsname}%
\let\auto@bib@innerbib\@empty
\bibitem [{\citenamefont {{\text For a recent review on electron-electron
  interaction in graphene, see}}\ \emph {et~al.}(2010)\citenamefont {{\text For
  a recent review on electron-electron interaction in graphene, see}},
  \citenamefont {Kotov}, \citenamefont {Uchoa}, \citenamefont {Pereira},
  \citenamefont {Castro-Neto},\ and\ \citenamefont {Guinea}}]{KG10}%
  \BibitemOpen
  \bibfield  {author} {\bibinfo {author} {\bibnamefont {{\text For a recent
  review on electron-electron interaction in graphene, see}}}, \bibinfo
  {author} {\bibfnamefont {V.}~\bibnamefont {Kotov}}, \bibinfo {author}
  {\bibfnamefont {B.}~\bibnamefont {Uchoa}}, \bibinfo {author} {\bibfnamefont
  {V.~M.}\ \bibnamefont {Pereira}}, \bibinfo {author} {\bibfnamefont {A.~H.}\
  \bibnamefont {Castro-Neto}}, \ and\ \bibinfo {author} {\bibfnamefont
  {F.}~\bibnamefont {Guinea}},\ }\href@noop {} {\  (\bibinfo {year} {2010})},\
  \Eprint {http://arxiv.org/abs/arXiv:1012.3484} {arXiv:1012.3484} \BibitemShut
  {NoStop}%
\bibitem [{\citenamefont {Kramberger}\ \emph {et~al.}(2008)\citenamefont
  {Kramberger}, \citenamefont {Hambach}, \citenamefont {Giorgetti},
  \citenamefont {R\"ummeli}, \citenamefont {Knupfer}, \citenamefont {Fink},
  \citenamefont {B\"uchner}, \citenamefont {Reining}, \citenamefont
  {Einarsson}, \citenamefont {Maruyama}, \citenamefont {Sottile}, \citenamefont
  {Hannewald}, \citenamefont {Olevano}, \citenamefont {Marinopoulos},\ and\
  \citenamefont {Pichler}}]{KP08}%
  \BibitemOpen
  \bibfield  {author} {\bibinfo {author} {\bibfnamefont {C.}~\bibnamefont
  {Kramberger}}, \bibinfo {author} {\bibfnamefont {R.}~\bibnamefont {Hambach}},
  \bibinfo {author} {\bibfnamefont {C.}~\bibnamefont {Giorgetti}}, \bibinfo
  {author} {\bibfnamefont {M.~H.}\ \bibnamefont {R\"ummeli}}, \bibinfo {author}
  {\bibfnamefont {M.}~\bibnamefont {Knupfer}}, \bibinfo {author} {\bibfnamefont
  {J.}~\bibnamefont {Fink}}, \bibinfo {author} {\bibfnamefont {B.}~\bibnamefont
  {B\"uchner}}, \bibinfo {author} {\bibfnamefont {L.}~\bibnamefont {Reining}},
  \bibinfo {author} {\bibfnamefont {E.}~\bibnamefont {Einarsson}}, \bibinfo
  {author} {\bibfnamefont {S.}~\bibnamefont {Maruyama}}, \bibinfo {author}
  {\bibfnamefont {F.}~\bibnamefont {Sottile}}, \bibinfo {author} {\bibfnamefont
  {K.}~\bibnamefont {Hannewald}}, \bibinfo {author} {\bibfnamefont
  {V.}~\bibnamefont {Olevano}}, \bibinfo {author} {\bibfnamefont {A.~G.}\
  \bibnamefont {Marinopoulos}}, \ and\ \bibinfo {author} {\bibfnamefont
  {T.}~\bibnamefont {Pichler}},\ }\href@noop {} {\bibfield  {journal} {\bibinfo
   {journal} {Phys. Rev. Lett.}\ }\textbf {\bibinfo {volume} {100}},\ \bibinfo
  {pages} {196803} (\bibinfo {year} {2008})}\BibitemShut {NoStop}%
\bibitem [{\citenamefont {Eberlein}\ \emph {et~al.}(2008)\citenamefont
  {Eberlein}, \citenamefont {Bangert}, \citenamefont {Nair}, \citenamefont
  {Jones}, \citenamefont {Gass}, \citenamefont {Bleloch}, \citenamefont
  {Novoselov}, \citenamefont {Geim},\ and\ \citenamefont {Briddon}}]{EB08}%
  \BibitemOpen
  \bibfield  {author} {\bibinfo {author} {\bibfnamefont {T.}~\bibnamefont
  {Eberlein}}, \bibinfo {author} {\bibfnamefont {U.}~\bibnamefont {Bangert}},
  \bibinfo {author} {\bibfnamefont {R.~R.}\ \bibnamefont {Nair}}, \bibinfo
  {author} {\bibfnamefont {R.}~\bibnamefont {Jones}}, \bibinfo {author}
  {\bibfnamefont {M.}~\bibnamefont {Gass}}, \bibinfo {author} {\bibfnamefont
  {A.~L.}\ \bibnamefont {Bleloch}}, \bibinfo {author} {\bibfnamefont {K.~S.}\
  \bibnamefont {Novoselov}}, \bibinfo {author} {\bibfnamefont {A.}~\bibnamefont
  {Geim}}, \ and\ \bibinfo {author} {\bibfnamefont {P.~R.}\ \bibnamefont
  {Briddon}},\ }\href@noop {} {\bibfield  {journal} {\bibinfo  {journal} {Phys.
  Rev. B}\ }\textbf {\bibinfo {volume} {77}},\ \bibinfo {pages} {233406}
  (\bibinfo {year} {2008})}\BibitemShut {NoStop}%
\bibitem [{\citenamefont {Gass}\ \emph {et~al.}(2008)\citenamefont {Gass},
  \citenamefont {Bangert}, \citenamefont {Bleloch}, \citenamefont {Wang},
  \citenamefont {Nair},\ and\ \citenamefont {Geim}}]{GG08}%
  \BibitemOpen
  \bibfield  {author} {\bibinfo {author} {\bibfnamefont {M.~H.}\ \bibnamefont
  {Gass}}, \bibinfo {author} {\bibfnamefont {U.}~\bibnamefont {Bangert}},
  \bibinfo {author} {\bibfnamefont {A.~L.}\ \bibnamefont {Bleloch}}, \bibinfo
  {author} {\bibfnamefont {P.}~\bibnamefont {Wang}}, \bibinfo {author}
  {\bibfnamefont {R.~R.}\ \bibnamefont {Nair}}, \ and\ \bibinfo {author}
  {\bibfnamefont {A.}~\bibnamefont {Geim}},\ }\href@noop {} {\bibfield
  {journal} {\bibinfo  {journal} {Nature Nanotechnology}\ }\textbf {\bibinfo
  {volume} {3}},\ \bibinfo {pages} {676} (\bibinfo {year} {2008})}\BibitemShut
  {NoStop}%
\bibitem [{\citenamefont {Reed}\ \emph {et~al.}(2010)\citenamefont {Reed},
  \citenamefont {Uchoa}, \citenamefont {Joe}, \citenamefont {Gan},
  \citenamefont {Casa}, \citenamefont {Fradkin},\ and\ \citenamefont
  {Abbamonte}}]{RA10}%
  \BibitemOpen
  \bibfield  {author} {\bibinfo {author} {\bibfnamefont {J.~P.}\ \bibnamefont
  {Reed}}, \bibinfo {author} {\bibfnamefont {B.}~\bibnamefont {Uchoa}},
  \bibinfo {author} {\bibfnamefont {Y.~I.}\ \bibnamefont {Joe}}, \bibinfo
  {author} {\bibfnamefont {Y.}~\bibnamefont {Gan}}, \bibinfo {author}
  {\bibfnamefont {D.}~\bibnamefont {Casa}}, \bibinfo {author} {\bibfnamefont
  {E.}~\bibnamefont {Fradkin}}, \ and\ \bibinfo {author} {\bibfnamefont
  {P.}~\bibnamefont {Abbamonte}},\ }\href@noop {} {\bibfield  {journal}
  {\bibinfo  {journal} {Science}\ }\textbf {\bibinfo {volume} {330}},\ \bibinfo
  {pages} {805} (\bibinfo {year} {2010})}\BibitemShut {NoStop}%
\bibitem [{\citenamefont {Mak}\ \emph {et~al.}(2011)\citenamefont {Mak},
  \citenamefont {Shan},\ and\ \citenamefont {Heinz}}]{MSH11}%
  \BibitemOpen
  \bibfield  {author} {\bibinfo {author} {\bibfnamefont {K.~F.}\ \bibnamefont
  {Mak}}, \bibinfo {author} {\bibfnamefont {J.}~\bibnamefont {Shan}}, \ and\
  \bibinfo {author} {\bibfnamefont {T.~F.}\ \bibnamefont {Heinz}},\ }\href@noop
  {} {\bibfield  {journal} {\bibinfo  {journal} {Phys. Rev. Lett.}\ }\textbf
  {\bibinfo {volume} {106}},\ \bibinfo {pages} {046401} (\bibinfo {year}
  {2011})}\BibitemShut {NoStop}%
\bibitem [{\citenamefont {Bostwick}\ \emph {et~al.}(2010)\citenamefont
  {Bostwick}, \citenamefont {Speck}, \citenamefont {Seyller}, \citenamefont
  {Horn}, \citenamefont {Polini}, \citenamefont {MacDonald},\ and\
  \citenamefont {Rothenberg}}]{BR10}%
  \BibitemOpen
  \bibfield  {author} {\bibinfo {author} {\bibfnamefont {A.}~\bibnamefont
  {Bostwick}}, \bibinfo {author} {\bibfnamefont {F.}~\bibnamefont {Speck}},
  \bibinfo {author} {\bibfnamefont {T.}~\bibnamefont {Seyller}}, \bibinfo
  {author} {\bibfnamefont {K.}~\bibnamefont {Horn}}, \bibinfo {author}
  {\bibfnamefont {M.}~\bibnamefont {Polini}}, \bibinfo {author} {\bibfnamefont
  {A.~H.}\ \bibnamefont {MacDonald}}, \ and\ \bibinfo {author} {\bibfnamefont
  {E.}~\bibnamefont {Rothenberg}},\ }\href@noop {} {\bibfield  {journal}
  {\bibinfo  {journal} {Science}\ }\textbf {\bibinfo {volume} {328}},\ \bibinfo
  {pages} {999} (\bibinfo {year} {2010})}\BibitemShut {NoStop}%
\bibitem [{\citenamefont {Liu}\ \emph {et~al.}(2008)\citenamefont {Liu},
  \citenamefont {Willis}, \citenamefont {Emtsev},\ and\ \citenamefont
  {Seyller}}]{LS08}%
  \BibitemOpen
  \bibfield  {author} {\bibinfo {author} {\bibfnamefont {Y.}~\bibnamefont
  {Liu}}, \bibinfo {author} {\bibfnamefont {R.~F.}\ \bibnamefont {Willis}},
  \bibinfo {author} {\bibfnamefont {K.~V.}\ \bibnamefont {Emtsev}}, \ and\
  \bibinfo {author} {\bibfnamefont {T.}~\bibnamefont {Seyller}},\ }\href@noop
  {} {\bibfield  {journal} {\bibinfo  {journal} {Phys. Rev. B}\ }\textbf
  {\bibinfo {volume} {78}},\ \bibinfo {pages} {201403} (\bibinfo {year}
  {2008})}\BibitemShut {NoStop}%
\bibitem [{\citenamefont {Shung}(1986)}]{S86}%
  \BibitemOpen
  \bibfield  {author} {\bibinfo {author} {\bibfnamefont {K.~W.~K.}\
  \bibnamefont {Shung}},\ }\href {\doibase 10.1103/PhysRevB.34.979} {\bibfield
  {journal} {\bibinfo  {journal} {Phys. Rev. B}\ }\textbf {\bibinfo {volume}
  {34}},\ \bibinfo {pages} {979} (\bibinfo {year} {1986})}\BibitemShut
  {NoStop}%
\bibitem [{\citenamefont {Wunsch}\ \emph {et~al.}(2006)\citenamefont {Wunsch},
  \citenamefont {Stauber}, \citenamefont {Sols},\ and\ \citenamefont
  {Guinea}}]{WSSG06}%
  \BibitemOpen
  \bibfield  {author} {\bibinfo {author} {\bibfnamefont {B.}~\bibnamefont
  {Wunsch}}, \bibinfo {author} {\bibfnamefont {T.}~\bibnamefont {Stauber}},
  \bibinfo {author} {\bibfnamefont {F.}~\bibnamefont {Sols}}, \ and\ \bibinfo
  {author} {\bibfnamefont {F.}~\bibnamefont {Guinea}},\ }\href@noop {}
  {\bibfield  {journal} {\bibinfo  {journal} {New Journal of Physics}\ }\textbf
  {\bibinfo {volume} {8}},\ \bibinfo {pages} {318} (\bibinfo {year}
  {2006})}\BibitemShut {NoStop}%
\bibitem [{\citenamefont {Hwang}\ and\ \citenamefont {Sarma}(2007)}]{HS07}%
  \BibitemOpen
  \bibfield  {author} {\bibinfo {author} {\bibfnamefont {E.~H.}\ \bibnamefont
  {Hwang}}\ and\ \bibinfo {author} {\bibfnamefont {S.~D.}\ \bibnamefont
  {Sarma}},\ }\href@noop {} {\bibfield  {journal} {\bibinfo  {journal} {Phys.
  Rev. B}\ }\textbf {\bibinfo {volume} {75}},\ \bibinfo {eid} {205418}
  (\bibinfo {year} {2007})}\BibitemShut {NoStop}%
\bibitem [{\citenamefont {Polini}\ \emph {et~al.}(2008)\citenamefont {Polini},
  \citenamefont {Asgari}, \citenamefont {Borghi}, \citenamefont {Barlas},
  \citenamefont {Pereg-Barnea},\ and\ \citenamefont {MacDonald}}]{PM08}%
  \BibitemOpen
  \bibfield  {author} {\bibinfo {author} {\bibfnamefont {M.}~\bibnamefont
  {Polini}}, \bibinfo {author} {\bibfnamefont {R.}~\bibnamefont {Asgari}},
  \bibinfo {author} {\bibfnamefont {G.}~\bibnamefont {Borghi}}, \bibinfo
  {author} {\bibfnamefont {Y.}~\bibnamefont {Barlas}}, \bibinfo {author}
  {\bibfnamefont {T.}~\bibnamefont {Pereg-Barnea}}, \ and\ \bibinfo {author}
  {\bibfnamefont {A.~H.}\ \bibnamefont {MacDonald}},\ }\href@noop {} {\bibfield
   {journal} {\bibinfo  {journal} {Phys. Rev. B}\ }\textbf {\bibinfo {volume}
  {77}},\ \bibinfo {eid} {081411} (\bibinfo {year} {2008})}\BibitemShut
  {NoStop}%
\bibitem [{\citenamefont {Pyatkovskiy}(2009)}]{P09}%
  \BibitemOpen
  \bibfield  {author} {\bibinfo {author} {\bibfnamefont {P.~K.}\ \bibnamefont
  {Pyatkovskiy}},\ }\href@noop {} {\bibfield  {journal} {\bibinfo  {journal}
  {Journal of Physics: Condensed Matter}\ }\textbf {\bibinfo {volume} {21}},\
  \bibinfo {pages} {025506} (\bibinfo {year} {2009})}\BibitemShut {NoStop}%
\bibitem [{\citenamefont {Gamayun}(2011)}]{G11b}%
  \BibitemOpen
  \bibfield  {author} {\bibinfo {author} {\bibfnamefont {O.~V.}\ \bibnamefont
  {Gamayun}},\ }\href@noop {} {\  (\bibinfo {year} {2011})},\ \Eprint
  {http://arxiv.org/abs/arXiv:1103.4597} {arXiv:1103.4597} \BibitemShut
  {NoStop}%
\bibitem [{\citenamefont {Tudorovskiy}\ and\ \citenamefont
  {Mikhailov}(2010)}]{TM10}%
  \BibitemOpen
  \bibfield  {author} {\bibinfo {author} {\bibfnamefont {T.}~\bibnamefont
  {Tudorovskiy}}\ and\ \bibinfo {author} {\bibfnamefont {S.~A.}\ \bibnamefont
  {Mikhailov}},\ }\href {\doibase 10.1103/PhysRevB.82.073411} {\bibfield
  {journal} {\bibinfo  {journal} {Phys. Rev. B}\ }\textbf {\bibinfo {volume}
  {82}},\ \bibinfo {pages} {073411} (\bibinfo {year} {2010})}\BibitemShut
  {NoStop}%
\bibitem [{\citenamefont {Rold\'an}\ \emph {et~al.}(2009)\citenamefont
  {Rold\'an}, \citenamefont {Fuchs},\ and\ \citenamefont {Goerbig}}]{RFG09}%
  \BibitemOpen
  \bibfield  {author} {\bibinfo {author} {\bibfnamefont {R.}~\bibnamefont
  {Rold\'an}}, \bibinfo {author} {\bibfnamefont {J.-N.}\ \bibnamefont {Fuchs}},
  \ and\ \bibinfo {author} {\bibfnamefont {M.~O.}\ \bibnamefont {Goerbig}},\
  }\href@noop {} {\bibfield  {journal} {\bibinfo  {journal} {Phys. Rev. B}\
  }\textbf {\bibinfo {volume} {80}},\ \bibinfo {pages} {085408} (\bibinfo
  {year} {2009})}\BibitemShut {NoStop}%
\bibitem [{\citenamefont {Gangadharaiah}\ \emph {et~al.}(2008)\citenamefont
  {Gangadharaiah}, \citenamefont {Farid},\ and\ \citenamefont
  {Mishchenko}}]{GFM08}%
  \BibitemOpen
  \bibfield  {author} {\bibinfo {author} {\bibfnamefont {S.}~\bibnamefont
  {Gangadharaiah}}, \bibinfo {author} {\bibfnamefont {A.~M.}\ \bibnamefont
  {Farid}}, \ and\ \bibinfo {author} {\bibfnamefont {E.~G.}\ \bibnamefont
  {Mishchenko}},\ }\href@noop {} {\bibfield  {journal} {\bibinfo  {journal}
  {Phys. Rev. Lett.}\ }\textbf {\bibinfo {volume} {100}},\ \bibinfo {pages}
  {166802} (\bibinfo {year} {2008})}\BibitemShut {NoStop}%
\bibitem [{\citenamefont {Khveshchenko}(2001)}]{K01}%
  \BibitemOpen
  \bibfield  {author} {\bibinfo {author} {\bibfnamefont {D.~V.}\ \bibnamefont
  {Khveshchenko}},\ }\href@noop {} {\bibfield  {journal} {\bibinfo  {journal}
  {Phys. Rev. Lett.}\ }\textbf {\bibinfo {volume} {87}},\ \bibinfo {pages}
  {246802} (\bibinfo {year} {2001})}\BibitemShut {NoStop}%
\bibitem [{\citenamefont {Aleiner}\ \emph {et~al.}(2007)\citenamefont
  {Aleiner}, \citenamefont {Kharzeev},\ and\ \citenamefont {Tsvelik}}]{AKT07}%
  \BibitemOpen
  \bibfield  {author} {\bibinfo {author} {\bibfnamefont {I.~L.}\ \bibnamefont
  {Aleiner}}, \bibinfo {author} {\bibfnamefont {D.~E.}\ \bibnamefont
  {Kharzeev}}, \ and\ \bibinfo {author} {\bibfnamefont {A.~M.}\ \bibnamefont
  {Tsvelik}},\ }\href@noop {} {\bibfield  {journal} {\bibinfo  {journal} {Phys.
  Rev. B}\ }\textbf {\bibinfo {volume} {76}},\ \bibinfo {pages} {195415}
  (\bibinfo {year} {2007})}\BibitemShut {NoStop}%
\bibitem [{\citenamefont {Gamayun}\ \emph {et~al.}(2009)\citenamefont
  {Gamayun}, \citenamefont {Gorbar},\ and\ \citenamefont {Gusynin}}]{GGG09}%
  \BibitemOpen
  \bibfield  {author} {\bibinfo {author} {\bibfnamefont {O.~V.}\ \bibnamefont
  {Gamayun}}, \bibinfo {author} {\bibfnamefont {E.~V.}\ \bibnamefont {Gorbar}},
  \ and\ \bibinfo {author} {\bibfnamefont {V.~P.}\ \bibnamefont {Gusynin}},\
  }\href@noop {} {\bibfield  {journal} {\bibinfo  {journal} {Phys. Rev. B}\
  }\textbf {\bibinfo {volume} {80}},\ \bibinfo {pages} {165429} (\bibinfo
  {year} {2009})}\BibitemShut {NoStop}%
\bibitem [{\citenamefont {Wang}\ \emph {et~al.}(2010)\citenamefont {Wang},
  \citenamefont {Fertig},\ and\ \citenamefont {Murthy}}]{WFM10}%
  \BibitemOpen
  \bibfield  {author} {\bibinfo {author} {\bibfnamefont {J.}~\bibnamefont
  {Wang}}, \bibinfo {author} {\bibfnamefont {H.~A.}\ \bibnamefont {Fertig}}, \
  and\ \bibinfo {author} {\bibfnamefont {G.}~\bibnamefont {Murthy}},\
  }\href@noop {} {\bibfield  {journal} {\bibinfo  {journal} {Phys. Rev. Lett.}\
  }\textbf {\bibinfo {volume} {104}},\ \bibinfo {pages} {186401} (\bibinfo
  {year} {2010})}\BibitemShut {NoStop}%
\bibitem [{\citenamefont {Gonz\'alez}(2010)}]{G10}%
  \BibitemOpen
  \bibfield  {author} {\bibinfo {author} {\bibfnamefont {J.}~\bibnamefont
  {Gonz\'alez}},\ }\href@noop {} {\bibfield  {journal} {\bibinfo  {journal}
  {Phys. Rev. B}\ }\textbf {\bibinfo {volume} {82}},\ \bibinfo {pages} {155404}
  (\bibinfo {year} {2010})}\BibitemShut {NoStop}%
\bibitem [{\citenamefont {Stauber}\ \emph {et~al.}(2010)\citenamefont
  {Stauber}, \citenamefont {Schliemann},\ and\ \citenamefont {Peres}}]{SSP10}%
  \BibitemOpen
  \bibfield  {author} {\bibinfo {author} {\bibfnamefont {T.}~\bibnamefont
  {Stauber}}, \bibinfo {author} {\bibfnamefont {J.}~\bibnamefont {Schliemann}},
  \ and\ \bibinfo {author} {\bibfnamefont {N.~M.~R.}\ \bibnamefont {Peres}},\
  }\href {\doibase 10.1103/PhysRevB.81.085409} {\bibfield  {journal} {\bibinfo
  {journal} {Phys. Rev. B}\ }\textbf {\bibinfo {volume} {81}},\ \bibinfo
  {pages} {085409} (\bibinfo {year} {2010})}\BibitemShut {NoStop}%
\bibitem [{\citenamefont {Hill}\ \emph {et~al.}(2009)\citenamefont {Hill},
  \citenamefont {Mikhailov},\ and\ \citenamefont {Ziegler}}]{HMZ09}%
  \BibitemOpen
  \bibfield  {author} {\bibinfo {author} {\bibfnamefont {A.}~\bibnamefont
  {Hill}}, \bibinfo {author} {\bibfnamefont {S.~A.}\ \bibnamefont {Mikhailov}},
  \ and\ \bibinfo {author} {\bibfnamefont {K.}~\bibnamefont {Ziegler}},\ }\href
  {http://stacks.iop.org/0295-5075/87/i=2/a=27005} {\bibfield  {journal}
  {\bibinfo  {journal} {Europhysics Letters}\ }\textbf {\bibinfo {volume}
  {87}},\ \bibinfo {pages} {27005} (\bibinfo {year} {2009})}\BibitemShut
  {NoStop}%
\bibitem [{\citenamefont {Yang}\ \emph {et~al.}(2009)\citenamefont {Yang},
  \citenamefont {Deslippe}, \citenamefont {Park}, \citenamefont {Cohen},\ and\
  \citenamefont {Louie}}]{YL09}%
  \BibitemOpen
  \bibfield  {author} {\bibinfo {author} {\bibfnamefont {L.}~\bibnamefont
  {Yang}}, \bibinfo {author} {\bibfnamefont {J.}~\bibnamefont {Deslippe}},
  \bibinfo {author} {\bibfnamefont {C.-H.}\ \bibnamefont {Park}}, \bibinfo
  {author} {\bibfnamefont {M.~L.}\ \bibnamefont {Cohen}}, \ and\ \bibinfo
  {author} {\bibfnamefont {S.~G.}\ \bibnamefont {Louie}},\ }\href@noop {}
  {\bibfield  {journal} {\bibinfo  {journal} {Phys. Rev. Lett.}\ }\textbf
  {\bibinfo {volume} {103}},\ \bibinfo {pages} {186802} (\bibinfo {year}
  {2009})}\BibitemShut {NoStop}%
\bibitem [{\citenamefont {Koshino}(2010)}]{K10}%
  \BibitemOpen
  \bibfield  {author} {\bibinfo {author} {\bibfnamefont {M.}~\bibnamefont
  {Koshino}},\ }\href@noop {} {\bibfield  {journal} {\bibinfo  {journal} {Phys.
  Rev. B}\ }\textbf {\bibinfo {volume} {81}},\ \bibinfo {pages} {125304}
  (\bibinfo {year} {2010})}\BibitemShut {NoStop}%
\bibitem [{\citenamefont {Castro-Neto}\ \emph {et~al.}(2009)\citenamefont
  {Castro-Neto}, \citenamefont {Guinea}, \citenamefont {Peres}, \citenamefont
  {Novoselov},\ and\ \citenamefont {Geim}}]{CG07}%
  \BibitemOpen
  \bibfield  {author} {\bibinfo {author} {\bibfnamefont {A.~H.}\ \bibnamefont
  {Castro-Neto}}, \bibinfo {author} {\bibfnamefont {F.}~\bibnamefont {Guinea}},
  \bibinfo {author} {\bibfnamefont {N.~M.~R.}\ \bibnamefont {Peres}}, \bibinfo
  {author} {\bibfnamefont {K.}~\bibnamefont {Novoselov}}, \ and\ \bibinfo
  {author} {\bibfnamefont {A.~K.}\ \bibnamefont {Geim}},\ }\href@noop {}
  {\bibfield  {journal} {\bibinfo  {journal} {Rev. Mod. Phys.}\ }\textbf
  {\bibinfo {volume} {81}},\ \bibinfo {pages} {109} (\bibinfo {year}
  {2009})}\BibitemShut {NoStop}%
\bibitem [{\citenamefont {Kubo}(1957)}]{K57}%
  \BibitemOpen
  \bibfield  {author} {\bibinfo {author} {\bibfnamefont {R.}~\bibnamefont
  {Kubo}},\ }\href@noop {} {\bibfield  {journal} {\bibinfo  {journal} {Journal
  of the Physical Society of Japan}\ }\textbf {\bibinfo {volume} {12}},\
  \bibinfo {pages} {570} (\bibinfo {year} {1957})}\BibitemShut {NoStop}%
\bibitem [{\citenamefont {Yuan}\ \emph
  {et~al.}(2010{\natexlab{a}})\citenamefont {Yuan}, \citenamefont {De~Raedt},\
  and\ \citenamefont {Katsnelson}}]{YRK10}%
  \BibitemOpen
  \bibfield  {author} {\bibinfo {author} {\bibfnamefont {S.}~\bibnamefont
  {Yuan}}, \bibinfo {author} {\bibfnamefont {H.}~\bibnamefont {De~Raedt}}, \
  and\ \bibinfo {author} {\bibfnamefont {M.~I.}\ \bibnamefont {Katsnelson}},\
  }\href@noop {} {\bibfield  {journal} {\bibinfo  {journal} {Phys. Rev. B}\
  }\textbf {\bibinfo {volume} {82}},\ \bibinfo {pages} {115448} (\bibinfo
  {year} {2010}{\natexlab{a}})}\BibitemShut {NoStop}%
\bibitem [{\citenamefont {Hams}\ and\ \citenamefont {De~Raedt}(2000)}]{HR00}%
  \BibitemOpen
  \bibfield  {author} {\bibinfo {author} {\bibfnamefont {A.}~\bibnamefont
  {Hams}}\ and\ \bibinfo {author} {\bibfnamefont {H.}~\bibnamefont
  {De~Raedt}},\ }\href {\doibase 10.1103/PhysRevE.62.4365} {\bibfield
  {journal} {\bibinfo  {journal} {Phys. Rev. E}\ }\textbf {\bibinfo {volume}
  {62}},\ \bibinfo {pages} {4365} (\bibinfo {year} {2000})}\BibitemShut
  {NoStop}%
\bibitem [{\citenamefont {Giuliani}\ and\ \citenamefont
  {Vignale}(2005)}]{GV05}%
  \BibitemOpen
  \bibfield  {author} {\bibinfo {author} {\bibfnamefont {G.~F.}\ \bibnamefont
  {Giuliani}}\ and\ \bibinfo {author} {\bibfnamefont {G.}~\bibnamefont
  {Vignale}},\ }\href@noop {} {\emph {\bibinfo {title} {Quatum Theory of the
  Electron Liquid}}}\ (\bibinfo  {publisher} {CUP, Cambridge},\ \bibinfo {year}
  {2005})\BibitemShut {NoStop}%
\bibitem [{\citenamefont {Stauber}\ \emph {et~al.}(2008)\citenamefont
  {Stauber}, \citenamefont {Peres},\ and\ \citenamefont {Geim}}]{SPG08}%
  \BibitemOpen
  \bibfield  {author} {\bibinfo {author} {\bibfnamefont {T.}~\bibnamefont
  {Stauber}}, \bibinfo {author} {\bibfnamefont {N.~M.~R.}\ \bibnamefont
  {Peres}}, \ and\ \bibinfo {author} {\bibfnamefont {A.~K.}\ \bibnamefont
  {Geim}},\ }\href@noop {} {\bibfield  {journal} {\bibinfo  {journal} {Phys.
  Rev. B}\ }\textbf {\bibinfo {volume} {78}},\ \bibinfo {pages} {085432}
  (\bibinfo {year} {2008})}\BibitemShut {NoStop}%
\bibitem [{\citenamefont {Shyu}\ and\ \citenamefont {Lin}(2000)}]{SL00}%
  \BibitemOpen
  \bibfield  {author} {\bibinfo {author} {\bibfnamefont {F.~L.}\ \bibnamefont
  {Shyu}}\ and\ \bibinfo {author} {\bibfnamefont {M.~F.}\ \bibnamefont {Lin}},\
  }\href@noop {} {\bibfield  {journal} {\bibinfo  {journal} {Phys. Rev. B}\
  }\textbf {\bibinfo {volume} {62}},\ \bibinfo {pages} {8508} (\bibinfo {year}
  {2000})}\BibitemShut {NoStop}%
\bibitem [{\citenamefont {Das~Sarma}\ and\ \citenamefont {Quinn}(1982)}]{SQ82}%
  \BibitemOpen
  \bibfield  {author} {\bibinfo {author} {\bibfnamefont {S.}~\bibnamefont
  {Das~Sarma}}\ and\ \bibinfo {author} {\bibfnamefont {J.~J.}\ \bibnamefont
  {Quinn}},\ }\href@noop {} {\bibfield  {journal} {\bibinfo  {journal} {Phys.
  Rev. B}\ }\textbf {\bibinfo {volume} {25}},\ \bibinfo {pages} {7603}
  (\bibinfo {year} {1982})}\BibitemShut {NoStop}%
\bibitem [{\citenamefont {Wehling}\ \emph {et~al.}(2011)\citenamefont
  {Wehling}, \citenamefont {\ifmmode \mbox{\c{S}}\else \c{S}\fi{}a\ifmmode
  \mbox{\c{s}}\else \c{s}\fi{}\ifmmode \imath \else \i
  \fi{}o\ifmmode~\breve{g}\else \u{g}\fi{}lu}, \citenamefont {Friedrich},
  \citenamefont {Lichtenstein}, \citenamefont {Katsnelson},\ and\ \citenamefont
  {Bl\"ugel}}]{WB11}%
  \BibitemOpen
  \bibfield  {author} {\bibinfo {author} {\bibfnamefont {T.~O.}\ \bibnamefont
  {Wehling}}, \bibinfo {author} {\bibfnamefont {E.}~\bibnamefont {\ifmmode
  \mbox{\c{S}}\else \c{S}\fi{}a\ifmmode \mbox{\c{s}}\else \c{s}\fi{}\ifmmode
  \imath \else \i \fi{}o\ifmmode~\breve{g}\else \u{g}\fi{}lu}}, \bibinfo
  {author} {\bibfnamefont {C.}~\bibnamefont {Friedrich}}, \bibinfo {author}
  {\bibfnamefont {A.~I.}\ \bibnamefont {Lichtenstein}}, \bibinfo {author}
  {\bibfnamefont {M.~I.}\ \bibnamefont {Katsnelson}}, \ and\ \bibinfo {author}
  {\bibfnamefont {S.}~\bibnamefont {Bl\"ugel}},\ }\href {\doibase
  10.1103/PhysRevLett.106.236805} {\bibfield  {journal} {\bibinfo  {journal}
  {Phys. Rev. Lett.}\ }\textbf {\bibinfo {volume} {106}},\ \bibinfo {pages}
  {236805} (\bibinfo {year} {2011})}\BibitemShut {NoStop}%
\bibitem [{\citenamefont {Wehling}\ \emph {et~al.}(2010)\citenamefont
  {Wehling}, \citenamefont {Yuan}, \citenamefont {Lichtenstein}, \citenamefont
  {Geim},\ and\ \citenamefont {Katsnelson}}]{WK10}%
  \BibitemOpen
  \bibfield  {author} {\bibinfo {author} {\bibfnamefont {T.~O.}\ \bibnamefont
  {Wehling}}, \bibinfo {author} {\bibfnamefont {S.}~\bibnamefont {Yuan}},
  \bibinfo {author} {\bibfnamefont {A.~I.}\ \bibnamefont {Lichtenstein}},
  \bibinfo {author} {\bibfnamefont {A.~K.}\ \bibnamefont {Geim}}, \ and\
  \bibinfo {author} {\bibfnamefont {M.~I.}\ \bibnamefont {Katsnelson}},\
  }\href@noop {} {\bibfield  {journal} {\bibinfo  {journal} {Phys. Rev. Lett.}\
  }\textbf {\bibinfo {volume} {105}},\ \bibinfo {pages} {056802} (\bibinfo
  {year} {2010})}\BibitemShut {NoStop}%
\bibitem [{\citenamefont {Yuan}\ \emph
  {et~al.}(2010{\natexlab{b}})\citenamefont {Yuan}, \citenamefont {De~Raedt},\
  and\ \citenamefont {Katsnelson}}]{YRK10b}%
  \BibitemOpen
  \bibfield  {author} {\bibinfo {author} {\bibfnamefont {S.}~\bibnamefont
  {Yuan}}, \bibinfo {author} {\bibfnamefont {H.}~\bibnamefont {De~Raedt}}, \
  and\ \bibinfo {author} {\bibfnamefont {M.~I.}\ \bibnamefont {Katsnelson}},\
  }\href@noop {} {\bibfield  {journal} {\bibinfo  {journal} {Phys. Rev. B}\
  }\textbf {\bibinfo {volume} {82}},\ \bibinfo {pages} {235409} (\bibinfo
  {year} {2010}{\natexlab{b}})}\BibitemShut {NoStop}%
\bibitem [{\citenamefont {Rold\'an}\ \emph {et~al.}(2010)\citenamefont
  {Rold\'an}, \citenamefont {Goerbig},\ and\ \citenamefont {Fuchs}}]{RGF10}%
  \BibitemOpen
  \bibfield  {author} {\bibinfo {author} {\bibfnamefont {R.}~\bibnamefont
  {Rold\'an}}, \bibinfo {author} {\bibfnamefont {M.~O.}\ \bibnamefont
  {Goerbig}}, \ and\ \bibinfo {author} {\bibfnamefont {J.-N.}\ \bibnamefont
  {Fuchs}},\ }\href@noop {} {\bibfield  {journal} {\bibinfo  {journal}
  {Semicond. Sci. Technol.}\ }\textbf {\bibinfo {volume} {25}},\ \bibinfo
  {pages} {034005} (\bibinfo {year} {2010})}\BibitemShut {NoStop}%
\bibitem [{\citenamefont {Stauber}\ and\ \citenamefont
  {G\'omez-Santos}(2010)}]{SG10}%
  \BibitemOpen
  \bibfield  {author} {\bibinfo {author} {\bibfnamefont {T.}~\bibnamefont
  {Stauber}}\ and\ \bibinfo {author} {\bibfnamefont {G.}~\bibnamefont
  {G\'omez-Santos}},\ }\href@noop {} {\bibfield  {journal} {\bibinfo  {journal}
  {Phys. Rev. B}\ }\textbf {\bibinfo {volume} {82}},\ \bibinfo {pages} {155412}
  (\bibinfo {year} {2010})}\BibitemShut {NoStop}%
\bibitem [{\citenamefont {Hobson}\ and\ \citenamefont
  {Nierenberg}(1953)}]{HN53}%
  \BibitemOpen
  \bibfield  {author} {\bibinfo {author} {\bibfnamefont {J.~P.}\ \bibnamefont
  {Hobson}}\ and\ \bibinfo {author} {\bibfnamefont {W.~A.}\ \bibnamefont
  {Nierenberg}},\ }\href@noop {} {\bibfield  {journal} {\bibinfo  {journal}
  {Phys. Rev.}\ }\textbf {\bibinfo {volume} {89}},\ \bibinfo {pages} {662}
  (\bibinfo {year} {1953})}\BibitemShut {NoStop}%
\bibitem [{\citenamefont {Adler}(1962)}]{A62}%
  \BibitemOpen
  \bibfield  {author} {\bibinfo {author} {\bibfnamefont {S.~L.}\ \bibnamefont
  {Adler}},\ }\href@noop {} {\bibfield  {journal} {\bibinfo  {journal} {Phys.
  Rev.}\ }\textbf {\bibinfo {volume} {126}},\ \bibinfo {pages} {413} (\bibinfo
  {year} {1962})}\BibitemShut {NoStop}%
\bibitem [{\citenamefont {Pellegrino}\ \emph {et~al.}(2010)\citenamefont
  {Pellegrino}, \citenamefont {Angilella},\ and\ \citenamefont
  {Pucci}}]{PAP10}%
  \BibitemOpen
  \bibfield  {author} {\bibinfo {author} {\bibfnamefont {F.~M.~D.}\
  \bibnamefont {Pellegrino}}, \bibinfo {author} {\bibfnamefont {G.~G.~N.}\
  \bibnamefont {Angilella}}, \ and\ \bibinfo {author} {\bibfnamefont
  {R.}~\bibnamefont {Pucci}},\ }\href@noop {} {\bibfield  {journal} {\bibinfo
  {journal} {Phys. Rev. B}\ }\textbf {\bibinfo {volume} {82}},\ \bibinfo
  {pages} {115434} (\bibinfo {year} {2010})}\BibitemShut {NoStop}%
\bibitem [{\citenamefont {Guinea}(2007)}]{G07}%
  \BibitemOpen
  \bibfield  {author} {\bibinfo {author} {\bibfnamefont {F.}~\bibnamefont
  {Guinea}},\ }\href@noop {} {\bibfield  {journal} {\bibinfo  {journal} {Phys.
  Rev. B}\ }\textbf {\bibinfo {volume} {75}},\ \bibinfo {pages} {235433}
  (\bibinfo {year} {2007})}\BibitemShut {NoStop}%
\bibitem [{\citenamefont {Zhang}\ \emph {et~al.}(2008)\citenamefont {Zhang},
  \citenamefont {Li}, \citenamefont {Basov}, \citenamefont {Fogler},
  \citenamefont {Hao},\ and\ \citenamefont {Martin}}]{ZM08}%
  \BibitemOpen
  \bibfield  {author} {\bibinfo {author} {\bibfnamefont {L.~M.}\ \bibnamefont
  {Zhang}}, \bibinfo {author} {\bibfnamefont {Z.~Q.}\ \bibnamefont {Li}},
  \bibinfo {author} {\bibfnamefont {D.~N.}\ \bibnamefont {Basov}}, \bibinfo
  {author} {\bibfnamefont {M.~M.}\ \bibnamefont {Fogler}}, \bibinfo {author}
  {\bibfnamefont {Z.}~\bibnamefont {Hao}}, \ and\ \bibinfo {author}
  {\bibfnamefont {M.~C.}\ \bibnamefont {Martin}},\ }\href@noop {} {\bibfield
  {journal} {\bibinfo  {journal} {Phys. Rev. B}\ }\textbf {\bibinfo {volume}
  {78}},\ \bibinfo {pages} {235408} (\bibinfo {year} {2008})}\BibitemShut
  {NoStop}%
\bibitem [{\citenamefont {van Schilfgaarde}\ and\ \citenamefont
  {Katsnelson}(2011)}]{SK11}%
  \BibitemOpen
  \bibfield  {author} {\bibinfo {author} {\bibfnamefont {M.}~\bibnamefont {van
  Schilfgaarde}}\ and\ \bibinfo {author} {\bibfnamefont {M.~I.}\ \bibnamefont
  {Katsnelson}},\ }\href@noop {} {\bibfield  {journal} {\bibinfo  {journal}
  {Phys. Rev. B}\ }\textbf {\bibinfo {volume} {83}},\ \bibinfo {pages} {081409}
  (\bibinfo {year} {2011})}\BibitemShut {NoStop}%
\end{thebibliography}%

\end{document}